\documentclass[a4paper,11pt]{article}

\usepackage{amsmath}
\usepackage{amssymb}
\usepackage{arrayjobx}
\usepackage[nosort]{cite}
\usepackage{subfigure}
\usepackage{graphicx}   
\usepackage{color} 
\usepackage{bbm}
\usepackage[bulletsep]{collref}
\usepackage[margin=2cm]{caption}

\usepackage[latin1]{inputenc} 
\usepackage[T1]{fontenc}

\usepackage[bookmarks=true,hyperfigures=true]{hyperref}
 \usepackage[a4paper,text={470pt,700pt},centering]{geometry}
\makeatletter
\let\old@makecaption=\@makecaption
\def\@makecaption{\small\old@makecaption}
\makeatother
\makeatletter
\newlength{\apb@width}
\newcommand{\autoparbox}[2][c]{\settowidth{\apb@width}{#2}\parbox[#1]{\apb@width}{#2}}
\newcommand{\includegraphicsbox}[2][]{\autoparbox{\includegraphics[#1]{#2}}}
\makeatother
\providecommand{\hypersetup}[1]{}

\providecommand{\texorpdfstring}[2]{#1}
\providecommand{\pdfbookmark}[3][]{}

\hypersetup{plainpages=false}
\hypersetup{pdfpagemode=UseOutlines}
\hypersetup{bookmarksnumbered=true}
\hypersetup{bookmarksopen=true}
\hypersetup{pdfstartview=FitH}
\hypersetup{colorlinks=false}
 \hypersetup{citebordercolor={.5 1 .5}}
 \hypersetup{urlbordercolor={.5 1 1}}
 \hypersetup{linkbordercolor={1 .7 .7}}
\hypersetup{citebordercolor={.6 .9 .6}}
\hypersetup{urlbordercolor={.7 .8 1}}
\hypersetup{linkbordercolor={1 .7 .7}}
\hypersetup{pdfborder={0 0 1 [3]}}

\numberwithin{equation}{section}
\newcommand{\unit}{\mathbbm{1}}

\newcommand{\ket}[1]{\mathopen{|}#1\mathclose{\rangle}}
\newcommand{\bra}[1]{\mathopen{\langle}#1\mathclose{|}}
\newcommand{\braket}[2]{\mathopen{\langle}#1|#2\mathclose{\rangle}}

\newcommand{\Tr}{\text{Tr}}

\ifx\genfrac\sdflkaj

\else

\fi

\newcommand{\nn}{\nonumber}

\makeatletter
\def\mr@ignsp#1 {\ifx\:#1\@empty\else #1\expandafter\mr@ignsp\fi}%
\newcommand{\multiref}[1]{\begingroup
\xdef\mr@no@sparg{\expandafter\mr@ignsp#1 \: }%
\def\mr@comma{}%
\@for\mr@refs:=\mr@no@sparg\do{\mr@comma\def\mr@comma{,}\ref{\mr@refs}}%
\endgroup}
\makeatother

\newcommand{\hypref}[2]{\ifx\href\asklfhas #2\else\href{#1}{#2}\fi}

\newcommand{\secref}[1]{Sec.~\multiref{#1}}

\newcommand{\appref}[1]{App.~\multiref{#1}}

\newcommand{\tabref}[1]{Tab.~\multiref{#1}}

\newcommand{\figref}[1]{Fig.~\multiref{#1}}
\renewcommand{\eqref}[1]{(\multiref{#1})}

\def\[{\begin{equation}}
\def\]{\end{equation}}
\def\<{\begin{eqnarray}}
\def\>{\end{eqnarray}}
\newcommand{\beq}{\begin{equation}}
\newcommand{\eeq}{\end{equation}}
\newcommand{\beqa}{\begin{eqnarray}}
\newcommand{\eeqa}{\end{eqnarray}}

\newcommand{\bc}[1]{{\color{blue} {}#1{}}}
\definecolor{purple}{rgb}{0.5,0,0.5}
\newcommand{\pc}[1]{{\color{purple} {}#1{}}}
\newcommand{\nc}[1]{{\color{black} {}#1{}}}

\usepackage{enumitem}

\begin{document}

\thispagestyle{empty}


\begin{flushright}\footnotesize
\texttt{
 TCDMATH 20-06\\
 SAGEX-20-15
 }
\end{flushright}
\vspace{1cm}

\begin{center}%
{\Large\textbf{\mathversion{bold}%
One-Loop Non-Planar Anomalous Dimensions\\
 in Super Yang-Mills Theory
}\par}%

\vspace{15mm}

\textrm{
Tristan McLoughlin, Raul Pereira and Anne Spiering }\vspace{8mm} \\
\textit{%
School of Mathematics \& Hamilton Mathematics Institute \\
 Trinity College Dublin \\
Dublin, Ireland 
} \\

  \texttt{\small $\{$tristan, raul,  spiering$ \}$@maths.tcd.ie\phantom{\ldots}} \\ \vspace{4mm}

\par\vspace{14mm}

\textbf{Abstract} \vspace{5mm}

\begin{minipage}{14cm}
We consider non-planar one-loop anomalous dimensions in maximally supersymmetric Yang-Mills theory and its marginally deformed analogues. Using the basis of Bethe states, we compute matrix elements of the dilatation operator and find compact expressions in terms of off-shell scalar products and hexagon-like functions. We then use non-degenerate quantum-mechanical perturbation theory to compute the leading  $1/N^2$ corrections to operator dimensions and as an example compute the large $R$-charge limit for two-excitation states through subleading order in the $R$-charge. Finally, we numerically study the distribution of level spacings for these theories and show that they transition from the Poisson distribution for integrable systems at infinite $N$ to the GOE Wigner-Dyson distribution for quantum chaotic systems at finite $N$.  
\end{minipage}

\end{center}

\newpage
\tableofcontents

\vspace{10mm}
\hrule
\vspace{5mm}


\section{Introduction}
The eigenvalue problem for the dilatation operator, $\mathfrak{D}$,  acting on the set of gauge-invariant local operators, $\mathcal{O}_i$, in $\mathcal{N}=4$ super Yang-Mills (SYM) theory,
\<
\mathfrak{D}\cdot \mathcal{O}_i =\Delta_i \mathcal{O}_i
\>
has been of continued interest due to its role as a proving ground for novel calculational techniques and because of its importance in the AdS/CFT correspondence. The operator dimensions, $\Delta_i =\Delta_i(g_{\rm YM}, N)$, are non-trivial functions of the coupling $g_{\rm YM}$ and $N$, the rank of the gauge group. In perturbation theory we can expand the dilatation operator in powers of the 't Hooft coupling $\lambda=g_{\rm YM}^2N$
\<
\mathfrak{D}=\sum_k g^{2k}\mathfrak D_{2k}~~~\text{with}~~~~g^2=\frac{\lambda}{16\pi^2}
\>
and at each order in $g^2$ we can further consider the large-$N$ expansion of the operator dimensions. A key development  \cite{Minahan:2002ve} was the insight that for the $\mathfrak{so}(6)$ sector of operators the one-loop, $\mathcal{O}(g^2)$, leading large-$N$ anomalous dimensions can be computed by means of an integrable spin chain.  Single-trace operators composed of $L$ scalar fields were identified with closed spin chains of length $L$ and the planar dilatation operator with a spin-chain Hamiltonian that can be diagonalised by use of the Bethe ansatz. This was subsequently extended to the full one-loop theory \cite{Beisert:2003yb} and to higher orders in perturbation theory \cite{Beisert:2003tq} as well as being observed at strong coupling \cite{Bena_2004}. This prompted a great deal of work and ultimately lead to non-perturbative results for the spectrum of planar anomalous dimensions first through the thermodynamic Bethe ansatz and subsequently by means of the Quantum Spectral Curve (QSC), see \cite{Arutyunov:2009ga, Beisert:2010jr, Bombardelli_2016, Gromov:2017blm} for reviews.

While less is known about non-planar anomalous dimensions there are a number of impressive perturbative results for specific operators. For example, twist-two operators at four loops were studied using standard Feynman diagramatics \cite{Velizhanin_2011, Velizhanin:2014zla} as well as twistor methods \cite{Fleury:2019ydf}, and the four-loop non-planar correction to the cusp anomalous dimension was computed in \cite{Henn:2019swt}. Additionally, 
the Hexagon formalism for correlation functions, \cite{Basso:2015zoa, Fleury_2017, Eden:2016xvg, Bargheer_2018}, provides an integrability-based method to studying non-planar  $\mathcal{N}=4$ SYM. The framework can be used to evaluate higher-point functions at higher genus and one can then in principle extract non-planar dimensions from OPE limits. A second approach, which was applied at tree-level in  \cite{Eden:2017ozn}, is to directly compute the non-planar two-point function by considering the four-point function with two operators taken to be the identity.
Moreover, there are results for the complete dilatation operator including its non-planar parts which was found at one-loop for the $\mathfrak{so}(6)$ sector in \cite{Beisert:2002bb, Constable:2002hw, Beisert:2002ff} and for the full theory in \cite{Beisert_2004}. Also at one-loop,  a spin-bit model that captures some features of the string-bit formalism for interacting strings \cite{Verlinde_2003} was used in \cite{Bellucci:2004ru} to compute non-planar corrections for operators in the $\mathfrak{su}(2)$ and $\mathfrak{sl}(2)$ sectors. The full two-loop dilatation operator in the $\mathfrak{su}(2)$ sector was found in \cite{Beisert:2003tq} and extended to the non-compact $\mathfrak{su}(1,1|2)$ sector in \cite{Zwiebel:2005er} by using the superconformal symmetry of the theory.

The problem of diagonalising the non-planar dilatation operator has itself been studied using a number of techniques. One approach makes use of group-theoretic insights, initially developed in the context of 1/2-BPS operators \cite{Corley:2001zk}, to make an appropriate choice of basis operators diagonalising two-point functions. 
Alternatively, one may attempt to perturbatively compute $1/N$ corrections about the planar limit using quantum-mechanical perturbation theory. One splits the dilatation operator into a leading planar part and subleading off-diagonal terms, which mix single- and multi-trace operators,  and then computes matrix elements of the subleading terms in the basis of planar eigenstates. Such an approach was used to compute the large $R$-charge limit of non-planar dimensions of two-impurity BMN operators in the $\mathfrak{su}(2)$ sector at one-loop and at two-loop level  \cite{Beisert:2002bb,Constable:2002hw,Beisert:2002ff}. In this work we will generalise this approach for the one-loop dilatation operator in the $\mathfrak{su}(2)$ sector by making use of Bethe states describing arbitrary numbers of  excitations or magnons. 
\begin{figure}   	\begin{eqnarray}
	\begin{array}{cc}
	\includegraphicsbox[scale=0.25]{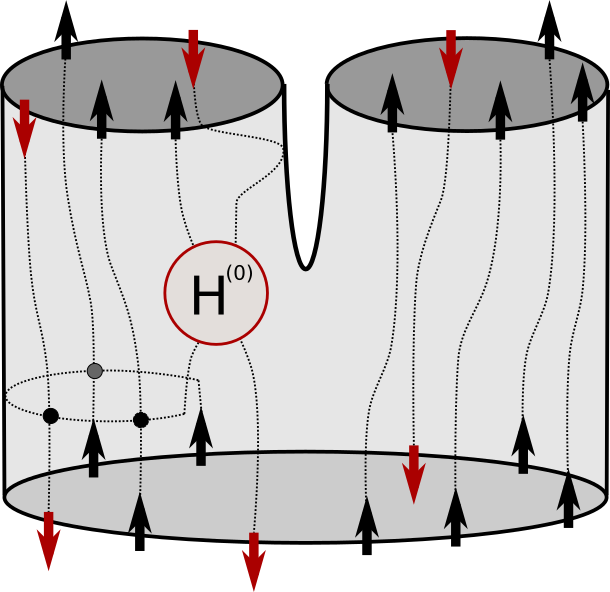}& ~~~~~~~~\includegraphicsbox[scale=0.25]{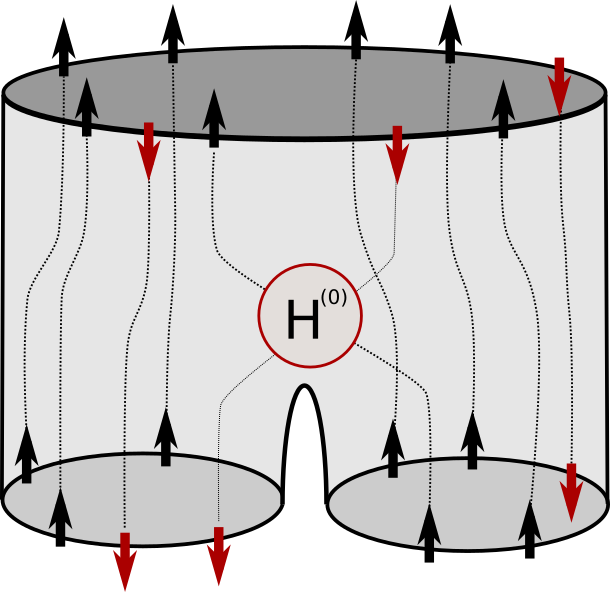}\nn\\
	~&~\nn\\
	(a)& ~~~~~~~~(b)\\
	\end{array}
	\end{eqnarray}
	\caption{The action of the non-planar dilatation operator on single-trace states (a) can be viewed as a simultaneous splitting of the spin-chain and an application of the planar Hamiltonian density, $H_j^{(0)}$, on pairs of non-adjacent spins.   The action on double-trace operators (b) is given by applying, $H_j^{(0)}$, to pairs of spins, one drawn from each spin-chain, while joining the two chains together.}
	\label{fig:Splitting}
\end{figure}
In \secref{sec:pertdim} we compute the action of the off-diagonal terms on Bethe states, schematically shown in \figref{fig:Splitting}$(a)$, and write the overlap of the resulting state with the tensor product of two Bethe states, corresponding to a double-trace operator, in terms of off-shell scalar products. The latter can be computed efficiently using the algebraic Bethe ansatz \cite{korepin1982calculation, slavnov1989calculation} or, equivalently at this order, a Hexagon-like formalism \cite{Basso:2015zoa}.  We similarly find the action on tensor products of Bethe states, see \figref{fig:Splitting}$(b)$, and compute the overlap with single-trace Bethe states in terms of off-shell scalar products. These overlaps can in principle be used to find corrected anomalous dimensions, or equivalently, spin-chain energies
\<
\Delta(g, N)=L+g^2 E(N)+\mathcal{O}(g^4)~,~~~\text{with}~~~E(N)=\sum_{k=0}^\infty \frac{1}{N^k} E^{(k)}
\>
by performing the sum over all such overlaps. Naively, as the overlaps are of order $1/N$, we expect the leading correction to be of order $1/N^2$. However, due to degeneracies in the planar spectrum, non-degenerate perturbation can fail and can result in order $1/N$ corrections to dimensions. 

The issue of planar degeneracies between operators with different numbers of traces was noted in \cite{Constable:2002vq, Beisert:2002ff, Beisert:2003tq}, see also \cite{Freedman_2003, Kristjansen:2003uy}, and requires solving a non-trivial mixing problem. We will instead consider deformations of $\mathcal{N}=4$ SYM theory for which these degeneracies are lifted. In particular, we consider $\beta$-deformed $\mathcal{N}=4$ SYM which is a marginal deformation of the  maximally supersymmetric  theory \cite{Leigh:1995ep} preserving $\mathcal{N}=1$ supersymmetry. We concentrate on the case $\beta\in\mathbb{R}$ for which the theory is exactly conformal to all loop orders in the planar limit \cite{Mauri:2005pa}. Most importantly, the planar spectral problem for the $\beta$-deformed theory is described by an integrable twist of the undeformed spin chain. Twisted asymptotic  Bethe equations at one and higher loops were found in \cite{Beisert:2005if}, while the twisted QSC was found in  \cite{Kazakov:2015efa} and used in \cite{Marboe:2019wyc} to study the anomalous dimensions of operators corresponding to the Konishi multiplet. In \secref{sec:def} we fix the form of the deformed non-planar dilatation operator, which includes additional double-trace terms, and then expand the action on operators in powers of $1/N$. In addition to the leading planar piece and the subleading $1/N$ terms, which are similar to those occuring in the undeformed theory, there are $1/N^2$ and $1/N^3$ terms which contribute to non-planar dimensions. We compute the matrix elements of the subleading dilatation operator in the basis of Bethe states and study the BMN limit of large $R$-charge, $J=L-2\to \infty$, where perturbation theory can be rewritten in terms of the effective loop- and genus-counting parameters \cite{Berenstein:2002jq}
\<
\label{gs}
g'=\frac{g^2_{\text{YM}} N}{16\pi^2 J^2}~~~\text{and}~~~ g_2=\frac{J^2}{N}~.
\>
The anomalous dimensions of two-impurity BMN operators, which additionally depend on an integer parameter $n$ and rescaled deformation parameter $b=\beta L/\pi$, can be written in terms of rescaled energies $\tilde{E}^{(k)}=J^{2-2k} E^{(k)}$  
\<
\Delta_n(g',g_2, b, J)=L+ g' \big[ \tilde E^{(0)}_n(b,J)+ g_2^2 \tilde E^{(2)}_n(b, J)+\mathcal{O}(g_2^4)\big]+\mathcal{O}((g')^2)~
\>
and we compute $\tilde E^{(2)}_n(b, J)$ through $\mathcal{O}(J^{-1})$.

Besides the lifting of degeneracies, there are a number of further advantages to considering the deformed theory. At a technical level it allows us to consider singular solutions to the undeformed Bethe equations. Such solutions correspond to finite energies but have singular wavefunctions and so matrix elements in the undeformed theory are not well defined. Instead they can be computed in the deformed theory, where the deformation parameter regularises singularities in the wavefunctions \cite{Nepomechie:2014hma}, and the limit of vanishing deformation parameter can be smoothly taken. More conceptually, the dependence of the spectrum on a continuous parameter allows for the phenomenon of level crossing whereby eigenvalues become degenerate at special values of the parameter. 
As was noted in the early days of quantum mechanics by von Neumann and Wigner \cite{von1929some}, given a generic theory depending on a number of parameters  it is necessary to tune at least two parameters to cause energy levels to cross and produce a degeneracy. Subsequently it was shown by Teller \cite{teller1937crossing} that the surfaces $E=E(\beta_1, \beta_2)$ representing energy levels depending on two such parameters $\beta_1, \beta_2$ are connected at points like the two sheets of a degenerate cone. In \cite{Korchemsky:2015cyx} the dimensions of local operators in $\mathcal{N}=4$ SYM were analysed as functions of the 't Hooft coupling and it was shown, by re-summing the large-$N$ expansion, that when $N$ is held fixed anomalous dimensions do not cross as $\lambda$ is varied. Studying the one-loop spectrum as a function of the deformation parameter $\beta$, we similarly find that at finite $N$ the anomalous dimensions repel and it is only at large $N$ that they cross.

Level repulsion is characteristic of a chaotic system where energy levels are correlated and so avoid each other, while in an integrable system they are uncorrelated and move independently, crossing on occasion. The phenomenon of (non-)repulsing energy levels can be studied by looking at the distribution of spacings between neighbouring energy levels. If one computes the probability, $P(s)ds$, that the normalised spacing between adjacent levels  lies in the interval between $s$ and $s+ds$, one finds that for a generic, chaotic, quantum system  $P(s)\to 0$ as $s\to 0$. For integrable systems it is generally the case that $P(s)$ goes to a constant as $s\to 0$ which reflects the presence of hidden symmetries in these models.
In \secref{sec:stats} we numerically study the spectrum of both the deformed and undeformed theories and show that in the planar limit the spectral distribution is Poisson, consistent with integrability, while at finite $N$ the distribution is Wigner-Dyson and corresponds to that of the Gaussian Orthogonal Ensemble (GOE) random matrix theory. We are thus able to numerically study the transition from quantum-integrable to quantum-chaotic systems in the context of interacting four-dimensional gauge theory as we vary $N$. An analysis of the transition from Poisson to Wigner-Dyson statistics in a context similar to planar $\mathcal{N}=4$ SYM was performed in \cite{Rabson} which considered and integrability breaking deformation of the XXZ spin-chain. The appearance of quantum chaos in the spectrum is in fact quite natural if we view the dilatation operator as the Hamiltonian of the theory defined on $\mathbb{R}\times S^3$ and multi-trace operators as defining states somewhat analogous to large nuclei in QCD. Indeed it was the work of Wigner \cite{wigner_1951} and Porter \& Rosenzweig \cite{Rosenzweig:1960zz} on the statistical properties of the energy levels of highly-excited nuclei that lead to much of the initial interest of physicists in random matrix theory \cite{mehta2004random}.
     
\subsection{Non-Planar Dilatation Operator}
In order to fix our conventions and notations we briefly review the one-loop dilatation operator of $\mathcal{N}=4$ SYM. We follow closely the work \cite{Beisert:2003tq} where more details regarding the calculations and generalisations to higher loops can be found. $\mathcal{N}=4$ SYM theory contains six scalar fields, $(\phi_I)^a{}_b$, $I=1,\dots 6$, $a,b=1,\dots N$ which transform as a vector of the $SO(6)\simeq SU(4)$ R-symmetry and in the adjoint representation of the $SU(N)$ gauge group. We will restrict ourselves to the $\mathfrak{su}(2)$ sector comprising operators made of products of traces of two complex scalar fields,  $Z=\tfrac{1}{\sqrt{2}}(\phi_{5}+i \phi_6)$ and $X=\tfrac{1}{\sqrt{2}}(\phi_{1}+i \phi_2)$, and so we consider operators such as
\<
\Tr(Z^{\ell_1})~,~~~\Tr(XZ^{\ell_1} XZ^{\ell_2})~,~~~\Tr(XZ^{\ell_1}XZ^{\ell_2} Z^{\ell_3})\Tr (X Z^{\ell_4})\Tr(XX)~.
\>
These operators can be organised into $SO(6)$ representations with Dynkin labels $[M,L-2M,M]$. This sector is known to be closed under the action of the dilatation operator and does not mix with operators containing other scalars, field strengths or fermions. To describe the action of the dilatation operator it is useful to 
make use of the notation for functional derivatives of fields, for example
\<
(\check{Z})^a{}_b\equiv \frac{\delta}{\delta(Z)^b{}_a} 
\>
such that
\<
(\check{Z})^a{}_b(Z)^c{}_d=\delta_b^c\delta^a_d-N^{-1}\delta^a_b\delta^c_d~, ~~~\text{and}~~~(\check{Z})^a{}_b (X)^c{}_d=0~.
\>
This can be used to derive the fusion and splitting formulae
\<
\label{eq:spitfus}
& &\Tr (A \check{Z})~ \Tr (B Z)=\Tr(AB)-N^{-1} \Tr A~\Tr B~,\nn\\
& & \Tr (A\check{Z}B Z)=\Tr A~ \Tr B -N^{-1}\Tr(AB)~,
\>
where it is assumed that $A$ and $B$ do not contain any $Z$'s.  The $N^{-1}$ terms are due to the fact that we are considering the $SU(N)$ gauge theory. This is not particularly important for $\mathcal{N}=4$ SYM and we could equally well consider a $U(N)$ gauge group, however it will become relevant when we subsequently consider the $\beta$-deformed theory.  
Using this notation, the tree-level dilatation operator in the $\mathfrak{su}(2)$ sector can be written as
\<
\mathfrak{D}_0=\Tr(Z\check{Z})+\Tr(X\check{X})
\label{eq:treedil}
\>
and simply counts the number of fields present in a given operator. 
 The one-loop correction to the dilatation operator is then given by \cite{Beisert:2002bb}
\<
\mathfrak{D}_2=-\frac{2}{N} :\text{Tr}([X,Z ][\check{X},\check{Z}]):
\label{eq:1loopdil}
\>
where the normal-ordering markers $:\,\,:$ indicate that the functional derivatives do not act on the fields in $\mathfrak{D}_2$ itself.  We can find the action on multi-trace operators by repeated use of identities \eqref{eq:spitfus}. For example on the length-six single-trace operator, $\Tr(X^2 Z^4)$, we have
\begin{align}
\mathfrak{D}_2\Tr(X^2 Z^4)=4\Big(\Tr (X^2Z^4)-\Tr(XZXZ^3)\Big)+\frac{4}{N}\Big(\Tr(X^2 Z^2)\Tr(Z^2)-\Tr(X ZXZ)\Tr(Z^2)\Big)~,
\end{align}
where we see that the leading term in a large-$N$ expansion corresponds to a superposition of single-trace operators and the subleading term is a double-trace contribution. 
In general, we can decompose the action of the one-loop dilatation operator on multi-trace operators into planar and non-planar pieces
\<
\mathfrak{D}_2=H^{(0)}+\frac{1}{N}H^-+\frac{1}{N}H^+~.
\label{eq:decomp}
\>
The planar piece, $H^{(0)}$,  leaves the number of traces in an operator unchanged, while the non-planar corrections, $H^{\pm}$, which are suppressed by a factor of $1/N$,  increase/reduce the number of traces in a given operator. In order to find the eigenvalues of $\mathfrak{D}_2$, one can first solve the planar problem using integrability and then attempt to use perturbation theory  to find the  $1/N^k$ corrections. 

\subsection{Planar Theory and Integrability}
\label{sec:plan_spin}
Mapping the problem of computing anomalous dimensions to that of computing integrable spin-chain energies proved to be an important step in solving the planar spectral problem. We will make use of the spin-chain notation and the results from integrability to organise the computation of non-planar corrections. We thus review the coordinate Bethe-ansatz approach to integrable spin chains here. Single-trace operators with $M$ insertions of $X$ fields in a background of $(L-M)$ $Z$'s will be denoted as
\<
\Tr(\overbrace{Z\dots Z}^{n_1-1}X\overbrace{Z\dots Z}^{n_2-n_1-1}X\dots)&\equiv&
\ket{\overbrace{\uparrow\dots \uparrow}^{n_1-1}\downarrow\overbrace{\uparrow\dots \uparrow}^{n_2-n_1-1}\downarrow \dots}_L\nn\\
&\equiv&
\ket{n_1, n_2,\dots,n_M}_{L}~.
\>
 Multi-trace operators with $K$ traces and $M$ insertions of $X$ fields, where $M=\sum_{k=1}^K M_{k}$, can be denoted by products of such states
\<
\prod_{k=1}^K\ket{n_1^{(k)},\dots, n_{M_{k}}^{(k)}}_{L_k}
\>
which is an element of the symmetrized tensor product. It will often be convenient to use the compressed notation $\ket{\{n\}}$ and
$\prod_k \ket{\{n^{(k)}\}}$. 

We can write general single-trace states as linear combinations of this basis 
in terms of a wavefunction $\psi_{\{n\}}$ which can depend on the quantum numbers describing the particular state. For example we will consider states with $M$ impurities characterised by the momenta $\{p\}=\{p_1,p_2, \dots, p_M\}$ of $M$ excitations, or magnons
\begin{align}
\ket{\{p\}}=\sum_{\{n\}}\psi_{\{n\}}^{\{p\}}\ket{\{n\}}~.
\label{eq:BetheES}
\end{align}
The sum is over the positions of excitations ranging over the nested values $1\leq n_1< n_2 <\dots <n_M\leq L$. 
We will compute overlaps of such spin-chain states and so we define the natural dual basis\footnote{This choice does not take into account the cyclicity of the traces which we thus need to impose as a  separate condition. }
\<
\braket{m_1, m_2, \dots, m_M}{n_1,n_2, \dots, n_M}=\prod_{j=1}^M\delta_{m_j, n_j}
\>
so that the scalar product for states $\ket{\tilde{\psi}}$ and $\ket{\psi}$ is given by
\<
\braket{\tilde{\psi}}{\psi}=\sum_{\{n\}} \tilde{\psi}^\ast_{\{n\}} {\psi}_{\{n\}}~.
\>

The action of the planar dilatation operator can be defined in this basis and is given by the well-known formula
\<
\label{eq:planar_dil}
H^{(0)}\ket{n_1,n_2,\dots}_L=2 \sum_{j=1}^M \Big(2\ket{ \dots, n_j, \dots }-\ket{\dots, n_j-1,\dots}-\ket{\dots,n_j+1, \dots}\Big)~.
\>
This spin-chain Hamiltonian can be diagonalised by means of the Bethe ansatz. The ferromagnetic vacuum state with no impurities
\<
\label{eq:vac}
\ket{\emptyset}=\ket{\uparrow\uparrow\dots \uparrow}
\>
is an eigenstate with zero energy.  Eigenstates with $M$ impurities have wavefunctions given as sums over  $\sigma\in \mathcal{S}_M$, permutations of $M$ objects,
\<
\label{eq:Bethe_wave}
\psi_{\{n\}}^{\{ p\}}\equiv\psi_{n_1,\dots, n_M} ^{p_1, \dots, p_M}=\frac{1}{\prod_{j<k}\sqrt{S(p_j,p_k)}}\sum_{\sigma\in \mathcal{S}_M}e^{i \sum_{j=1}^M p_{\sigma(j)}n_j}\prod_{\substack{j>k \\ \sigma(j)<\sigma(k)}} S(p_{\sigma(j)},p_{\sigma(k)})
\>
where $S(p_j,p_k)$ is the two-magnon S-matrix
\<
S(p_j,p_k)=-\frac{e^{ip_j+ip_k}+1-2e^{i p_k}}{e^{ip_j+ip_k}+1-2e^{i p_j}}~.
\>
In this definition we have made a particular choice for the overall, non-physical, phase of the wavefunction which is convenient for our subsequent purposes.
For these states to satisfy periodic boundary conditions the momenta must satisfy the Bethe equations, i.e.\ for each $j=1,\dots, M$
\<
\label{eq:momBE}
e^{i \phi_j}=1~,~~~~\text{where }~~~e^{i \phi_j}\equiv e^{ip_j L} \prod_{k\neq j}^M S(p_j, p_k)~,
\>
which implies that the wavefunctions satisfy the condition
\<
\psi^{\{p\}}_{n_1, n_2, \dots, n_M}=\psi^{\{p\}}_{n_2,\dots, n_M, n_1+L}~.
\>
Each eigenstate corresponds to a solution of the algebraic equations \eqref{eq:momBE} and the energy eigenvalue is given 
as a sum over individual magnon energies
\<
E^{(0)}(\{p\})= \sum_{j=1}^M \varepsilon(p_j)~,~~~~ \varepsilon(p_j)=8 \sin^2 \frac{p_j}{2}~.
\>
The cyclicity of the trace for gauge-theory operators becomes the condition that
the spin chain is invariant under the shift $n_j \to n_j+1$ and so we consider only states which satisfy the condition
\<
\prod_{j=1}^Me^{i p_j}=1~.
\>
It is useful to introduce rapidity variables  $u_j=\frac{1}{2}\cot \tfrac{p_j}{2}$, or
\<
\label{eq:rapmom}
e^{ip_j}=\frac{u_j+i/2}{u_j-i/2}~,
\>
for each excitation, which we can use to rewrite the S-matrix and the individual magnon energies as
\<
\label{eq:rapSE}
S(u_j, u_k)=\frac{u_j-u_k-i}{u_j-u_k+i}~~~\text{and}~~~\varepsilon(u_j)=\frac{2}{u_j^2+\tfrac{1}{4}}~.
\label{eq:Sinu}
\>
It will be useful to define the quantity
\begin{equation}
h(u_j,u_k)= \frac{u_j-u_k}{u_j-u_k+i}
\label{eq:hinu}
\end{equation}
so that the S-matrix is given as
\begin{equation}
S(u_j,u_k)= \frac{h(u_j,u_k)}{h(u_k,u_j)}\,.
\end{equation}
Additionally, it is convenient to define the normalisation factors for states involving momenta $p_1, p_2, \dots $ corresponding to rapidities $u_1, u_2, \dots$ as
\begin{equation}
\mathcal{N}(p(u))=\frac{\prod_{i<j} h(u_i, u_j)}{\prod_{j<k}\sqrt{S(u_j,u_k)}}
\end{equation}
and generalisations such as $\mathcal{N}(p, q)=\mathcal{N}(p)\mathcal{N}(q)$. 
Finally, similar to \cite{Escobedo:2010xs}, we will use the following short-hand notation for products
\begin{equation}
\label{eq:prodnot}
 f^{\{a\}} = \prod_{i} f(a_i)\,, \qquad  f^{\{a\}}_< = \prod_{i<j} f(a_i,a_j)\,, \qquad h^{\{a\}\{b\}} = \prod_{i,j} h (a_i,b_j)\,
\end{equation}
and 
\begin{equation}\label{NotationHat}
\{z\}_{\hat a} = \{z_1, \ldots, \hat z_a, \ldots, z_n\}=\{z_1, \ldots, z_{a-1}, z_{a+1}, \ldots, z_n\}
\end{equation}
for lists with a missing element.
Using this notation a Bethe state can be written as
\begin{equation}
\ket{\{p\}}=\mathcal{N}(p)\sum_{\{n\}}\sum_{\sigma \in S_M} \frac{1}{h^{\{p_\sigma\}}_< } e^{i p_\sigma \cdot n} \ket{\{n\}}\,
\label{eq:Betheansatz}
\end{equation}
and its conjugate as
\begin{equation}
\bra{\{p\}}=\mathcal{N}(p^\ast)\sum_{\{n\}}\sum_{\sigma \in S_M} \frac{1}{h^{\{p^*_\sigma\}}_> } e^{-i p^*_\sigma \cdot n} \bra{\{n\}}\,
\end{equation}
where the functions $h^{\{p\}}_<$ etc.\ should be understood as being defined in terms of the set of rapidities $\{u\}$ corresponding to the momenta $\{p(u)\}$. 

\section{Perturbative Non-Planar Anomalous Dimensions}
\label{sec:pertdim}

In this section we study the action of the non-planar dilatation operator on the planar eigenstates, the Bethe vectors, and subsequent overlaps with other Bethe states. We consider first the action of $H^-$ on a double-trace operator corresponding to the product of a length $L_q$ Bethe state $|\{q\} \rangle$ with $Q$ excitations and a length $L_r$ state $|\{r\} \rangle$ with $R$ excitations. There are $L_q L_r$ terms corresponding to the action of the dilatation operator on each pair of sites of the two spin chains. However, the terms where it acts on a $Z$ field at the  $i$-th site of the spin chain can be rewritten using the Bethe equations, so that they become equivalent to the action on a $Z$ at the first site. This can be seen by gathering all the terms in the state \eqref{eq:BetheES} with a $Z$ field at the $i$-th site and using the cyclicity and on-shell condition to write them with the $Z$ field at the first position:
\begin{align}
\sum_{l=0}^M \sum_{\substack{1\leq n_1<\ldots < n_l \leq i-1\\i+1\leq n_{l+1}<\ldots < n_M \leq L}}\psi^{\{p\}}_{\{n\}} |n_1,\ldots,n_l\rangle_{i-1} \otimes
\ket{Z}\otimes|n_{l+1},\dots, n_M\rangle_{L-i}&\nn \\
&\kern-100pt=   \sum_{2\leq n_1<\ldots< n_M\leq L } \psi_{\{n\}}^{\{p\}} \;\ket{Z}\otimes| \{n\}\rangle_{L-1}\,.
\label{eq:Zfirst}
\end{align} 
Analogously, the action on an $X$ field at the $i$-th site can be rewritten as the action on the same chain with the $X$ field placed at the first site
\begin{equation}
\sum_{l=1}^{M} \sum_{\substack{1\leq n_1<\ldots < n_l=i\\i=n_{l}<\ldots < n_{M} \leq L}}\psi^{\{p\}}_{n_1,\ldots,n_l=i, \ldots, n_{M}} |n_1,\ldots,n_l=i,\ldots, n_M\rangle=\sum_{1= n_1<\ldots< n_{M}\leq L} \psi_{\{n\}}^{\{p\}} \;| \{n\}\rangle\,.
\end{equation}
With these and similar simplifications the action of $H^-$ on the double-trace state can be written as
\begin{align}
H^-|\{q\} \rangle |\{r\}\rangle = 2 L_q L_r 
&\Bigg[~~\sum_{\substack{1\leq m_1 < \ldots < m_{Q}= L_q\\2\leq n_1 < \ldots < n_{R}\leq L_r}} \psi^{\{q\}}_{\{m\}} \psi^{\{r\}}_{\{n\}} |\{m\}_{\hat Q}\rangle \otimes|[X,Z]\rangle\otimes |\{n+L_q\}\rangle  \nonumber\\
&+\sum_{\substack{1\leq m_1 < \ldots < m_{Q}\leq L_q-1\\1= n_1 < \ldots < n_{R}\leq L_r}} \psi^{\{q\}}_{\{m\}} \psi^{\{r\}}_{\{n\}} |\{m \}\rangle \otimes|[Z,X]\rangle\otimes |\{n+L_q\}_{\hat{1}}\rangle \nonumber\\
&+\{\text{terms~with~} q \rightleftarrows r\}
~~ \Bigg]\,
\end{align}
where the terms on the right hand side all correspond to single-trace operators. 
The overlap with a dual state $\bra{\{p\}}$, of length $L_p=L_q+L_r$ and with $P=Q+R$ excitations, can then be computed
\begin{align}
\langle \{p\}| H^- | \{q\} \rangle | \{r\} \rangle&= 2L_q L_r \, \mathcal{N}(p^\ast, q, r)\sum_{\rho,\sigma,\tau} \frac{1}{h_>^{\{p^*_\rho\}} h_<^{\{q_\sigma\}} h_<^{\{r_\tau\}}}\nonumber\\
&\kern-35pt \times\Bigg( \delta_{Q\neq 0} (e^{i p^*_{\rho(Q)}}-1) e^{i L_q q_{\sigma(Q)}} e^{-i(L_q+1)(p^*_\rho)_{Q}^{Q+R}}  P_{L_q}\left(\{q_{\sigma}-p^*_\rho\}_1^{Q-1} \right) P_{L_r}\left(r_\tau-\{p^*_\rho\}_{Q+1}^{Q+R} \right)\nonumber\\
&\ +  \delta_{Q\neq 0} (1- e^{i p^*_{\rho(R+1)} } )e^{-i(L_r+1) (p_\rho^*)_{R+1}^{Q+R}}  P_{L_q}\left(\{q_{\sigma}\}_{2}^Q-\{p^*_\rho\}_{R+2}^{R+Q}\right) P_{L_r}\left(r_\tau-\{p^*_\rho\}_1^R\right) \nonumber\\
&\ +\{\text{terms~with~} q \rightleftarrows r\} 
\Bigg) \,, \label{eq:HmO}
\end{align}
where we define the sets $\{t_\lambda\}_a^b = \{t_{\lambda(a)},\ldots ,t_{\lambda(b)}\}$ and denote products of exponentials over such sets using the notation $e^{i L (t_\lambda)_a^b}=\prod_{i=a}^b e^{i L t_{\lambda(i)}}$.
The factors of $P_L(z)$ in \eqref{eq:HmO} correspond to the geometric sums of exponentials in the wavefunctions which can be rewritten as sums over ordered partitions
\begin{equation}
P_L(z)= \sum_{1\leq n_1 < \ldots < n_{|z|} \leq L-1} e^{i z\cdot n}  = \sum_{l=0}^{|z|} \prod_{k=1}^l \frac{1}{e^{-i \sum_{j=k}^l z_l} -1} \prod_{k=l+1}^{|z|} \frac{e^{i z_k L}}{e^{i \sum_{j=l+1}^k z_l}-1}\,.
\label{eq:opartitions}
\end{equation}
Using these notations we can write a similar expression for overlaps of $H^+$ as sums over ordered partitions:
\begin{align}
\langle  \{r\} |\langle \{q\} | H^+ | \{p\} \rangle &=2 L_p\,  \mathcal{N}_+(p,q^\ast, r^\ast) \sum_{\rho,\sigma,\tau} \frac{1}{h_<^{\{p_\rho\}} h_>^{\{q^\ast_\sigma\}} h_>^{\{r^*_\tau\}}} \nn \\
&\kern-60pt \times\Bigg( \delta_{Q\neq 0} (e^{i q^*_{\sigma(Q)}}-1) e^{-i L_q q^*_{\sigma(Q)}} e^{i(L_q-1) \{p_\rho\}_{Q}^{Q+R}}  P_{L_q-1}\left(\{p_\rho - q^*_{\sigma}\}_1^{Q-1} \right) P_{L_r+1}\left(\{p_\rho\}_{Q+1}^{Q+R} -r^*_\tau \right)\nonumber\\
&+  \delta_{Q\neq 0} (1- e^{i q^*_{\sigma(1)}} ) e^{i(L_r+1) (p_\rho)_{R+1}^{R+Q}}   P_{L_q-1}\left(\{p_\rho\}_{R+2}^{R+Q}-\{q_\sigma^*\}_2^Q\right) P_{L_r+1}\left(\{p_\rho\}_1^R-r^*_\tau \right) \nonumber\\
&+\{\text{terms~with~} q \rightleftarrows r\}
\Bigg) \,.\label{eq:Hp0}
\end{align}
The normalisation in this case is defined slightly differently with $\mathcal{N}_+=\mathcal N/S$, where $S$ is a symmetry factor that equals $2$ when the states in the double trace are equal and $1$ otherwise.
Carrying out the geometric sums via \eqref{eq:opartitions}  makes these formulae useful for analysing states of arbitrary lengths. However, while these expressions are reasonably compact, they involve sums over permutations for each of the sets of external momenta and so they rapidly become impractical as the number of excitations grows. The same growth is known from the computation of spin-chain scalar products in the coordinate Bethe ansatz 
and by making use of known results in this case we can find further simplifications. 

\subsection{Matrix Elements from Spin-Chain Scalar Products} 
\label{sec:SCOvlp}
The scalar product of two Bethe states
\<
\langle \{l\}| \{k\}\rangle_L&=& \sum_{1\leq n_1 < \dots n_{|k|} \leq L} \psi^\ast{}^{\{l\}}_{\{n\}} \psi^{\{k\}}_{\{n\}}= \mathcal{N}(k,l^\ast)  \sum_{\rho,\sigma} \frac{P_{L+1}(k_\rho-  l^*_\sigma)}{h_>^{\{l^*_\sigma\}}h_<^{\{k_\rho\}}}
\label{eq:Betheproduct}
\>
involves double-sums over permutations and so is generally complicated to evaluate. Fortunately, there are well-known formulae for such scalar products which were developed in the Algebraic Bethe Ansatz (ABA) approach to integrable spin chains (see \secref{sec:ABA} for a brief review). In the case where both sets of momenta $ \{k\}$  and $\{l\}$ do not satisfy the Bethe equations, i.e.\ they are off-shell, the scalar product can be written as a sum over partitions of the sets of momenta into subsets of equal cardinality \cite{Korepin:1982gg}, see \eqref{eq:OBIP}. Similar simplifications can be used to rewrite the expressions of the overlaps \eqref{eq:HmO} and \eqref{eq:Hp0}. Each term in the formulae for the overlaps  non-trivially involves one momentum of an excitation from the single-trace operator, which we label $p_j$, and one excitation momentum from the double-trace operator, i.e.\ from either $\{q\}$ or $\{r\}$, which we label as $q_i$ or $r_i$. The remaining momenta are simply contracted using a rescaled spin-chain scalar product
\begin{align}
(\{l\}|\{k\})_L\equiv\frac{\braket{\{l\}}{\{k\}}_L}{\mathcal{N}(k,l^\ast)}~.
\end{align}
We can thus write the overlaps \eqref{eq:HmO} and \eqref{eq:Hp0} in terms of the off-shell scalar products by splitting the single-trace excitation momenta into three subsets, 
$\{ p \}= s \cup t \cup \{ p_j \}$, with the cardinality of $s$ equal to that of 
$\{ q \}_{\hat{i}}$ (or $\{ r \}_{\hat{i}}$) 
and the cardinality of $t$ equal to that of $\{r\}$ (resp.\ $\{q\}$). In terms of off-shell scalar products, the overlap of $H^-$ can then be written as a sum over all such splittings 
\begin{align}
\label{eq:Hm1}
\langle \{p\}| H^- | \{q\} \rangle | \{r\} \rangle&=2 L_q L_r \, \mathcal{N}(p^*,q,r)
\sum_{\substack{i,j \\ s \cup t = \{p\}_{\hat j}}}
\frac{e^{i p^*_j}-1}{h^{q_i q_{\hat i}}}
\big[s^{L_q+1\, \ast}_\circlearrowleft-t^{L_r+1\, \ast}_\circlearrowright\big]  
(s| \{q\}_{\hat{i}})_{L_q-1}(t|\{r\})_{L_r-1}\nn\\
&\kern+150pt+\{\text{terms~with~} q \rightleftarrows r\}
\end{align}
and that of $H^+$ as 
\begin{align}
\label{eq:Hp1}
\langle \{r\}|\langle \{q\}|  H^+ | \{p\} \rangle&=2 L_p\, \mathcal{N}_+(p,q^\ast,r^*)   
\sum_{\substack{i,j \\ s \cup t = \{p\}_{\hat j}}}
\frac{e^{i q^*_i}-1}{h^{q^\ast_{\hat i}q_i^\ast}}
\big[s^{L_q-1}_{j\,\circlearrowleft}-t^{L_r+1}_{j\, \circlearrowright}\big] 
( \{ q \}_{\hat{i}}|s)_{L_q-2}
( \{r\} |t)_{L_r}\nn\\
&\kern+180pt+\{\text{terms~with~} q \rightleftarrows r\}~.
\end{align}
In addition to the scalar products of Bethe states these expressions involve $(e^{ip}-1)$ factors, which are essentially the same as arise in the planar dilatation operator, and ordering factors for which we introduced the notations
\begin{align}
\label{eq:ophases}
s^{L}_{j\, \circlearrowleft}= \frac{e^{-i L s} }{h^{ p_j t} h^{s p_j} h^{s t} }~,~~~ t^{L}_{j\, \circlearrowright} =\frac{e^{-i L t} }{h^{ p_j s } h^{t p_j} h^{ t s } }~.
\end{align}
These terms account for the phase acquired by the $p_j$ magnon as it is shifted around the chain before being contracted with a magnon on the double-trace operator. For each configuration there are two different ways to carry out this reordering and the overlap is a linear combination of both.

Spin-chain scalar products have previously appeared in the context of $\mathcal{N}=4$ SYM in the computation of structure constants. In the all-order hexagon approach, \cite{Basso:2015zoa}, structure constants are written as sums over partitions of the magnon excitations and it was noted that this formulation is related to the scalar-product formula of Korepin \cite{Korepin:1982gg}. It is therefore convenient to use a tree-level version of the hexagon formulation of scalar products
\begin{align}\label{OSSP}
(\{l(v)\}| \{k(u)\})
&=(-1)^M\prod_{j=1}^M (u_j+i/2)(v^*_j-i/2) \sum_{\substack{\alpha \cup\bar \alpha=\{k\}\\ \beta \cup \bar \beta= \{l^*\} }} \frac{e^{i L (\bar \alpha -\bar\beta)} G(\alpha,\beta) G(\bar\beta,\bar\alpha)}{h^{\alpha \bar \alpha}  h^{\bar \beta\beta} }  \,,
\end{align}
where
\begin{equation}\label{Hdet}
G(\alpha(u), \beta(v^\ast)) =  \frac{\det\left[ \frac{i}{(u_j-v^*_k)(u_j-v^*_k+i)}\right] \prod_{j,k}(u_j-v^*_k+i) }{\prod_{j<k} (u_k-u_j)(v^*_j-v^*_k)}\,,
\end{equation}
in order to rewrite the overlaps \eqref{eq:Hm1,eq:Hp1}.

\subsection{A Hexagon-like Formulation}

In the previous section we obtained the non-planar dilatation operator overlaps as sums over partitions of the rapidities, in a way that is reminiscent of the Hexagon formulation of three-point correlation functions \cite{Basso:2015zoa}. In that context, the partitions of the rapidities arise naturally in the large-volume regime where the correlation function is broken down to its simplest building blocks, the Hexagon form factors. Crucially, these form factors satisfy a set of axioms which, together with the diagonal symmetries and some educated guesswork, can be used to obtain an all-loop description of structure constants. A particular feature of the Hexagon is its conical defect which is associated with the existence of three asymptotic regions and corresponds to a monodromy composed by three crossing operations. In the context of non-planar overlaps between a single-trace and a double-trace operator, a similar role seems to be played by the three distinct traces. In this section we investigate the properties of the objects arising from the action of $H^+$ and $H^-$ and find that they satisfy the same form factor axioms as appear in the context of correlation functions.

The sum over determinants occuring in our rewriting of off-shell scalar products \eqref{OSSP} can be found in a straighforward way from the hexagonalization of three-point functions \cite{Basso:2015zoa}.  To be precise we consider the three-point function of two unprotected operators in the $SU(2)$ sector, one with $X$ excitations and the other with $\bar X$, and one rotated half-BPS operator. The $X$ and $\bar X$ fields must be Wick contracted at tree level in order to produce a non-vanishing contribution. If there are $l$ Wick contraction between the excited operators, then the structure constant is
\begin{equation}\label{3ptA}
C^{\bar X| X}_{\{p\}|\{q\}} \propto  \sum_{\substack{\alpha \cup \bar \alpha = \{p\}\\\ \beta \cup \bar \beta = \{q\}}} \omega_l(\alpha,\bar\alpha) \omega_{L_q-l}(\beta,\bar\beta) \mathcal  H (\alpha | \beta) \mathcal H (\bar \beta | \bar \alpha)\,,
\end{equation}
with the splitting factor defined as
\begin{equation}
\omega_l(\alpha,\bar \alpha) = e^{i \bar\alpha l } \prod_{\substack{u_i \in \bar\alpha, u_j \in \alpha\\
		i<j}} S(u_i,u_j)\,.
\end{equation}
The Hexagon function $\mathcal H$ in this particular configuration is simply related to our determinant expression \eqref{Hdet} by
\begin{equation}\label{HH}
\mathcal H(\alpha|\beta) = h_<^{\alpha} h_<^{\beta} G(\alpha, \beta) \,.
\end{equation}
The Hexagon description of three-point functions allows the evaluation of general configurations where all three operators are excited. If we now let two of the operators have $\bar X$ excitations, while the other is composed of $X$ fields, then the structure constant becomes
\begin{equation}\label{3ptB}
C^{\bar X|  X| \bar X}_{\{p\}|\{q\}|\{r\}} \propto \sum_{\substack{\alpha \cup \bar \alpha = \{p\}\\ \beta \cup \bar \beta = \{q\}\\ \gamma \cup \bar \gamma = \{r\}}} \omega_{l_{pq}}(\alpha,\bar \alpha) \omega_{l_{qr}}(\beta,\bar \beta)\omega_{l_{pr}}(\gamma,\bar \gamma)\mathcal H(\alpha | \beta | \gamma) \mathcal H(\bar \gamma | \bar \beta | \bar \alpha)\,,
\end{equation}
where $l_{ij}$ denote the number of Wick contraction between operators $i$ and $j$ at tree level  and the sum over partitions is further restricted by the fact that $\mathcal H(\alpha|\beta|\gamma)$ is non-vanishing only when the cardinality of $\beta$ matches that of $\alpha \cup \gamma$. It is interesting to note that while \eqref{3ptB} is given as a sum over partitions of three sets of rapidities, a naive tree-level evaluation would give rise to geometric sums naturally yielding a sum over five partitions
\begin{equation}\label{3ptC}
C^{\bar X|  X| \bar X}_{\{p\}|\{q\}|\{r\}} \propto \sum_{\substack{\alpha \cup \bar \alpha = \{p\}\\ \gamma \cup \bar \gamma = \{r\}}} \sum_{\substack{s \cup t = \{q\}\\\beta \cup \bar \beta = s \\ \delta \cup \bar \delta = t}} \frac{e^{i(\bar \alpha -\bar\beta)l_{12}} e^{i(\bar\gamma-\bar\delta)l_{23}} e^{i s l_{12}}}{h^{\alpha \bar \alpha}h^{\gamma\bar\gamma} h^{\bar \delta \delta} h^{\bar \beta \beta} h^{ts}} G(\alpha,\beta) G(\bar \beta,\bar \alpha)  G(\gamma,\delta) G(\bar \delta,\bar \gamma) \,.
\end{equation}
The equivalence of these descriptions follows from the following tree-level relation
\begin{equation}\label{Hrel}
\mathcal H(\alpha|\beta|\gamma)= h_<^\alpha h_<^\beta h_<^\gamma \sum_{\mu \cup \nu = \beta} \frac{G(\alpha,\mu)G(\nu,\gamma)}{h^{\mu\nu}}\,.
\end{equation}
We should stress that, computationally speaking, \eqref{3ptB} is not necessarily a more efficient version of \eqref{3ptC} as the objects $\mathcal H(\alpha|\beta|\gamma)$ do not have a known compact determinant description. Nevertheless, while the computational gain might not be considerable, there is a conceptual advantage due to the fact that the Hexagon functions can be bootstrapped. First, they obey the Watson equation 
\begin{equation}\label{Watson}
\mathcal H( \ldots | \ldots, \beta_i, \beta_{i+1},\ldots | \ldots) = S(\beta_i,\beta_{i+1}) \mathcal H(\ldots| \ldots, \beta_{i+1},\beta_i \ldots |\ldots)  \,,
\end{equation}
which holds also for the exchange of excitations in the other edges, and they also satisfy the decoupling conditions
\begin{align}\label{Decoupling}
-i \underset{\alpha_{|\alpha|} = \beta_1}{\mathrm{Res}} \left[\mathcal H(\ldots, \alpha_{|\alpha|} | \beta_1, \ldots | \ldots) \right] &= \mathcal H(\ldots,\alpha_{|\alpha|-1}| \beta_2,\ldots| \ldots)\,, \nonumber\\
-i \underset{\beta_{|\beta|} = \gamma_1}{\mathrm{Res}} \left[\mathcal H(\ldots | \ldots, \beta_{|\beta|} | \gamma_1, \ldots ) \right] &= \mathcal H(\ldots,| \ldots,\beta_{|\beta|-1}| \gamma_2,\ldots)\,.
\end{align}
Together with the diagonal symmetries of three-point functions, these form-factor axioms allowed the determination of the hexagon functions at any value of the coupling \cite{Basso:2015zoa}.

With this in mind, we can attempt a similar rewriting of the dilatation operator overlaps. It is useful to work with normalised spin-chain states where we divide by the norms of on-shell Bethe states $\|\{p\}\|=\sqrt{\braket{\{p\}}{\{p\}}}$. These can be conveniently calculated using the Gaudin formula \eqref{eq:Gaudin} which for the coordinate Bethe-ansatz normalisation is 
\<
\|\{p(u)\}\|^2=(-1)^M \prod_j (u_j+i/2)(u_j-i/2)\, \text{det}\, \partial_u \phi(u)
\>
with $\phi$ defined in \eqref{eq:momBE}. This can be combined with the normalisation factors of the overlaps to define a new normalisation factor
\< 
\widetilde{\mathcal{N}}(p(u),q(v),r(w))= \frac{\mathcal{N}(p,q,r)}{h_<^{\{p\}} h_<^{\{q\}}  h_<^{\{r\}} \sqrt{\text{det}\, \partial_u \phi(u)\, \text{det}\, \partial_v \phi(v)\, \text{det}\, \partial_w \phi(w)} }~,
\>
where we used the fact that solutions of the Bethe equations are invariant under complex conjugation, e.g.\ $\{u^\ast\}=\{u\}$, \cite{vladimirov1986proof}, and the cyclicity condition to simplify the expressions\footnote{There is a potential ambiguity in our simplifications arising from square roots of S-matrices in $\mathcal{N}$. There are in principle combinations of rapidities such that products of S-matrices cross the square-root branch cut resulting in additional minus signs in the normalisation. The same ambiguity seems to appear in the Gaudin norm and so will cancel. Moreover, these signs will appear symmetrically in the overlaps of $H^-$ and $H^+$ and thus certainly cancel in the calculation of energies.}. 
The overlap with normalised external states can then be written as
\begin{align}\label{eq:Vm}
V^-(q,r;p) =~& 2 L_q L_r \,\widetilde{\mathcal  N}(p,q,r)  \sum_{\substack{\alpha \cup \bar \alpha = \{q\}\\ \beta \cup \bar \beta = \{p\}\\ \gamma \cup \bar \gamma = \{r\}}} \omega_{L_q}(\alpha, \bar\alpha) \omega_{L_r}(\beta, \bar\beta) \omega_0(\gamma, \bar\gamma) \times \nonumber \\
&\kern-30pt\times \Bigg[\mathcal H(\alpha|\beta|\gamma) \Big(\mathcal H^-_1(\bar\gamma|\bar\beta|\bar\alpha) +\mathcal H^-_2(\bar\gamma|\bar\beta|\bar\alpha)\Big)+ \Big( \mathcal H^-_1(\alpha|\beta|\gamma) +\mathcal H^-_2(\alpha|\beta|\gamma)\Big)\mathcal H(\bar\gamma|\bar\beta|\bar\alpha)\Bigg]\,,
\end{align}
where $\mathcal H$ is the same as in \eqref{Hrel}, and we define the new functions $\mathcal H^-_i$ as 
\begin{align}\label{NewHm}
\mathcal H^-_1(\alpha|\beta|\gamma) &= h_<^{\alpha} h_<^{\beta} h_<^{\gamma}  \sum_{\substack{i,j \\\mu\cup \nu=\beta_{\hat j}}} \frac{(e^{i\beta_j}-1)(e^{-i\beta_j}-1)(e^{i \alpha_i}-1)G(\alpha_{\hat i},\mu) G(\nu,\gamma)}{e^{i (\mu+\gamma-\alpha_{\hat i}-\nu)}h^{\alpha_{i}  \alpha_{\hat i}} h^{\mu \beta_j} h^{\beta_j \nu} h^{\mu \nu}}\,,\nonumber\\
\mathcal H^-_2(\alpha|\beta|\gamma) &=  h_<^{\alpha} h_<^{\beta} h_<^{\gamma} \sum_{\substack{i,j\\ \mu \cup \nu=\beta_{\hat j}}} \frac{(e^{i\beta_j}-1)(e^{-i\beta_j}-1)(e^{-i \gamma_{i}}-1)G(\alpha,\mu) G(\nu,\gamma_{\hat i})}{e^{i (\mu+\gamma_{\hat i}-\alpha-\nu)}h^{\gamma_{\hat i} \gamma_i} h^{\mu \beta_j} h^{\beta_j \nu} h^{\mu \nu}}\,.
\end{align}
We remind the reader that $\mu_{\hat k}$ denotes the set of rapidities $\mu$ without $\mu_k$, following the notation introduced earlier in \eqref{NotationHat}, while the short-hand notations for products are defined in \eqref{eq:prodnot}. This decomposition of the overlap in \eqref{eq:Vm} seems to fit the splitting of Figure \ref{fig:Splitting}(b) particularly well. By cutting the pair of pants depicted in that figure, one would naively expect the side facing away to be represented by the original hexagon $\mathcal H$ of \eqref{Hrel}, while the side facing forward should lead to something new as it contains the action of the commutator from the dilatation operator.
Similarly, the overlap $V^+$ can be rewritten as
\begin{align}\label{eq:Vp}
V^+(p;q,r) =~& 2 L_p \,\widetilde{\mathcal  N}_+(p,q,r)  \sum_{\substack{\alpha \cup \bar \alpha = \{q\}\\ \beta \cup \bar \beta = \{p\}\\ \gamma \cup \bar \gamma = \{r\}}} \omega_{L_q}(\alpha, \bar\alpha) \omega_{L_r}(\beta, \bar\beta) \omega_0(\gamma, \bar\gamma) \times \nonumber \\
&\kern-50pt\times \Bigg[\mathcal { H}^+_0(\alpha|\beta|\gamma) \Big(\mathcal { H}^+_1(\bar\gamma|\bar\beta|\bar\alpha) +\mathcal { H}^+_2(\bar\gamma|\bar\beta|\bar\alpha)\Big)+ \Big( \mathcal { H}^+_1(\alpha|\beta|\gamma) +\mathcal { H}^+_2(\alpha|\beta|\gamma)\Big)\mathcal { H}^+_0(\bar\gamma|\bar\beta|\bar\alpha)\Bigg]\,,
\end{align}
where the normalization is now $\widetilde{\mathcal N}_+ = \widetilde{\mathcal N}/S$, with $S$ a symmetry factor that equals 2 when the states in the double-trace are the same, and 1 otherwise, and we have further defined the functions $\mathcal H^+_i$
\begin{align}\label{NewHp}
\mathcal { H}^+_0(\alpha|\beta|\gamma) &=  h_<^{\alpha} h_<^{\beta} h_<^{\gamma}  \sum_{\mu\cup \nu=\beta} e^{i (\alpha-\mu)}\frac{G(\alpha,\mu) G(\nu,\gamma)}{h^{\mu\nu}}\,,\nonumber\\
\mathcal { H}^+_1(\alpha|\beta|\gamma) &= h_<^{\alpha} h_<^{\beta} h_<^{\gamma} \, e^{i(\alpha+\gamma-\beta)}\sum_{\substack{i,j\\\mu\cup \nu=\beta_{\hat j}}} \frac{(e^{i \alpha_i}-1)(e^{-i \alpha_i}-1)(e^{-i \beta_j}-1)G(\alpha_{\hat i},\mu) G(\nu,\gamma)}{e^{i (\mu-\alpha_{\hat i})}h^{\alpha_i  \alpha_{\hat i}} h^{\mu \beta_j} h^{\beta_j \nu} h^{\mu \nu}}\,,\nonumber\\
\mathcal { H}^+_2(\alpha|\beta|\gamma) &=  h_<^{\alpha} h_<^{\beta} h_<^{\gamma} \sum_{\substack{i,j\\\mu\cup \nu=\beta_{\hat j}}} \frac{(e^{i \gamma_i}-1)(e^{-i \gamma_i}-1)(e^{i \beta_j}-1)G(\alpha,\mu) G(\nu,\gamma_{\hat i})}{e^{i (\gamma_{\hat i}-\nu)}h^{ \gamma_{\hat i} \gamma_i} h^{\mu \beta_j} h^{\beta_j \nu} h^{\mu \nu}}\,.
\end{align}
Unfortunately, in this case we are not able to write any of the new objects in terms of the original hexagon function $\mathcal H$, since the partitions of the rapidities $\beta$ into $\mu$ and $\nu$ appear with a distinct structure. The decomposition is however very similar to that of $V^-$, and seems to match once again the intuition derived from Figure \ref{fig:Splitting}(a), with the cutting producing a product between a simpler structure with a more complex ones.
We wish to emphasize that while these formulae appear quite involved, they can be straightforwardly evaluated once the rapidities are known, by  using, for example, Mathematica.

\begin{figure}[t]
	\centering
	\includegraphics[width=0.6\linewidth]{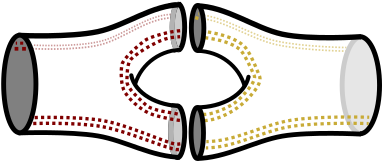}
	\caption{Both $V^+$ and $V^-$ can be seen as a pair of pants where the asymptotic regions correspond to the three distinct traces involved in the overlap. We have found that each of them can be decomposed into hexagon-like objects satisfying the Watson and decoupling conditions. By glueing them together one can reconstruct the torus, thus finding the non-planar corrections to two-point functions.}\label{torus}
\end{figure}

While these expressions are a \textit{post hoc} massaging of the expressions in \eqref{eq:Hm1,eq:Hp1}, when written in this form they clearly resemble the formulas for structure constants. Importantly, the new objects $\mathcal H_i^+$ and $\mathcal H^-_j$ also obey the Watson equations \eqref{Watson} and decoupling conditions \eqref{Decoupling}. Note that these properties of $\mathcal H^+_i$ and $\mathcal H^-_i$ follow from those of the object $\mathcal H(\alpha|\beta)$ defined in \eqref{HH}. This is non-trivial nevertheless, as it occurs only for certain functions of the rapidities in the summands of \eqref{NewHm,NewHp}. This hints at the possibility that the non-planar dilatation operator overlaps can be written in terms of Hexagon-like objects and potentially determined even at higher orders in perturbation theory. 
The corrections to the energies of single-trace operators are obtained through the action of $H^+$ and $H^-$ and a sum over intermediate double-trace operators, which has the natural representation of cutting the torus into two pairs of pants. The fact that the overlaps themselves have a decomposition into hexagon-like objects therefore seems to indicate a possible tesselation of the torus as depicted in Figure \ref{torus}. 

There is an implicit notion of crossing that comes with the decoupling condition. It is natural to imagine that, once such an operation is defined, the excitations can be moved around, so that we relate the hexagon-like objects to a single function where all rapidities are on the same edge. It is upon crossing of the excitations in \eqref{Decoupling} to the same edge that a particle-antiparticle pair $\bar X(u^{2\gamma}) X(u)$ can form in a manifest way and decouple from the corresponding form factor. Such a formulation of $\mathcal H^+_i$ and $\mathcal H^-_i$ with all excitations on the same edge would also be the ideal setup for implementing a bootstrap of those objects. Unfortunately, crossing operations do not commute with the perturbative expansion, and since our one-loop analysis gives access only to the more complicated form of these objects, we were not able to explore further the possibility of such a bootstrap programme. 

\subsection{Anomalous Dimensions from Overlaps}

Our main goal in calculating the above overlaps is to perturbatively compute the leading non-planar correction to operator anomalous dimensions. The general idea is to apply first-order quantum-mechanical perturbation theory. We denote the planar energies as $E^{(0)}$ and their non-planar corrections at order $N^{-k}$ as $E^{(k)}$. Given a single-trace operator characterised by momenta $\{p(u)\}$ solving the Bethe equations and with planar energy $E^{(0)}(\{p\})$, the non-planar correction is
\<
\label{eq:pert_coor}
E^{(2)}(\{p\})=\sum_{\{I\}}
\frac{ V^-(p; I)
	V^+(I; p)}{E^{(0)}(\{p\})-E^{(0)}(\{ I\})}\,.
\>
The sum over $I$ is taken over all intermediate double-trace states
\<
\ket{I}=\ket{\{q\}}_{L_q}\ket{\{r\}}_{L_r}
\>
where we must sum over all lengths $1<L_q<L_p-1$ and for each length sum also over all solutions $\{q\}$, $\{r\}$ of the Bethe equations corresponding to operators of lengths $L_q$ and $L_r$ with planar energy $E^{(0)}(\{I\})=E^{(0)}(\{q\})+E^{(0)}(\{r\})$.

As a simple example let us consider the unprotected operators of length six in the $[2,2,2]$  $SO(6)$ representation. There are two single-trace operators with planar energies and rapidities given by
\<
E^{(0)}_{(6,2a)}&=&2(5+ \sqrt{5})~, ~~~ u_{(6,2a),1}=-{u_{(6,2a),2}}=\frac{1}{2}\sqrt{1-\frac{2}{\sqrt{5}}}~,\nn\\
E^{(0)}_{(6,2b)}&=&2(5- \sqrt{5})~, ~~~ u_{(6,2b),1}=-{u_{(6,2b),2}}=\frac{1}{2}\sqrt{1+\frac{2}{\sqrt{5}}}~,
\>
both of which mix with the double-trace operator with rapidities $u_{(4,2),1}=-u_{(4,2),2}=1/2\sqrt{3}$ and planar energy $E_{(4,2)}^{(0)}=12$. The overlaps can be simply found from the general formulae \eqref{eq:Vm} and \eqref{eq:Vp} 
\<
& & V^-({ u_{(6,2a)}}; u_{(4,2)}, \emptyset)=\frac{4}{3}(5+3\sqrt{5})~,~~~V^-({u_{(6,2b)}}; u_{(4,2)}, \emptyset)=\frac{4}{3}(5-3\sqrt{5})
\>
and $V^+({u_{(4,2)}}, \emptyset;u_{(6,2a)})=V^+({u_{(4,2)}}, \emptyset;u_{(6,2b)})=6\sqrt{2}$. The resulting non-planar corrections are 
\<
E_{(6,2a)}^{(2)}=8(5+ 2\sqrt{5})~~~\text{and}~~~
E_{(6,2b)}^{(2)}=8(5- 2\sqrt{5})~.
\>
These results are in agreement with those found by direct calculation \cite{Bianchi:2002rw}, and follow  from diagonalising the dilatation operator \cite{Beisert:2003tq}. 

In the case where there are only two magnons we can in fact solve the Bethe equation for any length, $L$, if we consider only cyclic solutions with $\prod_j e^{i p_j}=1$. In terms of the momenta such solutions are given by 
\<
\label{eq:Bethetwo}
p_1=-p_2=\frac{2 \pi n}{L-1}~,~~~
\>
with $n\in \mathbb{Z}$ and $0<n<\frac{L-1}{2}$. Given such a complete set of solutions it is possible to numerically carry out the sum over intermediate states so that we can compute quite efficiently the corrections to energies even for states with quite long lengths, e.g.\ $n=1$ for $L=100, 250, 400$, which to six digits gives
\<
E_{L=\{100,250,400\}}^{(2)}=L^2 \{0.758732, 0.770021, 0.772582\}~.
\>
From this and similar numerical examples it can be seen that the corrections to the energies of long operators scale as $L^2/N^2$. This is essentially the well-known BMN limit \cite{Berenstein:2002jq}  where one considers operators with large R-charge, $J$.  The non-planar corrections to two-magnon states in the BMN limit were computed in \cite{Beisert:2002bb, Constable:2002vq}, see also \cite{Janik:2002bd , Beisert:2002ff} and shown to be 
\<
\label{eq:BMN_two}
\Delta_n=L+g'\Big[16\pi n^2+g_2^2\left(\frac{1}{3}+\frac{35}{8\pi^2 n^2}\right)\Big]~.
\> 
It is straightforward to check that our general expressions reproduce this result by substituting the two-magnon rapidities, $u_{n,1}=-u_{n, 2}=\tfrac{L-1}{2\pi n}$, into \eqref{eq:Vm} and \eqref{eq:Vp} and taking the large $L$ limit. We must consider the overlaps with all double-trace operators consisting of a vacuum state of length $(1-r)L$ and two-magnon states with rapidities  $u_{m,1}=-u_{m, 2}=\tfrac{r L-1}{2\pi m}$. Following \cite{Beisert:2002ff}, we then expand in $L$, sum over $m=0, \dots, \infty$ and approximate the sum over intermediate lengths by an integral over $r$ from $0$ to $1$. At leading order in $J=L-2$ this reproduces \eqref{eq:BMN_two}, while at subleading orders we find
the same result but with $J$ replaced with $L-1=J+1$ which is the natural parameter from the perspective of the Bethe equations. 

It is naturally interesting to consider higher numbers of excitations. For example at $L=7$ with three excitations, i.e.\ for states in the $[3,1,3]$ representation, we have two single-trace operators with planar dimensions $E^{(0)}_{(7,3a/b)}=10$. Due to the degeneracy of the states, a naive application of relation \eqref{eq:pert_coor} will fail as it is not clear which linear combination of the Bethe states to use as planar eigenstates. We may use the fact that the two degenerate states are distinguished by their transformation under the parity operation \cite{Beisert:2003tq}. This operator, $\mathcal{P}$, reverses the order of fields within each trace, for example
\<
\label{eq:parity}
\mathcal{P}:\Tr(XZXXZZ)\mapsto \Tr(ZZXXZX)~,
\>
and commutes with the complete non-planar dilatation operator. Thus the non-planar eigenoperators must have definite parity, and consequently also their planar limits. The rapidities for the two $L=7$ and $M=3$ Bethe solutions $u_{(7,3a)}$ and $u_{(7,3b)}$ can be easily found using the method (and Mathematica programme) of \cite{Marboe:2016yyn}. They can be seen to transform into each other under parity which acts on finite rapidities by $u_i\to -u_i$ while rapidities at infinity are left invariant. The two parity eigenstates can then be formed from the corresponding Bethe eigenstates as $\ket{\pm}=\tfrac{1}{\sqrt{2}}(\ket{u_{(7,3a)}}\pm\ket{{u_{(7,3b)}}} )$. Having identified the proper planar linear combinations, we can proceed by computing the mixing with double-trace operators. We choose as our basis of double-trace operators 
\<
\ket{ u_{(5,3)} }_5\ket{ \emptyset }_2~,~~~ \ket{ u_{(5,2)} }_5\ket{ \infty }_2~,~~~
\ket{ u_{(4,2)} }_4\ket{ \infty}_3
\>
where we have labelled the Bethe states by the magnon rapidities rather than the momenta and $u_{(5,3)}=\{\tfrac{1}{2}, -\tfrac{1}{2}, \infty\}$, and  $u_{(5,2)} =\{ \tfrac{1}{2}, -\tfrac{1}{2} \}$. Both of these operators have positive parity and the linear combination $\sqrt{\tfrac{2}{3}}\ket{ u_{(5,3)} }_5\ket{ \emptyset }_2-\sqrt{\tfrac{1}{3}}\ket{ u_{(5,2)} }_5\ket{ \infty }_2$ is the remaining non-protected operator in the $[3,1,3]$ representation. The other linear combination is a descendant of a two-excitation double-trace operator from $[2,3,2]$. The non-vanishing overlaps following from \eqref{eq:Vm} and \eqref{eq:Vp} are
\<
& & V^-(u_{(7,3a)}; u_{(5,3)},\emptyset)=V^-(u_{(7,3b)}; u_{(5,3)},\emptyset)=2\sqrt{\tfrac{14}{3}}\nn\\
& & V^-(u_{(7,3a)}; u_{(5,2)},\infty)=V^-(u_{(7,3b)}; u_{(5,2)},\infty )=-2\sqrt{\tfrac{7}{3}}\nn\\
& & V^+(u_{(5,3)},\emptyset;u_{(7,3a)})=V^+(u_{(5,3)},\emptyset;{u_{(7,3b)}})=40\sqrt{\tfrac{2}{21}}\nn\\
& & V^+(u_{(5,2)},\infty;u_{(7,3a)})=V^+(u_{(5,2)},\infty ;{u_{(7,3b)}})=-40\sqrt{\tfrac{1}{21}}~,
\>
while the overlaps involving $\ket{ u_{(4,2)} }_4\ket{ \infty}_3$ are all zero, which is expected since primary operators cannot mix with descendants.
Now by applying \eqref{eq:pert_coor} for the parity eigenstates, we find that the non-planar corrections arise from the mixing of the positive-parity eigenstate $\ket{+}$ with the double-trace state (which has planar energy $E^{(0)}=8$) and are given by
\<
{ E_{(7,3+)}^{(2)}}=80~,~~~{E_{(7,3-)}^{(2)}}=0
\>
which agrees with \cite{Ryzhov:2001bp, Beisert:2003tq}.

  The occurrence of degenerate parity pairs in the planar limit is quite general and so to use non-degenerate perturbation theory we must work within sectors of definite parity.
 Unfortunately, as has been noted by several authors, for example in \cite{Constable:2002vq},  \cite{Beisert:2002ff}, \cite{Beisert:2003tq}, the energies following from the Bethe equations demonstrate an additional degeneracy which is relevant to the mixing problem between multi-trace operators. For example, if we consider the two-excitation states with Bethe solution \eqref{eq:Bethetwo}, states with different lengths, $L_a$, $L_b$, and mode numbers, $n_a$, $n_b$, but equal ratios $\tfrac{n_a}{L_a-1}=\tfrac{n_b}{L_b-1}$ will have equal energies. Correspondingly, in the planar limit, the single-trace state corresponding to the spin-chain state 
$\ket{\{ \tfrac{2\pi m}{L-1},-\tfrac{2\pi m}{L-1}\}}_L$
is degenerate with the double-trace state
\<
\label{eq:degen}
\ket{ \{\tfrac{2\pi \tilde{m}}{L-L_1-1},\tfrac{2\pi \tilde{m}}{L-L_1-1} \}}_{ L-L_1} \ket{\emptyset}_{L_1}
\>
with $\tfrac{ \tilde{m}}{L-L_1-1}=\tfrac{ m}{L-1}$. 

While it is less straightforward to show, analogous degeneracies generally also occur for higher excitation numbers. As just one example, if we consider the $L=8$ operators with $M=3$, we see that there are three solutions to the Bethe equations. Two of which, whose rapidities we denote $u_{(8,3a)}$ and $u_{(8,3b)}$, are degenerate parity pairs with energy ${E_{(8,3a/b)}^{(0)}}=8$, while the third is a singular solution with energy ${E_{(8,3s)}^{(0)}}=12$. There is in this case a positive parity double-trace state which is degenerate
\<
\sqrt{\tfrac{3}{4}}\ket{ u_{(5,3)} }_5\ket{ \emptyset }_3-\sqrt{\tfrac{1}{4} }\ket{ u_{(5,2)} }_5\ket{ \infty }_3~
\>
and which mixes with the positive-parity linear combination of single traces. The mixing matrix can be computed from the overlaps and is 
\<
\begin{pmatrix}
	0 & -4\sqrt{15} \\
	-\tfrac{32}{\sqrt{15}} & 0
\end{pmatrix}
\>
from which we can compute the leading corrections to the energies ${ E^{(1)}}=\pm 8\sqrt{2}$. We can now proceed to use the corresponding eigenstates to find the subleading $1/N^2$ corrections. As we proceed to longer lengths and more impurities, the need to diagonalise the mixing matrix will rapidly become difficult. One way to avoid this problem is to deform the theory to remove such degeneracies. In principle, if we can completely solve the deformed problem, one can then hope to remove the deformation parameter however as this requires resumming the $1/N$ corrections before removing the deformation we will only be able to make preliminary steps in this direction. 

There is another reason for considering the deformed theory which has to do with the singular solutions of the Bethe equations. Already at $L=6$ and $M=3$ there is a solution $u_1=i/2$, $u_2=-i/2$ and $u_3=0$ for which the Bethe wavefunction is singular and naive application of the above formulae will lead to unphysical infinities. It is possible to regularize the Bethe equations by the introduction of a twist, see \cite{Nepomechie:2014hma} for a useful discussion and further references, which is equivalent to the deformation parameter we introduce below. We can use the solutions of the twisted Bethe equations and the overlaps of the deformed theory to compute non-planar energies which reproduce the undeformed results in the limit of vanishing deformation. 

\section{\texorpdfstring{$\beta$}{Beta}-deformed SYM Theory}
\label{sec:def}
We now turn to the $\beta$-deformed $\mathcal{N}=4$ SYM theory preserving $\mathcal{N}=1$ supersymmetry. The theory's Lagrangian may be obtained from the undeformed Lagrangian by replacing all products of fields by a Moyal-like $\star$-product where the non-commutativity occurs in the internal $SU(4)$ R-symmetry directions \cite{Lunin:2005jy}. Using $\mathcal{N}=1$ superspace formalism this corresponds to adding a single-trace deformation to the superpotential, however when written in terms of the component fields this results in both single-trace and double-trace deformations of the Lagrangian \cite{Jin:2012np,Fokken:2013aea}. For this theory the $U(N)$ gauge group is no longer conformal at the quantum level due to the couplings of $U(1)$ scalars. These degrees of freedom decouple at the infrared fixed point corresponding to the $SU(N)$ theory and we will thus consider only the $SU(N)$ gauge group. 

In the remainder of this section we use \bc{blue} colour to denote how the $\beta$-deformation changes terms existing in the undeformed theory, and use \pc{purple} to emphasize terms that are new.

\subsection{\texorpdfstring{$\beta$}{Beta}-deformed Dilatation Operator}

The planar dilatation operator for the deformed theory has been previously studied using both integrable methods \cite{Beisert:2005if} and direct field theory computations \cite{Fokken:2013mza}. The non-planar dilatation operator can in principle be directly computed from the deformed Lagrangian using
standard Feynman diagrammatics or perhaps more efficiently using on-shell methods \cite{Caron-Huot:2016cwu}. We instead fix its form by using symmetries and known one-loop results. 
The form of the single-trace part of the dilatation operator is simply inherited from the undeformed theory and is fixed by the planar theory.  It is found by replacing the commutators in \eqref{eq:1loopdil} by the $\beta$-deformed commutator $[.,.]_\beta$ defined via the R-charges of the fields. In the $\mathfrak{su}(2)$ sector spanned by $X=\phi_{14}$ and $Z=\phi_{12}$ the only relevant commutator is
\begin{align}
[X,Z]_\bc{\beta}=\bc{e^{i\beta}}XZ-\bc{e^{-i\beta}}ZX.
\end{align}
This is supplemented by a double-trace term which is necessary to make the theory exactly conformal \cite{Fokken:2013aea}. The form of this term follows from the deformed action \cite{Jin:2012np,Fokken:2013aea} and in the $\mathfrak{su}(2)$ sector it becomes
\<
:\text{Tr}[X,Z]_\bc{\beta} \text{Tr} [\check{X},\check{Z}]_\bc{\beta}:~.
\>
 We fix the coefficient of this term by imposing that the operator $\Tr(XZ)$ is a protected operator
\<
\mathfrak{D}_2 \Tr(XZ)=0~.
\>
This has been shown perturbatively at one- and two-loop level by direct calculation \cite{Freedman:2005cg, Penati:2005hp}. Using these conditions we find that the deformation leaves the tree-level dilatation operator \eqref{eq:treedil} unchanged, while the one-loop correction \eqref{eq:1loopdil} gets deformed to
\<
\mathfrak{D}_2=-\frac{2}{N}\left(:\text{Tr}([X,Z]_\bc{\beta} [\check{X},\check{Z}]_\bc{\beta}):\pc{-\frac{(e^{i \beta}-e^{-i \beta})^2}{N}:\text{Tr}(XZ) \text{Tr}(\check{X}\check{Z}):}\right)~.
\label{eq:dildef}
\>
It is important to note that although the double-trace term is suppressed by $1/N$, it can be relevant at leading order when acting on short operators and results in the vanishing anomalous dimension of $\Tr(XZ)$. For longer operators in the planar limit this term is not relevant, however it is essential in understanding the non-planar corrections.
The fusion and splitting formulas \eqref{eq:spitfus} imply for the action of the one-loop dilatation operator on single-trace states
\begin{align}
\mathfrak{D}_2 \Tr(XAZB)&= \frac{2}{N}\Big( \bc{e^{-i \beta}}~\Tr(A)~\Tr([X,Z]_\bc{\beta} B)-\bc{e^{i\beta}}~\Tr([X,Z]_\bc{\beta} A)~\Tr(B)\Big)\nn\\
&\quad\pc{+\frac{2(e^{i\beta}-e^{-i \beta})}{N^2}\Big(\Tr ([X,Z]_{\beta} \{A,B\})+\Tr([X,Z]_{\beta}) ~\Tr(A)~\Tr(B)\Big)}\nn\\
&\quad\pc{-\frac{4(e^{i\beta}-e^{-i\beta})}{N^3}~\Tr ([X,Z]_{\beta})~\Tr (AB)}~,
\label{eq:dil1}
\end{align}
where the double-trace part of the dilatation operator contributes the triple-trace term at order $1/N^2$.  For the action of the dilatation operator on double-trace states we find
\begin{align}
\mathfrak{D}_2\Tr(XA)\Tr(ZB)&=\frac{2}{N}\Big(\bc{e^{-i\beta}}\Tr([X,Z]_\bc{\beta} BA)-\bc{e^{i\beta}}\Tr([X,Z]_\bc{\beta} AB)\Big)\nn\\
&\quad \pc{+\frac{2(e^{i\beta}-e^{-i\beta})}{N^2}\Big(\Tr(A)~\Tr([X,Z]_{\beta} B)+\Tr([X,Z]_{\beta} A)~\Tr(B)}\nn\\
&\quad\pc{+\Tr([X,Z]_{\beta})~\Tr(AB)\Big)-\frac{4(e^{i\beta}-e^{-i\beta})}{N^3}\Tr([X,Z]_{\beta})~\Tr(A)~\Tr(B)}~.
\label{eq:dil2}
\end{align}

Relations \eqref{eq:dil1} and \eqref{eq:dil2} suggest that the deformed one-loop dilatation operator can be decomposed into planar and non-planar pieces similar to the undeformed case \eqref{eq:decomp}, however we now find subleading contributions and so we  decompose \eqref{eq:dildef} as
\<
\mathfrak{D}_2=H^{(0)}_\bc{\beta}+\frac{1}{N}H^-_\bc{\beta}+\frac{1}{N}H^+_\bc{\beta} \pc{+\frac{1}{N^2}H_\beta^{(2)}+\frac{1}{N^3}H_\beta^{(3)} }.
\label{eq:decombeta}
\>
As for the undeformed case $H_\beta^{(0)}$ leaves the number of traces in an operator unchanged while $H^{\pm}_\beta$ increases/reduces the number of traces. $H_\beta^{(2)}$ and $H_\beta^{(3)}$ are subleading terms which only arise in the deformed theory.  In particular, $H_\beta^{(2)}$ has a contribution which leaves the number of traces unchanged and so we have diagonal overlaps which we consider in \secref{sec:defovlps}.


\subsection{Deformed Planar Theory}
\label{sec:def_plan_spin}

The action of the planar dilatation operator on single-trace operators of length $L>2$ is quite similar to the undeformed action and is given by
\<
\label{eq:def_planar_dil}
H_{\beta}^{(0)}\ket{n_1,n_2,\dots}_L=2\sum_{j=1}^M \Big(2\ket{\dots, n_j, \dots }-\bc{e^{2i\beta}}\ket{\dots, n_j-1,\dots}-\bc{e^{-2i\beta}}\ket{\dots,n_j+1, \dots}\Big)~.
\>
It can be related to the integrable deformation of the Heisenberg XXX-Hamiltonian \cite{Roiban:2003dw, Berenstein:2004ys}
\<
H_D=\sum_{i=1}^L \big[ \unit_{i,i+1}-\sigma_i^z \sigma_{i+1}^z-2\bc{ e^{2i \beta}}\sigma_i^- \sigma_{i+1}^+ -2 \bc{e^{-2i \beta}}\sigma_i^+ \sigma_{i+1}^-\big]~,
\>
 so that the planar spectrum can still be solved using integrability. As the Hamiltonian $H_D$ commutes with $\sum_i \sigma_i^z$ we can still consider sectors with fixed excitation number $M=L-\sum_i \sigma_i^z$. The vacuum corresponding to $M=0$ is the same as in the undeformed theory, i.e.\ \eqref{eq:vac}, and has energy $E^{(0)}(\emptyset)=0$. Similarly, the one-excitation eigenstate is given by the usual Bethe state, but its energy becomes 
\<\label{E0}
E^{(0)}(p)=4(1-\cos(p+\bc{2\beta}))~
\>
which we see is no longer degenerate with $E_0$ when $p=0$. For more excitations the spectrum and wavefunctions are still given by the Bethe ansatz \eqref{eq:Betheansatz} but now with the deformed S-matrix
\<
S_{\beta}(p_j,p_k)=-
\frac{e^{i (p_j+p_k)} \bc{e^{2i\beta}}+\bc{e^{-2i\beta}}-2 e^{i p_k}}{e^{i (p_j+p_k)} \bc{e^{2i\beta}}+\bc{e^{-2i\beta}}-2 e^{i p_j}}~.
\>
The Bethe equations determining the momenta are as in the undeformed theory \eqref{eq:momBE} but with the S-matrix replaced with $S_\beta$ and the trace cyclicity  condition still requires that the momenta satisfy  ${\rm exp}(i \sum_{j}p_j)=1$.
The dependence of the S-matrix on the deformation parameter can be removed by defining the shifted momenta
\<
\tilde p_j=p_j+2\beta
\>
so that
\<
S_{\beta}(p_j,p_k)=-
\frac{e^{i (\tilde p_j+\tilde p_k)}+1-2 e^{i \tilde p_k}}{e^{i (\tilde p_j+\tilde p_k)}+1-2 e^{i \tilde p_k}}.
\>
However, this new parametrisation makes the parameter $\beta$ manifest in the Bethe equations and cyclicity condition. Introducing the rapidity variable $u=\tfrac{1}{2}\cot \tfrac{\tilde p}{2}$ they are given by
\<
\left(\frac{u_j+\tfrac{i}{2}}{u_j-\tfrac{i}{2} }\right)^L \prod_{k\neq j}^M \frac{u_j-u_k-i}{u_j-u_k+i}=\bc{e^{2 i L \beta}}~~~~\text{and}~~~~\prod_{j=1}^M\frac{u_j+\tfrac{i}{2}}{u_j-\tfrac{i}{2}}=\bc{e^{2 i M\beta }}~,
\> 
thus in terms of the rapidity variable both $S_\beta$ and the corresponding function $h_\beta$  can be defined as in the undeformed case, i.e.\ via \eqref{eq:Sinu} and \eqref{eq:hinu}.

One consequence of the deformation is that the degeneracy occurring in the undeformed theory between single-trace and double-trace operators \eqref{eq:degen} is lifted. This can be seen directly in the case of single-trace operators corresponding to two-magnon states, $\ket{\{k,-k\}}_L$, by solving the Bethe equations to the first non-vanishing order in the deformation parameter
\<
k(m,L)=\frac{2\pi m}{L-1} \pc{-\frac{2\beta^2}{L-1}\cot \left(\frac{m \pi}{L-1}\right)+\mathcal{O}(\beta^4)}~, ~~~m\in  \mathbb{Z}~.
\>
For generic real values of $\beta$ there will be no integers $\tilde{m}$ and $L_1$ such that $k(\tilde{m}, L-L_1)=k(m, L)$ and hence no double-trace operator will be degenerate with the single-trace operator.  While we do not have a similar proof for states with more excitations, direct diagonalisation of the dilatation matrix for operators with short lengths shows that the degeneracy of excited states is lifted in all cases of operators which were unprotected in the undeformed theory.  This reduced degeneracy increases the number of operators for which we can compute the non-planar corrections to the energies by using non-degenerate perturbation theory. 


\subsection{Matrix Elements and Dimensions}
\label{sec:defovlps}
The action of the non-planar dilatation operator on Bethe states and the corresponding overlaps can be computed by essentially the same method as for the undeformed theory. For the deformed theory, if we wish to compute the corrections to the energies to order $\mathcal{O}(1/N^2)$ we must consider not only the off-diagonal contributions from $H^\pm_\beta$ but also the diagonal contributions from $H^{(2)}_\beta$.

\paragraph{Off-diagonal Overlaps.}
We can write the overlaps of $H^\pm_\beta$ using the notation of \secref{sec:SCOvlp}. As for the undeformed theory, the solutions of the deformed Bethe equations are invariant under complex conjugation which can be used to simplify the expressions. For $H^-_\beta$ the overlaps are almost identical to \eqref{eq:Hm1}
\begin{align}
\label{eq:Hmdef}
\kern-10pt\langle \{p\}| H_\beta^- | \{q\} \rangle | \{r\} \rangle&=2 L_q L_r \, \mathcal{N}(p^\ast,q,r)\times \\
&\kern-80pt \Bigg[
\sum_{\substack{i,j \\ s \cup t = \{p\}_{\hat j}}}
 \frac{e^{i p^\ast_j} \bc{e^{ 2i\beta}}-1}{h^{q_i q_{\hat i}}}
\big[ \bc{e^{-2i \beta} }s^{L_{q+1}\, \ast}_{j\, \circlearrowleft} - t^{L_r+1\, \ast}_{j\, \circlearrowright}\big]  (s|\{q\}_{\hat{i}})_{L_q-1} (t|\{r\})_{L_r-1}+\{\text{terms~with~} q \rightleftarrows r\}\Bigg]\nn~.
\end{align}
The function $h$ in this formula has exactly the same form as in the undeformed theory \eqref{eq:hinu} when written in terms of the on-shell rapidities, and similarly for the S-matrices implicit in the scalar products. We should remember however that the  rapidities themselves depend on the deformation parameter through the Bethe equations.  The overlaps of $H_\beta^+$ involve additional contributions in the case where one of the traces has length two and are given by
\begin{align}
\label{eq:Hpdef}
\langle \{r\}|\langle \{q\}|  H_\beta^+ | \{p\} \rangle&= 2 L_p\, \mathcal{N}_+ (p,q^\ast,r^\ast) \, \Bigg[
\sum_{\substack{i,j \\ s \cup t = \{p\}_{\hat j}}}
\frac{1}{h^{q^\ast_{\hat i}q_i^\ast}}\Big[
(e^{iq^*_i} \bc{e^{2i\beta}}-1)( \bc{e^{-2i \beta} } s^{L_q-1}_{j\,\circlearrowleft}-t^{L_r+1}_{j\, \circlearrowright})\\
&\kern-30pt \pc{-4\delta_{Q,1} \delta_{L_q,2} \sin^2 \beta(e^{i q^*_i} s^{L_q-1}_{j\,\circlearrowleft}+ t^{L_r+1}_{j\, \circlearrowright}) } \Big] 
( \{ q \}_{\hat{i}}|s)_{L_q-2}
( \{r\} |t)_{L_r}+\{\text{terms~with~} q \rightleftarrows r\}\Bigg]~.\nn
\end{align}
As in the undeformed theory, dividing by the norms of the external states we can define the normalised overlaps $V_\beta^\pm$.

\paragraph{Diagonal Overlaps.}

The contribution $H_\beta^{(2)}$, which does not occur in the undeformed theory, to the dilatation operator \eqref{eq:decombeta} contains both length-preserving and -changing parts. Here we are interested in the former, since the computation of non-planar corrections at order $1/N^2$ to the anomalous dimensions requires solely the diagonal overlap $\braket{\{p\}|H^{(2)}_\beta}{\{p\}}$. 
Using \eqref{eq:Zfirst} one finds for the action of $H_\beta^{(2)}$ on a Bethe state \eqref{eq:BetheES}
\begin{align}
&\pc{H^{(2)}_\beta\ket{\{p\}}}=2L_p(e^{i\beta}-e^{-i\beta})\sum_{x=2}^{L_p}\sum_{l=1}^P\sum_{\substack{2\leq n_1<...<n_{l-1}<n_l= x \\ x<n_{l+1}<...<n_P\leq L_p}}\psi^{\{p\}}_{\{n\}}~\nn\\
&\hspace{1cm}\times\Big(\ket{[X,Z]_\beta}\otimes\ket{n_1+1,...,n_{l-1}+1}_{x-2}\otimes\ket{n_{l+1},...,n_P}_{L_p-x}\nn\\
&\hspace{2.0cm}+\ket{n_1-1,...,n_{l-1}-1}_{x-2}\otimes\ket{[X,Z]_\beta}\otimes\ket{n_{l+1},...,n_P}_{L_p-x}-2\delta_{L_p,2}\ket{[X,Z]_\beta}\Big)\nn\\
&\hspace{1cm}+\{\text{double-trace~terms}\},
\end{align}
where the $\delta_{L_p,2}$-term arises from the enhanced contribution of the last double-trace term in \eqref{eq:dil1}. The diagonal overlap can then be written in terms of ordered partitions \eqref{eq:opartitions} as
\begin{align}
&\pc{\braket{\{p\}|H^{(2)}_\beta}{\{p\}}}=2L_p\delta_{L_p\neq 2}~\mathcal N(p^*,p)(e^{i\beta}-e^{-i\beta})\sum_{\rho,\sigma}\frac{e^{i\beta} e^{i p_{\rho(1)}^*}-e^{-i\beta}}{h_{>}^{\{p_\rho^*\}}h_{<}^{\{p_\sigma\}}}\nn\\
&\hspace{3.3cm}\times\sum_{x=2}^{L_p}\sum_{l=1}^P \left(e^{i(x-1) p_{\sigma(1)}}\prod_{k=2}^l S_\beta\left(p_{\sigma(1)},p_{\sigma(k)}\right)+1\right)e^{i(x-1)(p_\sigma)_{l+1}^P}e^{-i(x-2)(p^*_\rho)_{l+1}^P}\nn\\
&\hspace{3.3cm}\times P_{x-1}\left(\{p_{\sigma}-p^*_{\rho}\}_2^l\right)P_{L_p-x+1}\left(\{p_{\sigma}-p^*_{\rho}\}_{l+1}^P\right)~.
\end{align}
Note that $P_L(z)$ vanishes for $|z|\geq L$, cf.\ \eqref{eq:opartitions}.
In terms of scalar products of normalised Bethe states \eqref{eq:Betheproduct} the overlap can be written as
\begin{align}
\label{eq:Hd2}
&\pc{\braket{\{p\}|H^{(2)}_\beta}{\{p\}}}=2L_p\delta_{L_p\neq 2}~\mathcal N(p^*, p)(e^{i\beta}-e^{-i\beta})\nn\\
&\hspace{1.2cm}\times\sum_{\substack{k,l=1\\\kappa\cup\bar\kappa=\{p\}_{\hat k}\\ \lambda\cup\bar\lambda=\{p^*\}_{\hat l}}}^P\sum_{x=2}^{L_p}
\frac{e^{i\beta} e^{ip_l^*}-e^{-i\beta}}{h^{p_k\bar\kappa}h^{\kappa\bar\kappa}} 
\left(\frac{e^{-i(x-1) \kappa}}{h^{\kappa p_k}}+\frac{e^{i(x-1) \bar\kappa}}{h^{p_k\kappa}}
\right)\frac{ e^{-i(x-2)\bar\lambda}}{h^{p^*_{\hat l}p^*_l}h^{\bar\lambda\lambda}}(\lambda|\kappa)_{x-2}(\bar\lambda|\bar\kappa)_{L_p-x}~,
\end{align}
where $\kappa$ and $\lambda$ are of the same cardinality, which must be smaller than $x-1$. Finally we can divide by the square of the norm of the external state, $\|p\|^2$, to defined normalised overlaps $\pc{V^{(2)}_\beta(\{p\})}$ which can be then used to compute the energy corrections of single trace states. 

\paragraph{Anomalous Dimensions.} As the deformation lifts many of the degeneracies present in $\mathcal{N}=4$ SYM we can use the overlaps in the deformed theory to compute the corrections to energies for a wide range of states by using the deformed analogue of \eqref{eq:pert_coor}
\<
\label{eq:pert_coor_def}
E_{\nc{\beta}}^{(2)}(\{p\})=\sum_{\{I\}}
\frac{ V_{\nc{\beta}}^-(p; I)
	V_{\nc{\beta}}^+(I; p)}{E_{\nc{\beta}}^{(0)}(\{p\})-E_{\nc{\beta}}^{(0)}(\{ I\})}+\pc{V^{(2)}_\beta(\{p\})}\,.
\>
The additional input to such a calculation are the solutions to the deformed Bethe equations. Solving the deformed Bethe equations is generally a non-trivial task, however for short lengths it can be done either for specific numerical values of $\beta$ or by starting with the undeformed result and perturbatively solving for $\beta \ll 1$. The latter is particularly useful when we wish to use the deformation as a regulator of singular solutions of the undeformed Bethe equations. One must be careful with the order of limits as the one-loop anomalous dimensions are functions of both $\beta$ and $N$ and we may choose to first expand in large $N$ and then small $\beta$, $E(\beta \gg N^{-1})$, or alternatively first in small $\beta$ and then large $N$, $E(\beta \ll N^{-1})$. In general these expansions will not commute. 

For example, let us consider the $L=6$, $M=3$ single-trace operator described in the planar undeformed theory by the roots $\{u_1=0, u_2=-i/2, u_3=i/2\}$ with planar energy $E^{(0)}=12$. This solution is singular as it has rapidities separated by $i$. It has a vanishing $1/N^2$ correction to the energy due to the $\mathfrak{su}(2)$ symmetry which ensures there is no other operator with which it can mix.  We can study the same operator in the deformed theory where the mixing problem is non-trivial and we find from direct diagonalisation that through $\mathcal{O}(N^{-4})$, and keeping only the leading terms in the $\beta$-expansion, we have
\<
E(\beta \gg N^{-1})=(12-72 \beta^2 +\mathcal{O}(\beta^4))+\frac{1}{N^2}\big[-\frac{2304}{23} +\mathcal{O}(\beta^2)\big]+\frac{1}{N^4}\big[\frac{400896}{12167\beta^2}+\mathcal{O}(\beta^0)\big]~.    
\>
In this expression we can see that the leading non-planar term does not reduce to the vanishing $1/N^2$ undeformed answer in the $\beta\to 0$ limit and in fact the $1/N^4$ term is singular. There will be additional singular terms at subsequent powers in the $1/N$ expansion that would need to be resummed to recover the smooth limit.  

As the wavefunction in the deformed theory is perfectly regular we can use \eqref{eq:Hmdef, eq:Hpdef} and \eqref{eq:Hd2} to compute the overlaps, then take the $\beta \to 0$ limit and use these expressions to perturbatively compute the undeformed non-planar correction. To be explicit, we need to solve for the deformed rapidities to $\mathcal{O}(\beta^6)$ to find a non-singular wavefunction since
\<
u^\beta_3-u^\beta_2=i+24576 i \beta^6+\mathcal{O}(\beta^{8})~
\>
and in general for length $L$ singular solutions we need $\mathcal{O}(\beta^L)$ to resolve the singularity. This solution mixes with the double-trace operator $\ket{ u_{(4,2)} }_4\ket{u_{(2,1)}}_2$ of planar energy $E^{(0)}_{(4, 2)}=12-32/3\beta^2+\mathcal{O}(\beta^4)$ with normalised overlaps
\<
& & V^+_\beta=-48\sqrt{2}\beta~~~\text{and}~~~ V^-_\beta=-64 \sqrt{2}\beta~
\>
where we have kept only the first leading term in the $\beta$-expansion. The deformation is particularly important when calculating the norm of the singular state, which diverges in the $\beta \to 0$ limit. However the overlaps themselves are smoothly vanishing in this limit and so, consistent with the symmetries, there is no mixing in the undeformed theory. If we instead use these overlaps in the perturbative formula \eqref{eq:pert_coor} we find a cancellation between the powers of $\beta$ in the overlaps and the energy differences. As the diagonal contribution is of order $\beta^2$, it gives no leading contribution and we find $E^{(2)}(\beta \gg N^{-1})= -\frac{2304}{23}+\mathcal{O}(\beta^2)$ in agreement with the result from direct diagonalisation.

Let us now turn to the $L=8, M=3$ singular Bethe state $\ket{u_{(8,3s)}}$ with planar energy $E^{(0)}_{(8,3s)}=12$. This operator is not protected by symmetry in the undeformed theory. Instead it mixes with the double-trace operator $\ket{u_{(6,3s)}}_6\ket{\emptyset}_2$ made up of the length-$6$ singular state and length-$2$ vacuum. From directly diagonalising the corrected energies are
\<
E_\pm=12\pm \frac{12}{N}~,
\>
where we see that due to the degeneracy the correction is $\mathcal{O}(N^{-1})$. Due to the singular nature of the Bethe solution we cannot directly use the overlap formulae of $\mathcal{N}=4$ SYM but again we can compute the mixing matrix using the regularised singular states in the deformed theory. We find that in this case they are non-vanishing in the $\beta \to 0$ limit
\begin{align}
V^+_\beta=4\sqrt{6}+\mathcal{O}(\beta^2)~~~\text{and}~~~ V^-_\beta=6\sqrt{6}+\mathcal{O}(\beta^2)~
\end{align}
and give the correct $1/N$ corrections. This corresponds to the case where $\beta \ll N^{-1}$. 

Alternatively, in the deformed theory as the degeneracy between the two states is lifted, with the planar energy of the single-trace state becoming $E^{(0)}_{(8,3s)}=12-36\beta^2+\mathcal{O}(\beta^4)$, we can use the same overlaps in non-degenerate perturbation theory in the small $\beta$-limit with $\beta\gg N^{-1}$ to find the $1/N^2$ corrections in the deformed theory. The contribution of the overlaps between the two regularised singular states to $E_{(8,3s)}^{(2)}$ is $+4/\beta^2$, i.e.\ it is singular in the limit $\beta\rightarrow 0$. There are additional overlaps with other double trace states however they are subleading as is the diagonal contribution which is $\mathcal{O}(\beta^4)$. Thus for $\beta\gg N^{-1}$ the non-planar corrections start at order $N^{-2}$ but are singular in the $\beta\to 0$ limit. This demonstrates that, in general, the two limits $\beta\rightarrow 0$ and $N^{-1}\rightarrow 0$ do not commute. 

\subsection{BMN Limit}

\begin{figure}[t]
	\centering
	\includegraphics[width=0.6\linewidth]{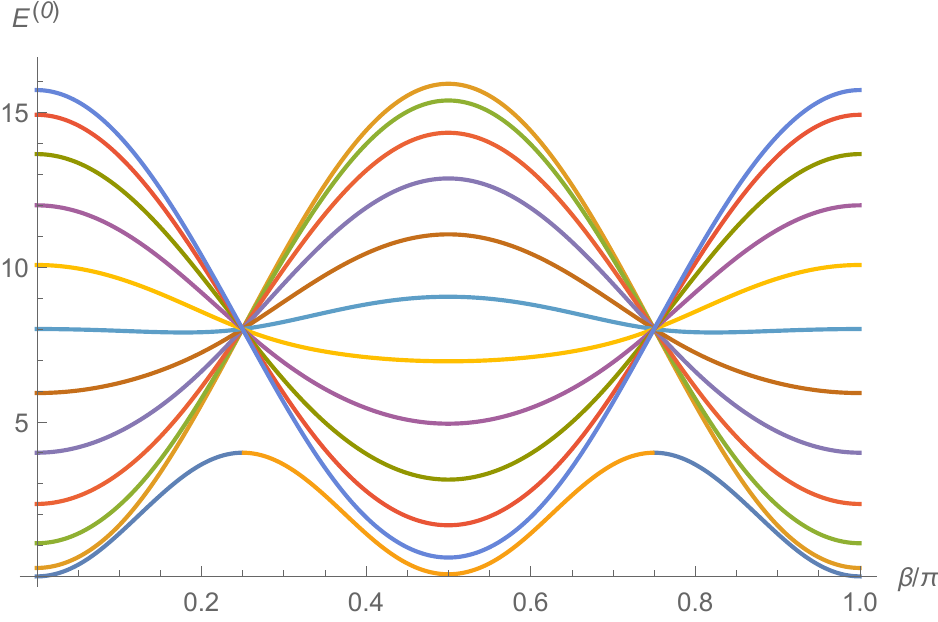}
	\caption{Planar energy levels for $L=25$ and two excitations in the deformed theory. While in the BMN regime we take the limit of small $\beta$, this plot already hints at the different nature of the zero-mode solution corresponding to the lowest energy.} \label{EnergiesM2}
\end{figure}

We will now look at the two-excitation single-trace solutions and their non-planar corrections in the BMN limit \cite{Berenstein:2002jq} of the deformed theory. First of all, we need to analyse carefully the rapidities that solve the Bethe equations. 
The solutions are periodic in the deformation parameter with period $\pi$ and they are symmetric around $\beta=\pi/2$. In general the solutions are parametrized by an integer $n$ which is given as
\begin{equation}
2 \pi n =L p_1 - i \log(S_\bc{\beta}(p_1,p_2))  \,.
\end{equation}
As can be seen in Figure \ref{EnergiesM2}, all but one of the energy levels become degenerate when the deformation parameter equals $\pi/4$.
Before that point the mode number $n$ is in the range $[0,\lfloor L/2 \rfloor -1]$. The solutions with positive mode numbers correspond to deformations of the primary operators in $\mathcal N=4$ SYM, while the zero mode becomes a descendant in the undeformed theory. After the crossing point the mode number $n$ takes values in the range $[1,\lfloor L/2\rfloor]$, with the lowest-energy state now corresponding to $n=\lfloor L/2\rfloor$.

Here we focus on the BMN limit where the deformation parameter scales as
\begin{equation}\label{ScaledBeta}
\beta=\pi b /L\,, \qquad \mbox{with $b$ fixed.}
\end{equation}
Effectively we thus concentrate on a regime of small deformations, with the mode number $n$ in the range $[0,\lfloor L/2 \rfloor -1]$ as described above. We can solve the Bethe equations perturbatively, and find that the rapidity for a strictly positive mode number $n$ is given by
\begin{equation}
u_n = \frac{L}{2(\bc{b}+n)\pi} \left(1+\frac{\bc{b}-n}{n L} + \frac{(\bc{b}+n)\left(\bc{3b(b-n)}-(\bc{b}+n)n^3\pi^2\right)}{3 n^3 L^2} +\ldots\right)\,,
\end{equation}
with momentum conservation requiring the other excitation in the solution to be $u_{-n}$. Meanwhile, the zero-mode solutions have a distinct expression where the expansion parameter becomes the square root of the length
\begin{equation}
\pc{u_0^\pm} = \frac{L}{2 {b} \pi} \left(1 \pm \frac{i }{L^{1/2}} - \frac{1}{L} \pm\frac{i({b^2} \pi^2-3)}{6 L^{3/2}}-\frac{2 {b^2} \pi^2}{3 L^2}\ldots \right) \,.
\end{equation}
The planar energies in the BMN limit can then be computed through \eqref{E0}, yielding
\begin{align}\label{Es2}
E_n^{(0)} &=  \frac{16 \pi^2(\bc{b^2}+n^2)}{L^2}\left(1 +\frac{2(n^2\bc{-b^2})}{(\bc{b^2}+n^2)L}+ \frac{3(3n^4\bc{-2b^2n^2-b^4})-(n^4\bc{+6b^2n^2+b^4})n^2\pi^2}{3(\bc{b^2}+n^2)n^2 L^2}+\ldots \right)\,,\nonumber\\
\pc{E_0^{(0)}} & \nc{= \frac{16 {b^2}\pi^2 }{L^2} \left( 1-\frac{1}{L}-\frac{(3+2{b^2}\pi^2)}{3L^2}+\ldots\right)}\,.
\end{align}
Note that despite the unusual expansion of the zero-mode rapidities $u_0^\pm$, the expansion of the corresponding energy is free of any square roots. 
Finally, while at the leading order the rapidities $u_0^\pm$ seem to be only a particular case of $u_{\pm n}$, it is important to note that the expression for the Gaudin norm, denoted here by $N_\psi=||\psi||^2$, differs already at the leading order by a factor of two 
\begin{align}
N_n &= L^2 \left(1-\frac{\bc{b^2} +n^2}{n^2 L} \pc{+\frac{2 b^2(n^2-b^2)}{n^4 L^2}+\ldots}\right)\,, \nonumber\\
\pc{ N_0\, }  & \nc{ = 2 L^2 \left(1 + \frac{ b^2 \pi^2-3}{3 L}+ \frac{2b^2\pi^2(4b^2 \pi^2-15)}{45 L^2}+\ldots\right)}\,.
\end{align}

\begin{figure}[t]
	\centering{
		\includegraphics[width=0.40\linewidth]{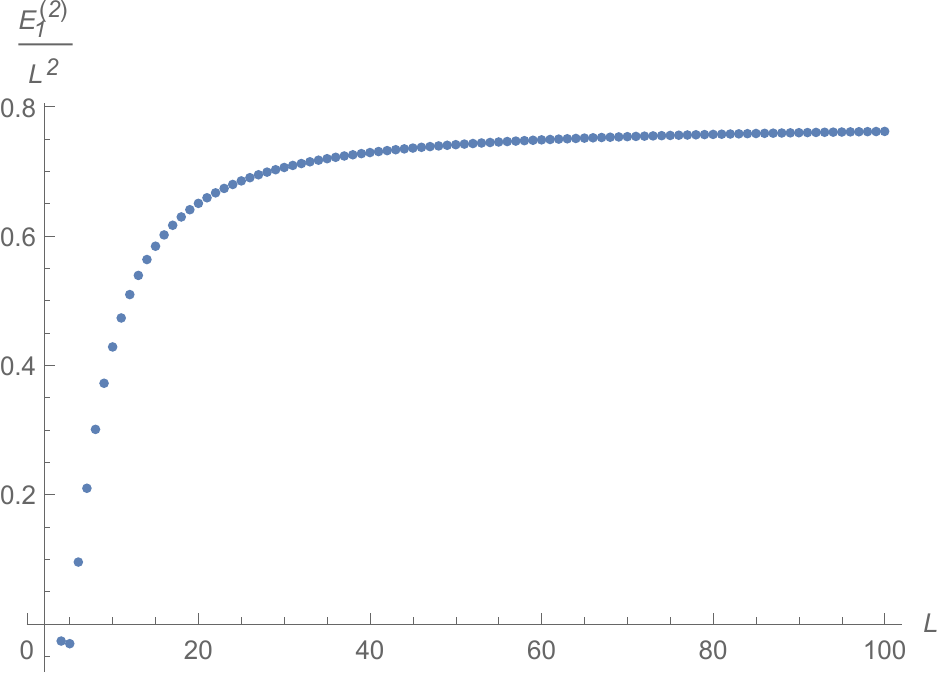} \hspace{5mm}\includegraphics[width=0.40\linewidth]{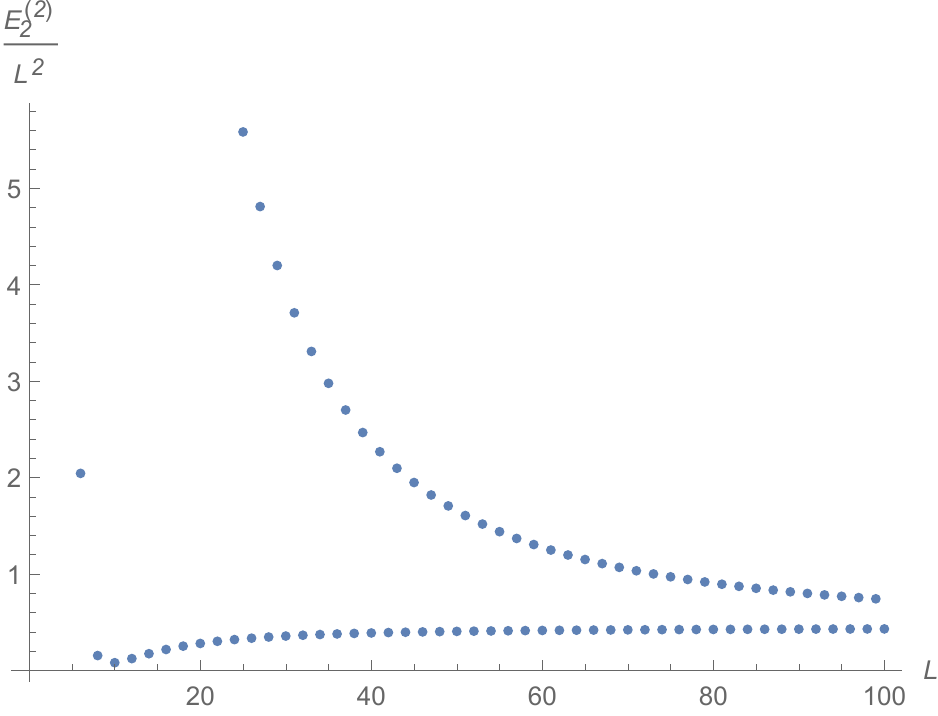}
	}	\caption{We compute the non-planar correction to the energies $E_n^{(2)}$ from the dilatation operator overlaps. We focus here on mode number $n=1,2$ for single-trace operators up to length 100. We  observe that for $n=2$ the large $L$ limit is approached differently for even- and odd-length operators. However, fitting the two curves with polynomials in $1/L$ we find that the mismatch in the coefficients starts only at the subsubleading order.} \label{EsBMN}
\end{figure}

Now that we understand the behaviour of the Bethe solutions, we can study the non-planar corrections to the energies in the BMN limit. The strategy is to expand the dilatation operator overlaps obtained in section \ref{sec:defovlps} and plug them into \eqref{eq:pert_coor_def} written explicitly as
\begin{equation}\label{formulaEs}
E_{\psi}^{(2)}= \sum_{\psi'}\frac{\langle \psi |H^-_\bc{\beta} | \psi'\rangle \langle \psi' |H^+_\bc{\beta} | \psi\rangle}{N_\psi N_{\psi'} \left(E_\psi^{(0)}- E_{\psi'}^{(0)} \right)}+\pc{\frac{\langle \psi| H_\beta^{(2)} | \psi \rangle}{N_\psi}} \,.
\end{equation}
In the deformed theory there are three contributions that we need to consider:
\begin{enumerate}
	\item Off-diagonal overlaps $H_\beta^\pm$ with double-trace operators where:
	\begin{enumerate}[label=(\Alph*), ref=\theenumi(\Alph*)]
		\item Both excitations end up on the same trace, \label{offdiagonalH1}
		\item The excitations split over the two traces. \label{offdiagonalH2}
	\end{enumerate}
	\item Diagonal overlap $H_\beta^{(2)}$. \label{diagonalH}
\end{enumerate}
In general the two-excitation overlaps corresponding to $H^-_\beta$ and $H^+_\beta$ scale at most as $L$ and $L^2$, respectively. Meanwhile, the sum over intermediate states $\psi'$ includes a sum over the splitting of the lengths $(L',L-L')$ in the double-trace operator, which can be approximated by the Euler-MacLaurin formula
\begin{equation}\label{EulerMacLaurin}
\sum_{L'=a}^b f(L') \approx L \int_{a/L}^{b/L} \mathrm d r f(r) + \frac{f(a)+f(b)}{2} + \ldots \,,
\end{equation}
thus leading to a further factor of $L$. Therefore, the one-loop non-planar energies $E_\psi^{(2)}$  scale at most as $L^2$ which, combined  with the colour factor $1/N^2$, produces at leading order in the BMN expansion a factor of $g' g_2^2$, with the relevant expansion parameters introduced in \eqref{gs}.
While in principle we can find the BMN corrections to the overlaps at any subleading order, the expansion of the energies eventually breaks down due to the approximation of the summation over intermediate states by an integration. More precisely, we find that for mode number $n>1$ the sub$^k$-leading BMN correction to the integrand with $k>1$  has simple poles at lengths $L'/L = n' /n$ with $n'=1,\ldots,n-1$. As explained in Figure \ref{EsBMN} this failure of the BMN expansion is in fact expected and agrees with the numerical experiments performed.

Let us then start with the configuration of uneven splitting \ref{offdiagonalH1}. We consider first an external state with positive mode number $n$ while the intermediate double-trace operator has two excitations on the trace of size $L' = r L$ and is described by another positive mode number $n'$. 
While $L'$ is smaller than $L$, the deformation parameter for the double-trace solution is still expanded in terms of the length $L$ of the single-trace operator as in \eqref{ScaledBeta}, so the rapidities, energies and norms of the double-trace states are written as 
\begin{equation}
u_{n'}'(L,b) = u_{n'}( r L, r b) \,, \qquad {E'}_{n'}^{(0)}(L,b)= E_{n'}^{(0)}(r L, r b)\,, \qquad N_{n'}'(L,b) = N_{n'}(r L, r b)\,,
\end{equation}
and analogously for the zero-mode expressions. The overlaps in the BMN limit then become
\begin{align}
H_{nn'}^- &= \frac{32 (r-1) r^3 n^2\sin^2(\pi r n)L^2}{(n'^2-r^2 n^2)} \left(1+ 2\frac{(r-1) n^2 n'^2 \bc{-r(n'^2-r n^2)b^2}}{rn^2(n'^2 - r^2 n^2 )L} \right. \nonumber\\
& \hspace{5.4cm}\left. {}+ \pi \cot(\pi r n)\frac{(2r-1)n^2\bc{+(3-2r) b^2}}{nL} + \ldots \right)\,,\nonumber\\
H_{n'n}^+ &= \frac{32 n'^2\sin^2(\pi r n)L}{(n'^2-r^2 n^2)} \left(1 + 2r\frac{(r-1) n^2 n'^2 \bc{-r(n'^2-r n^2)b^2}}{n'^2(n'^2 - r^2 n^2 )L} \right.  \nonumber\\
& \hspace{3.8cm} \left.{}+ \pi \cot(\pi r n)\frac{(2r-3)n^2\bc{+(1-2r)b^2}}{n L} + \ldots \right)\,.
\end{align}
On the other hand, if the intermediate double-trace operator consists of a zero-mode solution, then the $H_\beta^+$ overlap becomes suppressed and we have instead
\begin{align}
H_{n0}^- &= -32  (r-1) r\sin^2(\pi r n) L^2 + \ldots\,,\nonumber\\
\pc{H_{0n}^+} &\nc{=\frac{32 b^2 \sin^2(\pi r n)}{r n^2 }+\ldots}\,.
\end{align}
The contribution to the non-planar energies is the combination of those two cases, leading to
\begin{align}\label{Ena}
E_{n,A}^{(2)} &= L\int_{2/L}^{(L-2)/L} \mathrm d r \left(  \frac{H_{n0}^- H_{0n}^+}{N_n N'_0 \left(E_n^{(0)}-{E'}_0^{(0)}\right)}+ \sum_{n'=1}^\infty \frac{H_{nn'}^- H_{n'n}^+}{N_n N'_{n'} \left(E_n^{(0)}-{E'}_{n'}^{(0)}\right)} \right) \nonumber\\
&=\left(\frac{1}{3} + \frac{35}{8 \pi^2 n^2}\right) L^2\left(1 + \frac{\bc{4 b^2}-2 n^2}{n^2 L}+\ldots \right) \,.
\end{align}
Note that in this particular case the subleading correction of the Euler-McLaurin formula is vanishing. Furthermore, the contribution of the intermediate zero-mode is crucial for the simplicity of this formula, which would otherwise be plagued by more complex functions such as $\int_0^{z} \mathrm dt \sin(t)/t$. Finally, taking $b=0$ yields the non-planar correction to the two-excitation energies of $\mathcal N=4$ SYM. In that theory the rapidities are more naturally written as an expansion in even powers of $1/(L-1)$, thus matching the result obtained here. 
As explained above, the integral approximation at the subsubleading order is not well defined for $n>1$, and that is manifested here by the presence of poles in the integrand. However, the expression for $n=1$ appears to be well defined at any order, thus allowing us to obtain
\begin{align}
E_{1,A}^{(2)} =&\frac{105(1\bc{-12 b^2+15 b^4})-(1\bc{+144 b^2-9 b^4})\pi^2-8(3\bc{+4 b^2+b^4}) \pi^4-288 (1+\bc{b^2})^2 \pi^2\zeta_3}{24 \pi^2 } \nonumber\\
&+ \frac{(105+8 \pi^2)(\bc{2b^2}-1)}{12\pi^2}L +\left(\frac{1}{3} + \frac{35}{8 \pi^2}\right) L^2
\,.
\end{align}
This expression matches the coefficients in the fit to the data of Figure \ref{EsBMN} to 8 digits of precision.

Still considering the configuration of uneven splitting \ref{offdiagonalH1}, we focus now on the case when the external state is a zero-mode. If the intermediate operator is a zero-mode as well, then each of the overlaps is suppressed by a power of $L$ and given by
\begin{align}
\pc{H_{00}^-} & \nc{= 32 \pi^2(r-1)(r-2)r^2 b^2 L\left(1+\frac{6r^2-6(r-2)+\pi^2 r (r-2)(3+r(2r-3))}{6r(r-2)L}+ \ldots\right)}\,,\nonumber\\
\pc{H_{00}^+ }& \nc{=32 \pi^2 r b^2\left(1+\frac{-12+ \pi^2 r (3+r(2r-3))}{6 rL}+\ldots\right)}\,.
\end{align}
Notice that in this case the difference of planar energies $E_0^{(0)}-{E'}_0^{(0)}$ vanishes at the leading order, therefore enhancing the contribution of the intermediate zero-mode to $E_0^{(2)}$ by a factor of $L$.
In principle we would need to consider also the off-diagonal overlaps with a positive mode state, $H_{0n'}^-$ and  $H_{n'0}^+$, but each of them starts contributing at $L^0$. For that configuration the difference $E_0^{(0)}-{E'}_{n'}^{(0)}$ is once again $\mathcal O(L^2)$ and so we can safely ignore the contribution of these modes at the order we wish to consider.
The non-planar correction to the energy of the zero-mode solution is then given by
\begin{align}\label{E0a}
\pc{E_{0,A}^{(2)}} &= L\int_{2/L}^{(L-2)/L} \mathrm d r \left(  \frac{H_{00}^- H_{00}^+}{N_0 N'_0 \left(E_0^{(0)}-{E'}_0^{(0)}\right)}\right) +8 \pi^2 b^2\nonumber\\
&\nc{=\frac{20 \pi^2 b^2 L}{3} \left(1+\frac{2 \pi^2 b^2-225}{25 L}+\ldots \right) }\,,
\end{align}
where the second term of the first line corresponds to the subleading correction in the Euler-MacLaurin formula \eqref{EulerMacLaurin}. As expected $E_{0,A}^{(2)}$  vanishes in the undeformed theory, where the operator becomes a descendant of the chiral primary.

In order to study the second splitting configuration \ref{offdiagonalH2}, where $H_\beta^+$ leads to double-trace operators with an excitation in each of the traces, we need to consider the single-excitation solution in more detail. The rapidity in that case is given by
\begin{equation}
\pc{u} = \frac{L}{2 b \pi} \left(1- \frac{b^2 \pi^2}{3 L^2} +\ldots\right)\,.
\end{equation}
Note that $L$ here is again the length of the external single-trace operator as this solution turns out to be independent of the length of the trace it describes and is defined solely in terms of the scaled deformation parameter from \eqref{ScaledBeta}. The energy of this state is
\begin{equation}\label{E1}
\pc{E^{(0)}}=\frac{8 b^2\pi^2 }{L^2} \left(1-\frac{b^2 \pi^2}{3 L^2} +\ldots\right)\,,
\end{equation}
and the norm is simply the length of the operator. It is important to remember that the single-excitation solution of length 2, $\Tr(ZX)$, is an exception to this formula. The operator is in fact protected due to the contribution of the double-trace term to the planar dilatation operator.

When the single-trace operator corresponds to a non-zero mode, the overlaps are
\begin{align}
\pc{H_n^+ }& \nc{= \frac{16 \pi^2 b^2}{L} \left( \cos(2 \pi r n)-\frac{b^2}{n^2}  
	\right) + \ldots}\,,\nonumber\\
H_n^- & = 32 r (r-1) \sin^2(\pi r n)L^2  + \ldots
\,.
\end{align}
We can already see that this will contribute at the subsubleading order, so it suffices to consider the leading integral approximation which gives
\begin{align}\label{Enb}
\pc{E_{n,B}^{(2)}}&= \frac{L}{2} \int_{0}^{1} \mathrm d r  \frac{H_n^- H_n^+}{N_n r(1-r) L^2 (E_n^{(0)}-2 E^{(0)})} \nonumber\\
&\nc{= \frac{4 b^2}{n^4 } \left(n^2+2b^2 \right) + \ldots } \,.
\end{align}

Meanwhile, for an external zero mode, the overlaps are
\begin{align}
\pc{H_0^-} &\nc{= 16 \pi^2 b^2(1-r) r L\left(1+2r(r-1)+\frac{12 r (r-1) + (1+6r(r-1)+4 r^2(r-1)^2)b^2 \pi^2}{6L} +\ldots \right)}\,,\nonumber\\
\pc{H_0^+} & \nc{= 16 \pi^2 b^2 \left(1+\frac{\pi^2 b^2}{6 L}+ \ldots \right)}\,.
\end{align}
We now wish to perform the sum over operators as in equation \eqref{formulaEs}.
Looking at the expressions for the planar energies \eqref{Es2,E1}, we see that $E_0^{(0)}-2 E^{(0)}$ vanishes at leading order, which effectively enhances the leading order non-planar correction of the energy to $\mathcal O(L^1)$ (note that both  $H_0^+$ and $H_0^-$ are subleading).
However, this reasoning does not apply when one of the traces has length 2 due to the double-trace term of the dilatation operator, and so that particular double-trace operator contributes at a further subleading order. Therefore the non-planar correction to the energy from this splitting configuration becomes 
\begin{align}\label{E0b}
\pc{E_{0,B}^{(2)}}&=\frac{L}{2}\int_{3/L}^{(L-3)/L} \mathrm dr \frac{H_0^- H_0^+}{N_0 r(L-r) L^2\left(E_0^{(0)}-2 E^{(0)}\right)} -4 \pi^2 b^2 \nonumber\\
&\nc{=-\frac{8 \pi^2 b^2 L}{3} \left(1-\frac{120+7 \pi^2 b^2}{15L}+\ldots\right) }\,,
\end{align}
where, as before, the second term of the first line corresponds to a non-vanishing subleading contribution in the Euler-McLaurin approximation. Taking $b$ to zero we see that both $E_{n,B}^{(2)}$ and $E_{0,B}^{(2)}$ vanish as expected. In that limit the single-excitation solutions correspond to descendants of the undeformed theory, and so these splitting configurations are not expected to contribute to the non-planar corrections of energies in  $\mathcal N=4$ SYM.

Finally, in the deformed theory we should also consider the diagonal contribution of the dilatation operator \ref{diagonalH}. However, the overlap grows linearly with $L$, and so its normalized contribution to the non-planar energy starts only at $\mathcal O(1/L)$, which goes beyond the order we wish to consider here. Therefore, the non-planar correction to the energy of two-excitation single-trace operators is, at $\mathcal O(L^0)$ in the BMN limit, given by the sum of the off-diagonal uneven splittings from \eqref{Ena,E0a} with the symmetric ones in \eqref{Enb,E0b}
\begin{align}
E_{n}^{(2)} &= E_{n,A}^{(2)} + E_{n,B}^{(2)}\,, \nonumber\\
E_{0}^{(2)} &= E_{0,A}^{(2)} + E_{0,B}^{(2)}\,.
\end{align}
We can then write the scaling dimensions of two-excitation states in the BMN limit of large $R$-charge $J=L-2$, which for the non-zero modes yields
\begin{align}
\Delta_n = L + g' &\Bigg[16 \pi^2 (n^2\bc{+b^2} )\left(1 - \frac{2 (n^2\bc{+3b^2})}{(n^2 \bc{+b^2})J}+\mathcal O(J^{-2})\right)  \nonumber\\
&{}+ g_2^2 \left(\frac{1}{3}+\frac{35}{8\pi^2 n^2}\right)\left(1+\frac{2(n^2 \bc{+2b^2})}{n^2 J}+\mathcal O(J^{-2})\right)+ \mathcal O(g_2^4)\Bigg] + \mathcal O((g')^2)\,,
\end{align}
reproducing the result of \cite{Beisert:2002bb} at leading order, while for the zero mode we have
\begin{align}
\Delta_0 = L +{}&\pc{ g' \Bigg[16 \pi^2 b^2 \left(1 - \frac{5 }{J}+\frac{51-2 \pi^2 b^2}{3J^2}+\mathcal O(J^{-3})\right) }\nonumber\\ &\qquad\pc{+ 4 \pi^2 b^2 g_2^2 \left(\frac{1}{J}+\frac{4 \pi^2 b^2 -69}{9 J^2}+\mathcal O(J^{-3})\right)+ \mathcal O(g_2^4)\Bigg] + \mathcal O((g')^2)}\,.
\end{align}


\section{Level-Crossing and Spectral Statistics}
\label{sec:stats}
The lifting of the degeneracies in the spectrum of one-loop dimensions by the $\beta$-deformation is related to the phenomenon of level repulsion. In general, energy surfaces depending on a number of parameters are connected only at special points where multiple parameters are tuned and the energy surfaces possess a diabolo-like geometric structure (such points were thus called ``diabolic'' by Berry and Wilkinson \cite{berry1984diabolical}). The situation is quite different for systems with additional symmetries, such as integrable systems, where degeneracies occur even as only a single parameter is varied. For the spectrum of $\mathcal{N}=4$ SYM this implies that operator dimensions depending on parameters $\lambda$ and $N$ avoid crossing for generic fixed values of $N$ as $\lambda$ is varied, as was borne out in \cite{Korchemsky:2015cyx}. 

\begin{figure}
	\begin{eqnarray}
	\begin{array}{cc}
	\includegraphicsbox[width=7cm]{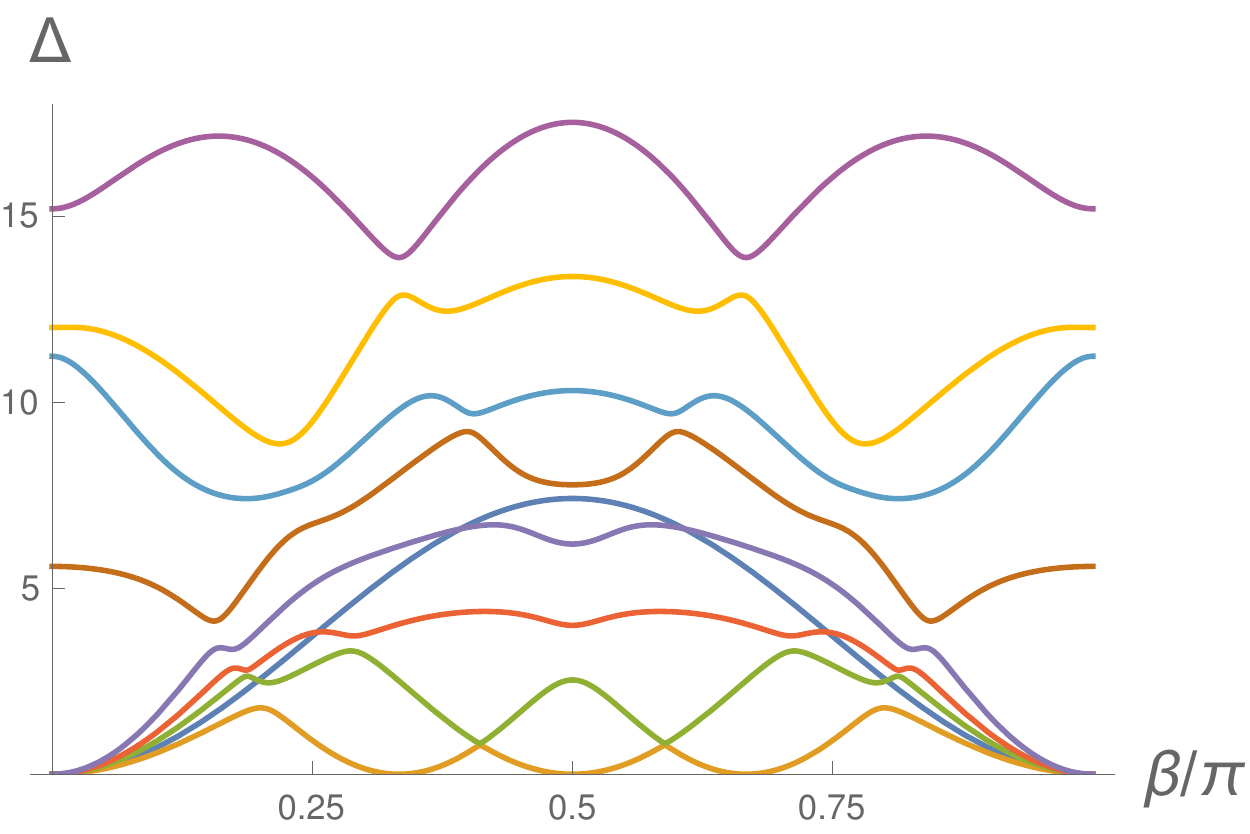}
	& ~~~~~~~	\includegraphicsbox[width=7cm]{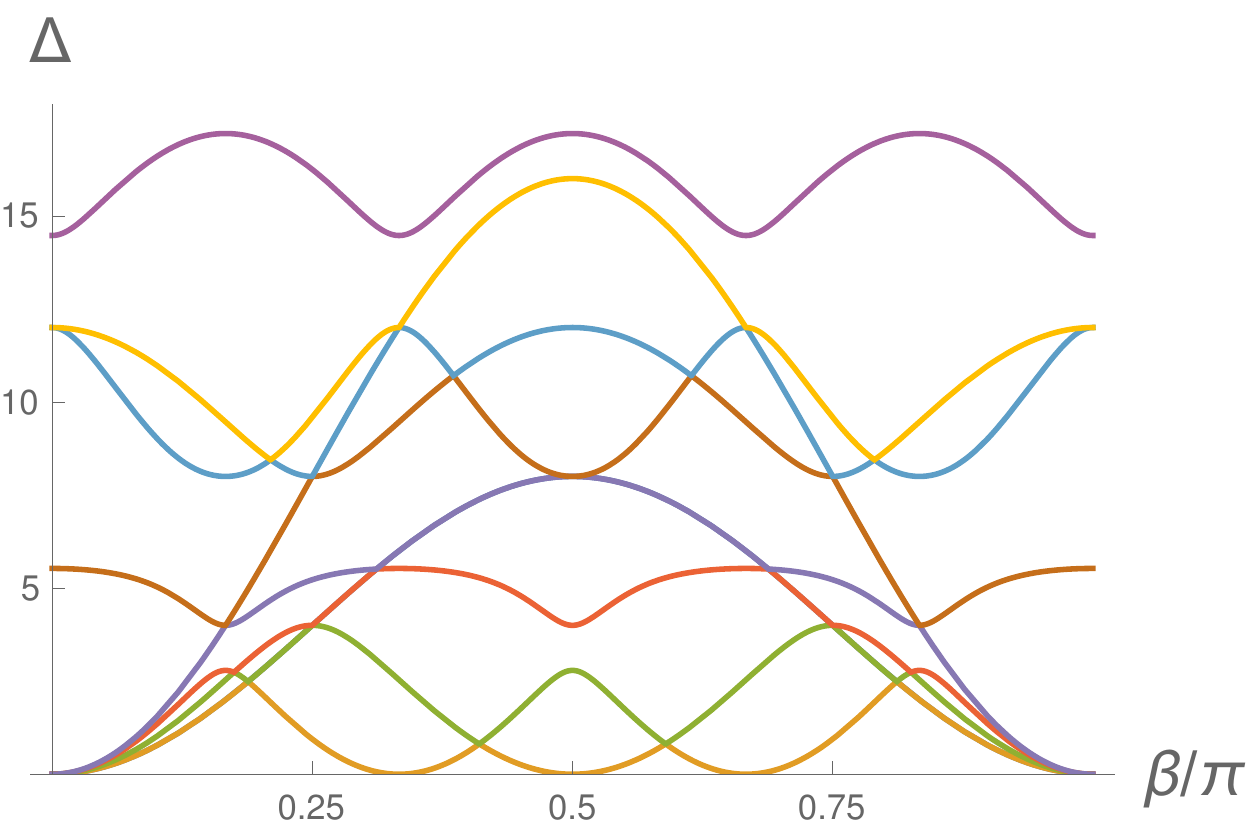}\nn\\
	& \nn\\
	~~~~~~~~~~~(\text{a}) &~~~~~~~~~~~~~~~~~(\text{b})
	\end{array}
	\end{eqnarray}
	\caption{Eight eigenvalues corresponding to states with $L=6$ and $M=3$ as functions of $\beta\in [0, \pi]$ with (\text{a})  $N=7$ (\text{b})  $N=10^6$. The top, purple, line and fourth from the top, brown, correspond in the undeformed, planar limit to descendants of two single trace two-impurity states. The second and third lines, yellow and light blue, correspond to the single trace three-impurity singular solution and a degenerate double trace operator. The remaining operators are protected in the undeformed theory but acquire non-vanishing anomalous dimensions for non-vanishing $\beta$.}
	\label{fig:LevelCross}
\end{figure}

In our case, being at one-loop, the  $\lambda$-dependence is trivial however we can study the spectrum as a function of both the deformation parameter, $\beta$, and the rank, $N$, of the gauge group. By numerically solving for the eigenvalues of specific families of operators we can see the behaviour of the scaling dimensions as we vary $\beta$ and as an example we consider the length-six, three impurity states in \figref{fig:LevelCross}. For finite values of $N$ the energy levels mostly repel and even at points where they appear to come close they do not in fact cross, maintaining a separation of $\sim 1/N^2$. There is one obvious exception which clearly does cross other levels at finite $N$. This is a double trace state that does not mix with other operators, receives no $1/N$ corrections and so is effectively uncorrelated with the other states. This is due to the fact that at half-filling the charge conjugation transformation $Z \leftrightarrow X$ combined with the parity transformation \eqref{eq:parity} is a symmetry which commutes with both the impurity number and the one-loop non-planar dilation operator. The double-trace operator is the only $L=6$, $M=3$ state with negative charge with respect to this transformation. This points to the fact that in order to avoid trivial crossings we must consider operators which have the same quantum numbers. At large values of $N$ we can see the appearance of crossings which occur at special values where $\beta/\pi \in \mathbb{Q}$; for example in  \figref{fig:LevelCross} there are crossings in the planar limit at $\beta=\pi/4$ and $\beta=\pi/6$. These points correspond to values where the $\beta$-deformed theory becomes equivalent to an orbifold of $\mathcal{N}=4$ SYM, e.g.\ \cite{Berenstein_2000}, which are known to have enhanced structures such as additional regions on the Coulomb branch.  

The level repulsion at finite-$N$ suggests the transistion from a quantum integrable model to a chaotic system. To further explore this it is interesting to compute the distribution of level spacings. Given a spectrum of energy levels one can easily show that if we assume they are uncorrelated the spacing of successive levels satisfies Poisson statistics, $P_P(s)=e^{-s}$. That this distribution is  a good description of integrable sytems has been numerically shown in a range of models including many-body systems such as the Heisenberg spin chain, the t-J model at its integrable supersymmetric point and the Hubbard model \cite{poilblanc1993poisson, hsu1993level}. There are significantly fewer analytical results, however  one important result by Berry and Tabour \cite{berry1977level} showed that for a ``generic'' integrable, semi-classical system the distribution of  energy levels is indeed Poisson.  There are known examples of integrable models for which this is not the case,  such as \cite{PhysRevE.70.026208} where by  considering finely tuned multi-parameter Richardson-Gaudin models, integrable models with non-integrable statistics were found. However in that case even small changes in the parameters resulted in a restoration of the integrable distribution.

It is well known that in Random Matrix Theory (RMT) the joint probability distribution for the eigenvalues, $x_1, x_2, \dots, x_S$, of $S\times S$ Hermitian random matrices is given by 
\<
P_\alpha(x_1, x_2, \dots, x_S)=C_{S_\alpha} \prod_{j<k} |x_j-x_k|^\alpha e^{-\tfrac{\alpha}{2} \sum_{j=1}^Sx_j^2}
\>
with $\alpha=1, 2, 4$ corresponding to orthogonal, unitary and symplectic ensembles, respectively. Furthermore, the distribution of spacings between eigenvalues, normalised so that the mean spacing is unity,  can be well approximated by the Wigner-Dyson distribution 
\<
\label{eq:WDdist}
P_{WD}(s)&=&A(\alpha) s^\alpha e^{-B(\alpha)s^2}~,
~~~~ A(\alpha)=2 \frac{\Gamma(1+\tfrac{\alpha}{2})^{1+\alpha}}{\Gamma(\tfrac{1+\alpha}{2})^{2+\alpha}}~,~~~~
B(\alpha)=\frac{\Gamma(1+\tfrac{\alpha}{2})^{2}}{\Gamma(\tfrac{1+\alpha}{2})^{2}}~.
\> 
A particularly important feature is that when approximating a physical system by random matrices the appropriate ensemble can be determined from the symmetries of the Hamiltonian, such as rotational or time-reversal symmetry, and does not depend on the specifics of the interactions. In particular, for a Hamiltonian with time-reversal and rotational symmetry it is appropriate to choose the Gaussian Orthogonal Ensemble (GOE)  corresponding to $\alpha=1$. 
\begin{figure}[h!]
	\begin{eqnarray}
	\begin{array}{ccc}
	\includegraphicsbox[scale=0.3]{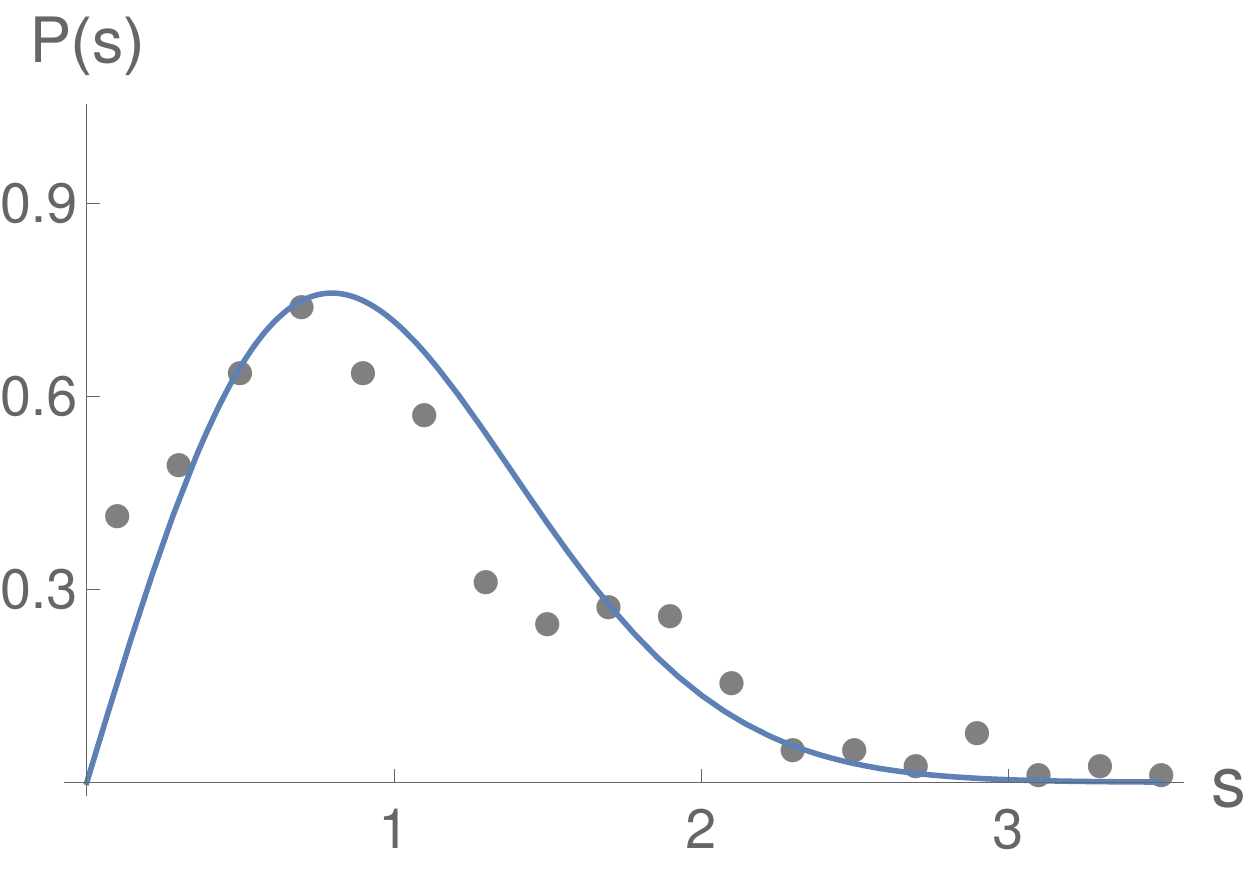}
	& ~~~~~	\includegraphicsbox[scale=0.3]{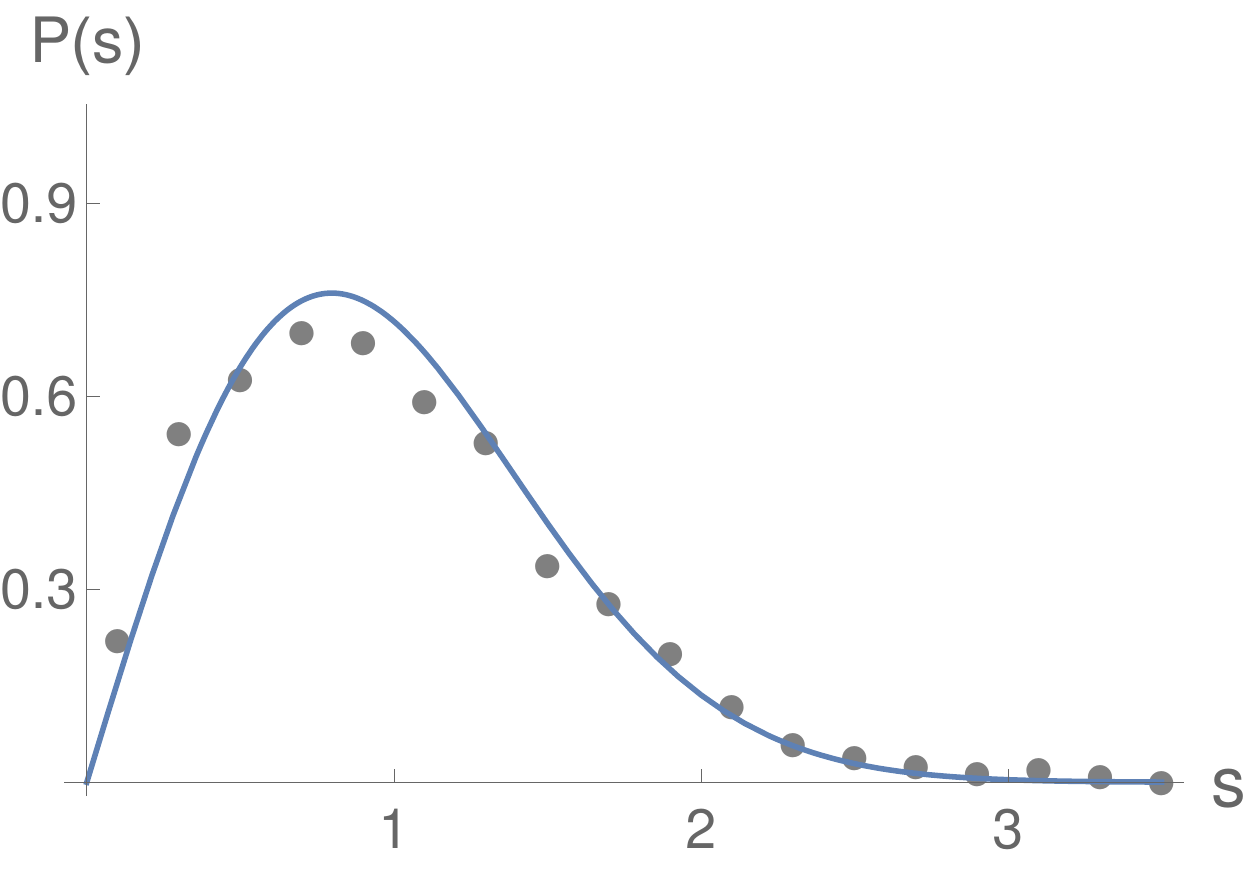}& ~~~~~	\includegraphicsbox[scale=0.3]{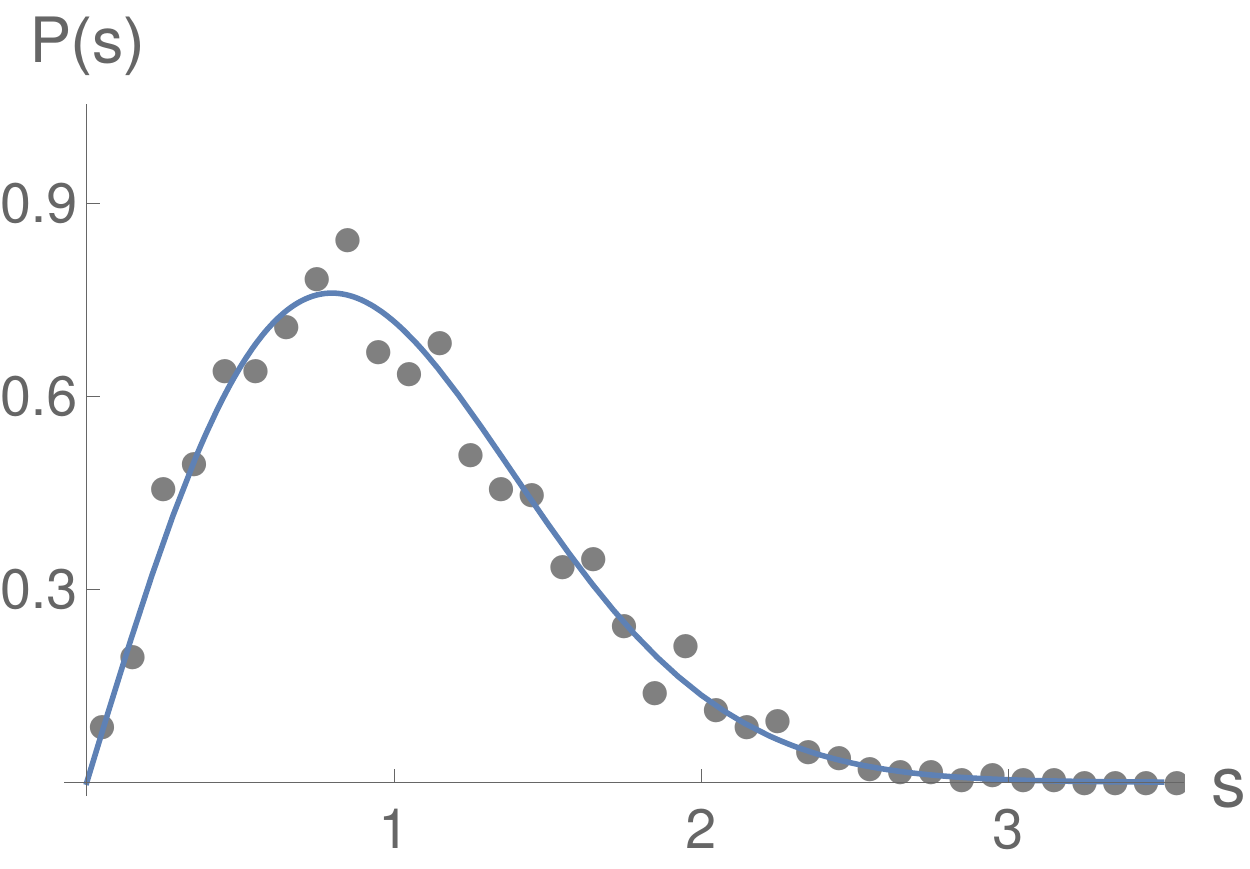}\nn\\
	& \nn\\
	~~~~~M=2 &~~~~~~~~~~~~~M=3&~~~~~~~~~~M=4\nn\\
	\includegraphicsbox[scale=0.3]{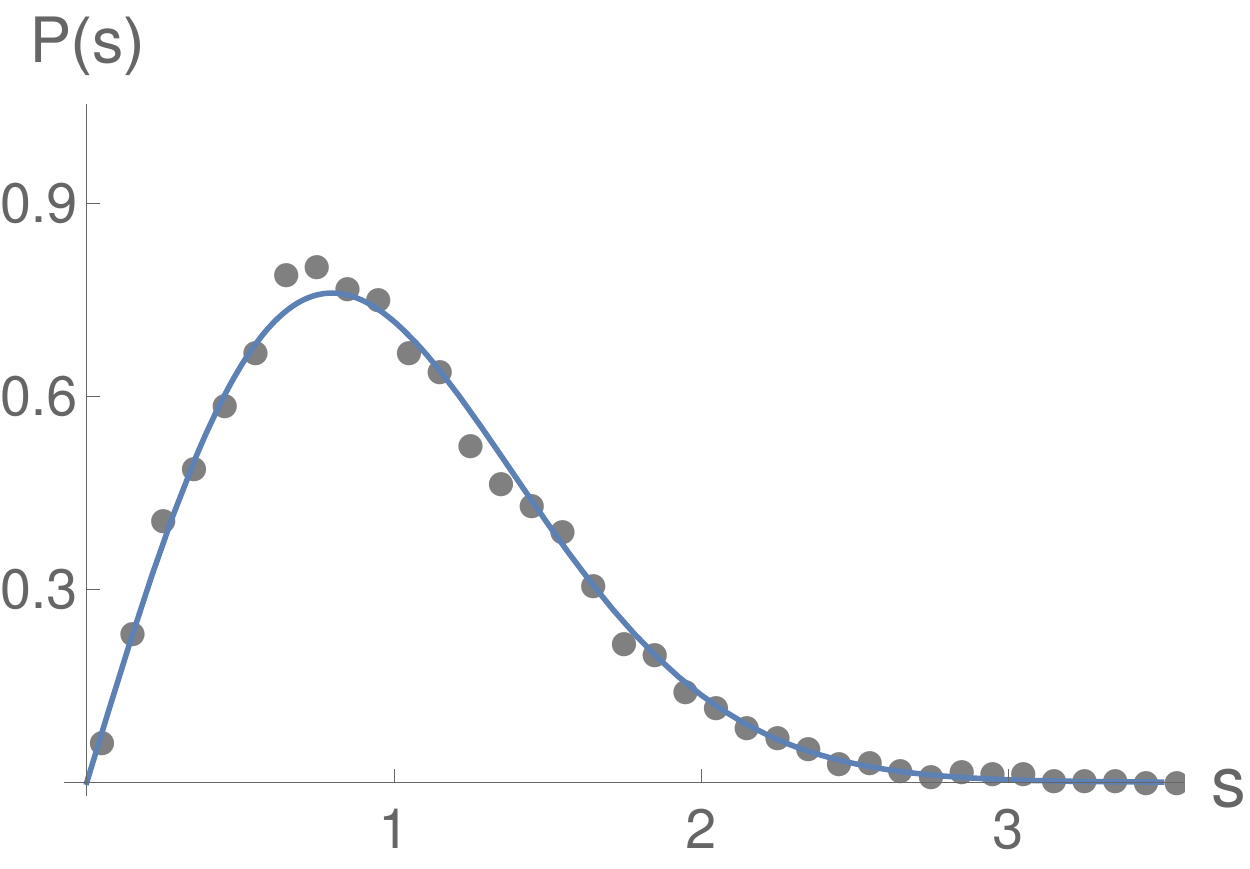}
	& ~~~~~~~	\includegraphicsbox[scale=0.3]{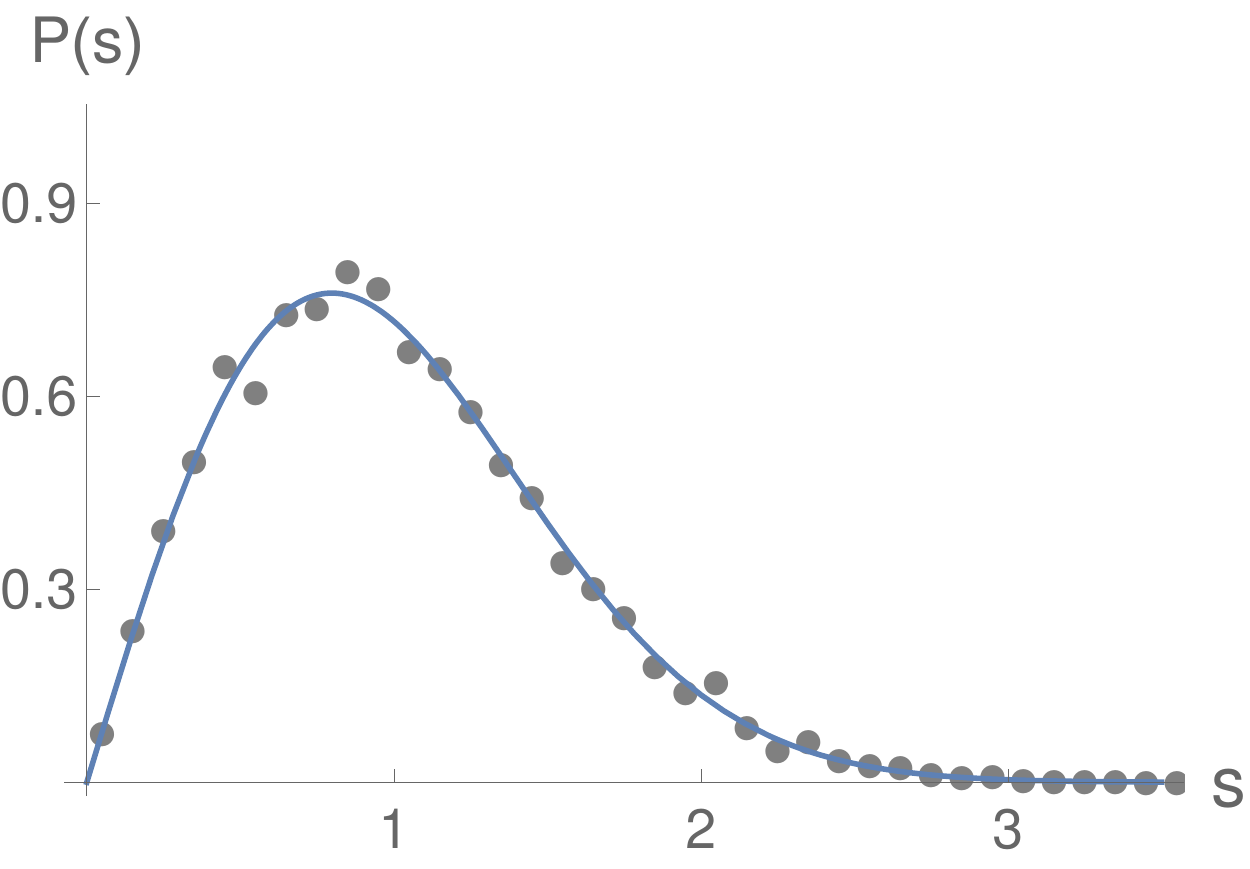}& ~~~~~~~	\includegraphicsbox[scale=0.3]{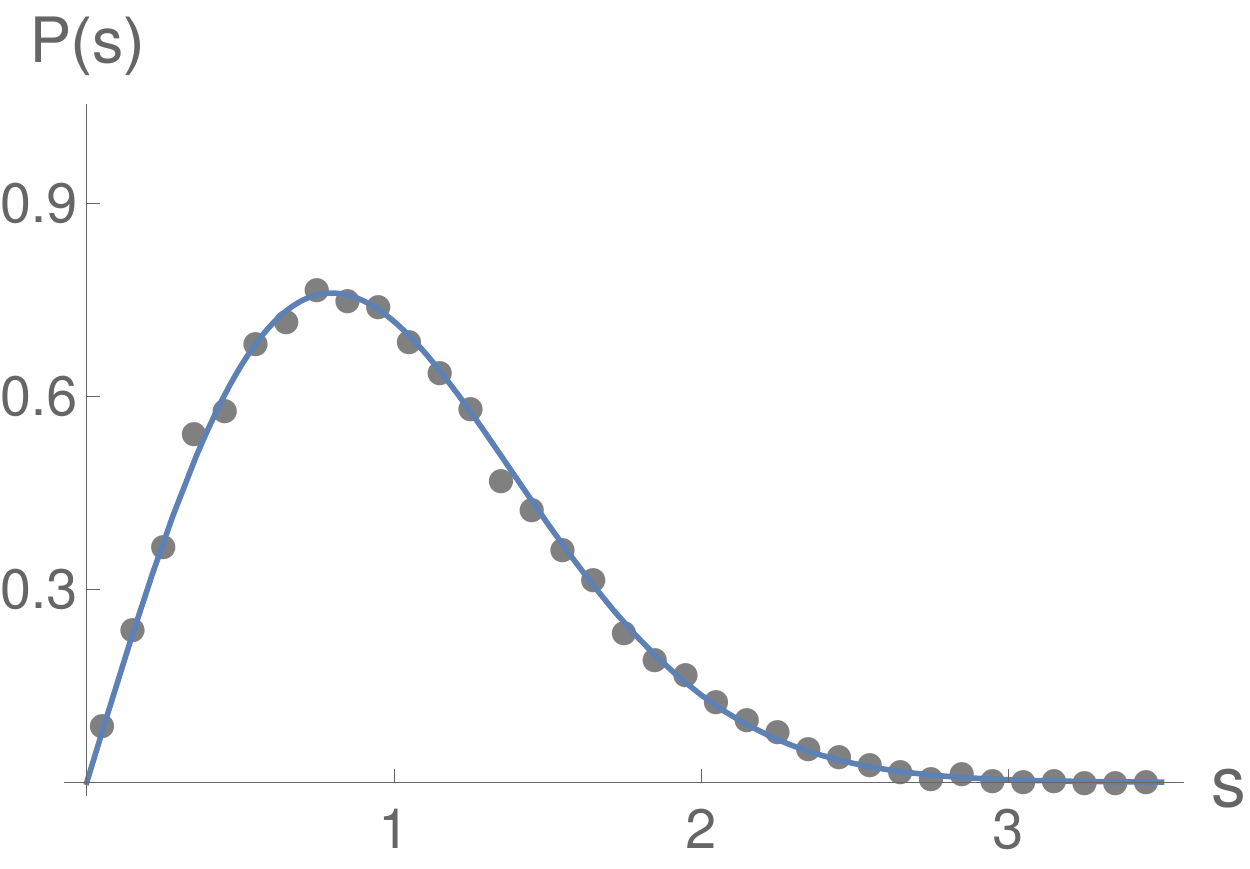}\nn\\
	& \nn\\
	~~~~~~~M=5 &~~~~~~~~~~~~~~~M=6&~~~~~~~~~~~~M=7\nn\\
	\end{array}
	\end{eqnarray}
	\caption{The level-spacing statistics for $L=16$, $\beta=0.9$ states at $N=17$. The grey dots are the numerically calculated values and the solid lines correspond to the Wigner-Dyson distribution, $P_{WD}(s)$. }
	\label{fig:NPStats}
\end{figure}

In order to similarly analyse the spectrum of one-loop anomalous dimensions we must first focus on a specific sector comprising states which have the same quantum numbers for any operators which commute with the dilatation operator. For the $\beta$-deformed theory we thus consider states with fixed bare dimension, $L$,  and excitation number, $M$.  Additionally we remove all zero energies and we do not consider the sector with $M=L/2$. The former correspond to protected states whose dimensions are fixed by supersymmetry and, as previously mentioned, the latter contains states which are related by symmetry  and so are degenerate. We then numerically compute the spectrum and order the dimensions in this sector $E_1, E_2, \dots, E_S$ so that $E_1\leq E_2 \leq \dots\leq E_S$~.   One important technicality is that the spectra of RMT are normalised so that the mean level density is unity. So in comparing with physical systems the spectrum must be rescaled to remove the overall dependence on the energy. This procedure is called unfolding and as we do not know the mean level distribution we find a way to approximate it. We describe our procedure in \appref{app:Unfold}, and find that the final result is relatively insensitive to the specific details of the unfolding.  After carrying out this step we label the unfolded spectrum of anomalous dimensions from smallest to largest: $x_1\leq  x_2\leq  \dots \leq x_S$.  
From the unfolded spectrum we estimate the distribution of level spacings between consecutive levels by computing $s_i=x_{i+1}-x_i$, then binning the data and calculating the fraction that occur in each bin. The results naturally depend on the bin size and a choice is made such that small changes do not significantly affect the overall results. The estimate of the probability distribution naturally improves with larger numbers of states and so one must compute the dimensions for relatively long operators. 

\begin{table}[h]
	\centering
	\begin{tabular}{|c c c c |} 
		\hline
		M  & No. of States & $\omega$ & $\alpha$ \\ [0.5ex] 
		\hline\hline
		2 & 400 & 0.55 & 0.48 \\ 
		\hline
		3 & 1035 & 0.79 & 0.70 \\
		\hline
		4 & 2316 & 0.95 & 0.91 \\
		\hline
		5 & 4198 & 0.99 & 1.03 \\
		\hline
		6 & 6539 & 0.99 & 0.98 \\
		\hline
		7 & 8431 & 0.98 & 0.96 \\ [0.5ex] 
		\hline
	\end{tabular}
	\caption{$L=16$ states with $M$ excitations for $N=17$ and $\beta=0.9$. The GOE Wigner-Dyson distribution corresponds to $\omega=1$ and $\alpha=1$.}
	\label{table:NPStats}
\end{table}

In \figref{fig:NPStats} and \tabref{table:NPStats} we present the results for $L=16$ states with $N=17$ and $\beta=0.9$. By visual inspection it is apparent that the GOE Wigner-Dyson distribution ($\alpha=1$) closely matches the data for most values of $M$. To be more quantitative, one can fit the data to the Brody distribution 
\<
\label{eq:brody}
P_B(s) =\Gamma\left( \tfrac{\omega+2}{\omega+1}\right)^{1+\omega}(1+\omega) s^\omega \text{exp}\left(-\Gamma\left( \tfrac{\omega+2}{\omega+1}\right)^{1+\omega} s^{1+\omega}\right)
\>
which is a one-parameter generalisation smoothly interpolating between the GOE Wigner-Dyson distribution ($\omega=1$) and the Poisson distribution 
($\omega=0$). In \tabref{table:NPStats} we show the best fit values of $\omega$ for different values of $M$ which are generally close to one suggesting that this is the appropriate value for the distribution at relatively small values of $N$. This fit captures the Gaussian behaviour of the exponential decay of the tail and the fact that the distribution goes to zero as $s\to 0$. If we assume the distribution is of the Wigner-Dyson form we can perform a fit to the general form \eqref{eq:WDdist} and find the best fit value of $\alpha$ which from \tabref{table:NPStats} can again be seen to be approximately one.   It is clear that the fit is better for higher excitation number as the values for $M=2$ are furthest from those of the GOE. The values for $M=0$, which are protected operators, and $M=1$, which are protected in the undeformed theory, clearly do not fit the Wigner-Dyson distribution but we do not  have a clear explanation for why the $M=2$ fit is so poor. In general however we find that the Gaussian Orthogonal Ensemble describes the non-planar distribution of energy levels in the $\mathfrak{su}(2)$ sector of deformed $\mathcal{N}=4$ SYM. 

\begin{figure}   
	\begin{eqnarray}
	\begin{array}{cc}
	\includegraphicsbox[scale=0.2]{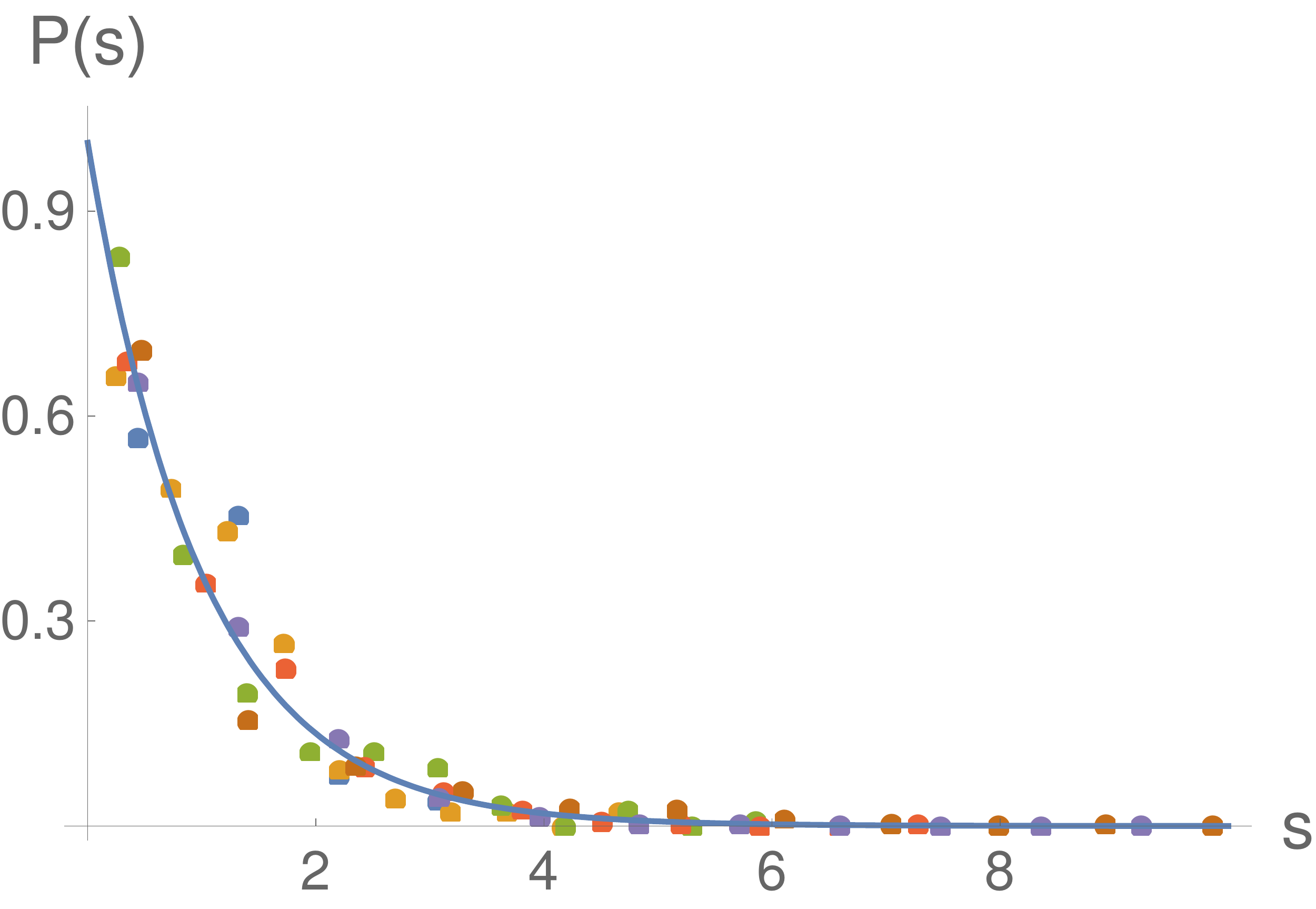}& ~~~~~\includegraphicsbox[scale=0.19]{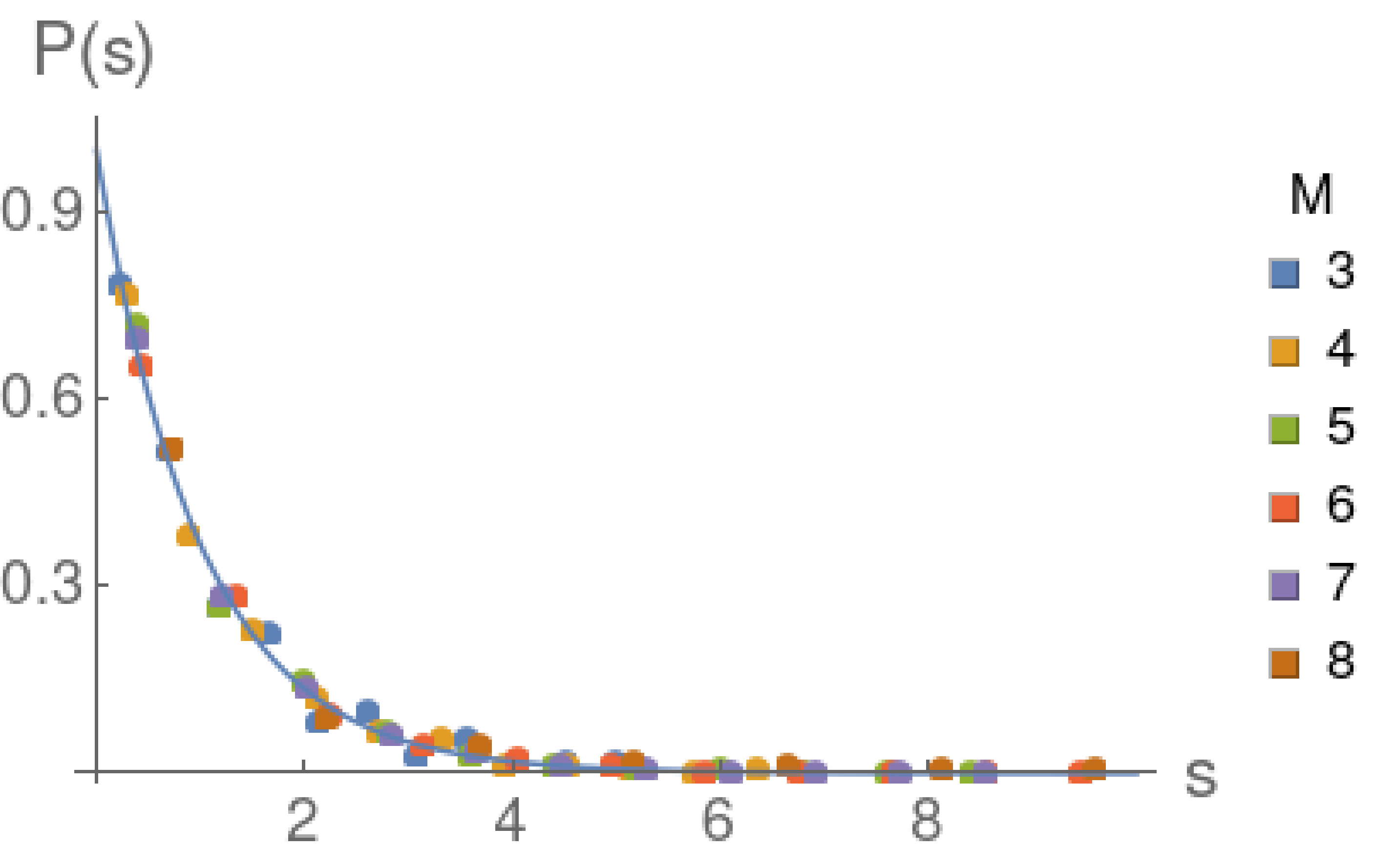}\nn
	\end{array}
	\end{eqnarray}
	\caption{The level-spacing statistics for $L=16$, $\beta=0.9$ single-trace (on the left) and double-trace (on the right) states in the planar limit. The blue markers are the numerically calculated values of unfolded spacings computed for states with $M=3$ and similarly for the other impurity numbers.  The solid line corresponds to the Poisson distribution, $P_{P}(s)$. }
	\label{fig:PStats}
\end{figure}

We can repeat the computation for the strict planar spectrum,  however in this case there are additional symmetries that we must account for. In particular the number of traces in a given operator is conserved under the action of the dilatation operator and so we must work at fixed number of traces. In the single-trace sector this reduces the problem to essentially that of the integrable twisted XXX  spin chain which is known to satisfy Poisson statistics while the multi-trace sectors are uncorrelated tensor products and so also have the same distribution, see \figref{fig:PStats}. We can again compare the planar spectrum to the Brody distribution \eqref{eq:brody} and we find that for most impurity numbers the fit is best for a value of $\omega\simeq 0$,  though there are a small number of cases where the value is larger. If we combine the distributions of single-trace states with $M\geq3$ together we find $-0.4 \leq \omega\leq 0.2$ depending on how we bin the data while  for the double-trace operators we find  $-0.4 \leq \omega\leq 0.4$ with the results generally close to zero. Thus  the spectrum appears to be well described by the Poisson distribution.

\begin{figure}
	\begin{eqnarray}
	\begin{array}{cc}
	\includegraphicsbox[scale=0.4]{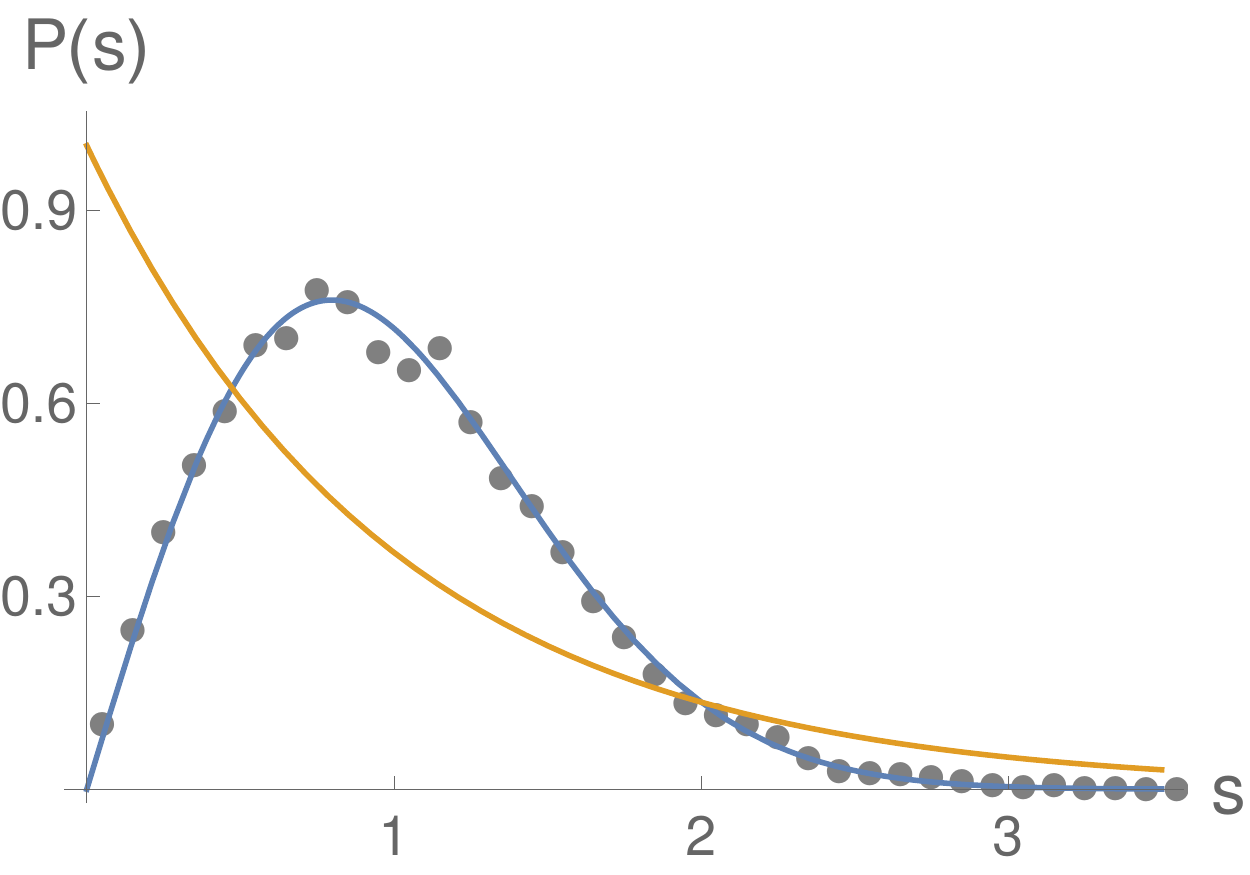}& ~~~~~\includegraphicsbox[scale=0.4]{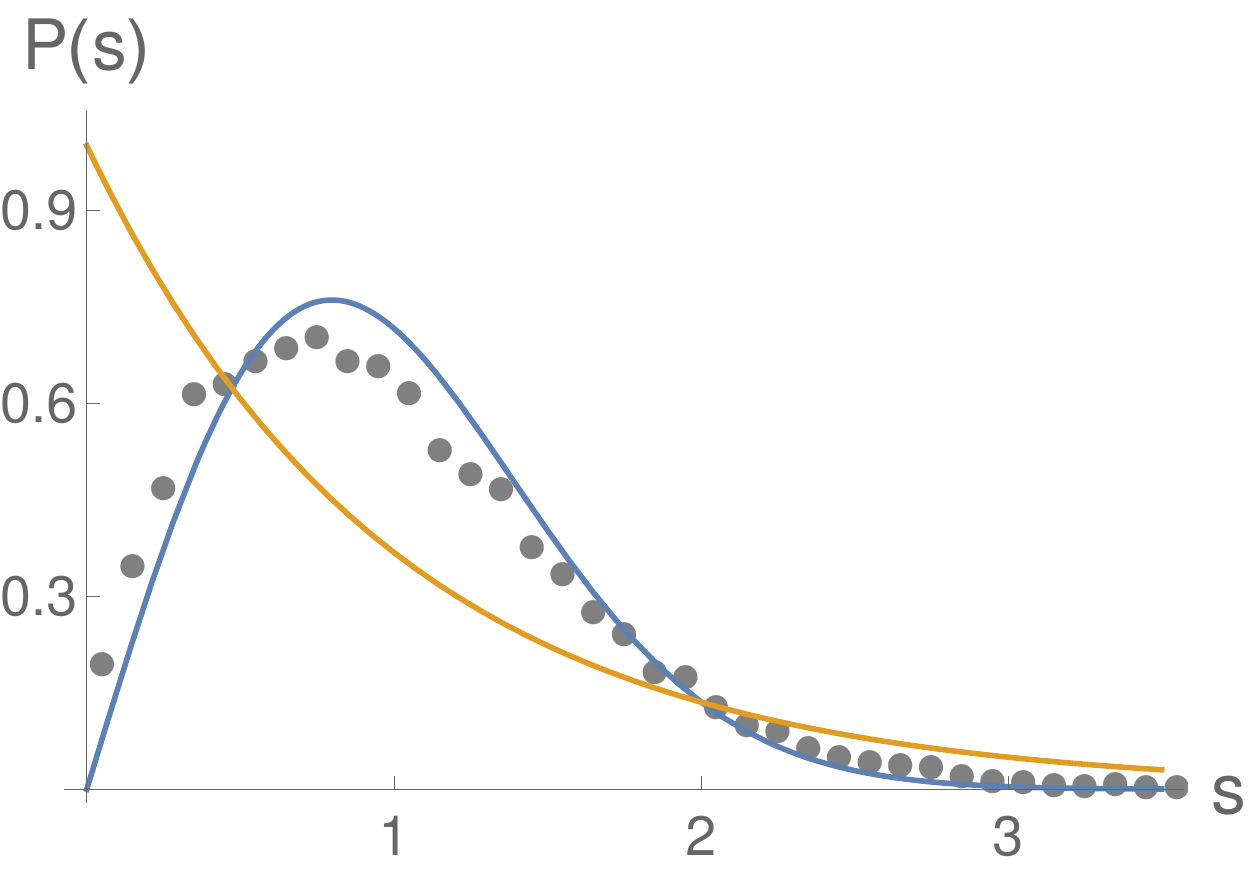}\nn\\
	~~~~~~~N=15 &~~~~~~~~~~~~~~~N=50\nn\\
	\includegraphicsbox[scale=0.4]{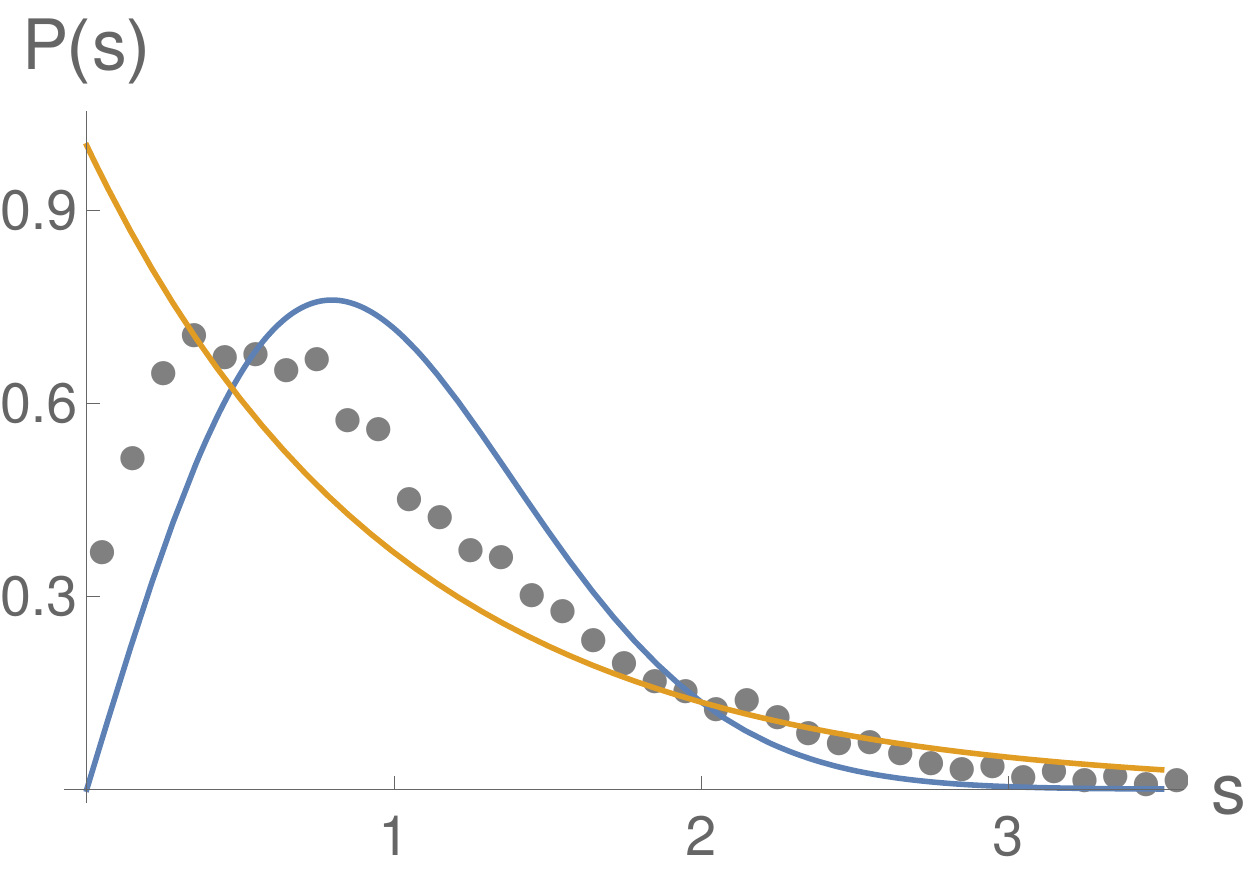}& ~~~~~\includegraphicsbox[scale=0.4]{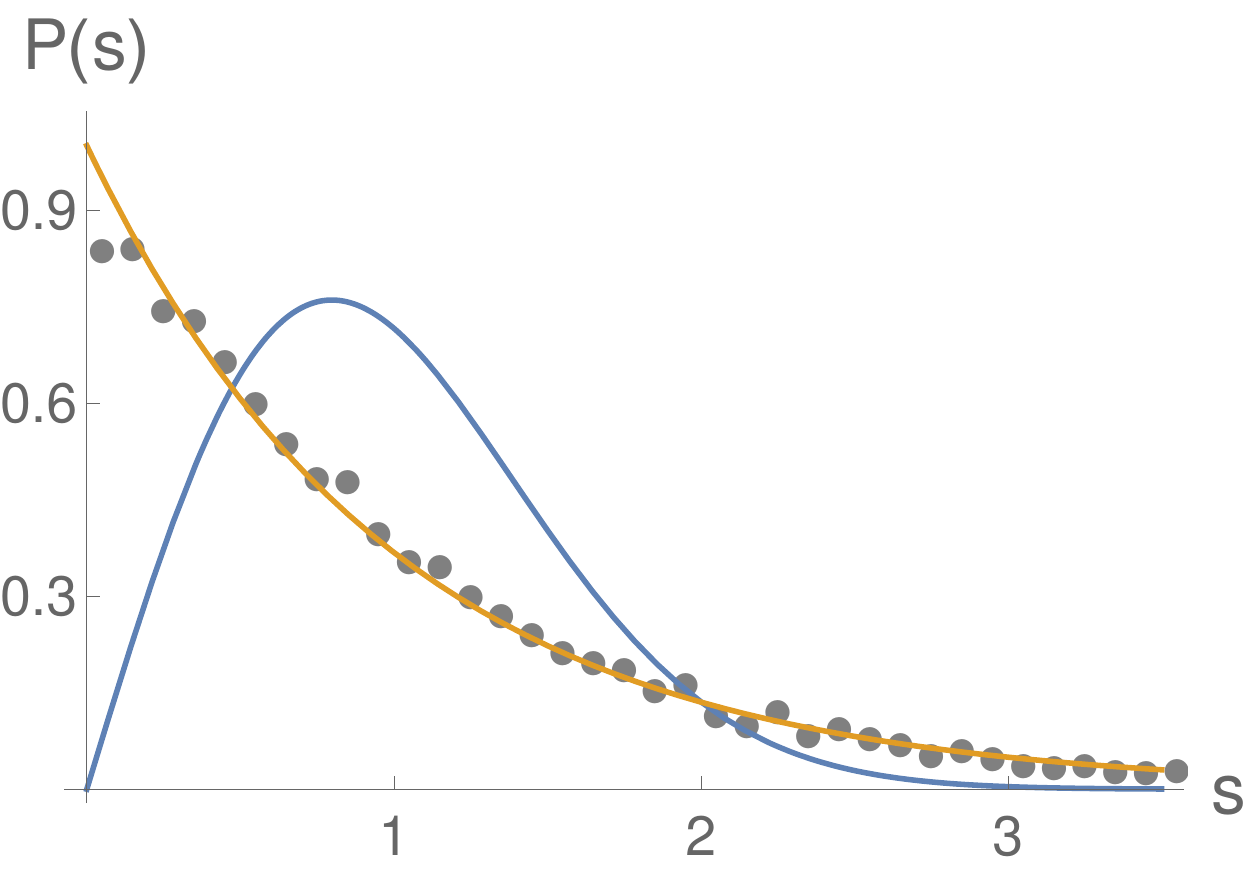}\nn\\
	~~~~~~~N=100 &~~~~~~~~~~~~~~~N=200\nn\\
	\end{array}
	\end{eqnarray}
	\caption{The level spacing statistics for $L=15$ states with $\beta=0.9$ and different values of $N$.  The spacings in each impurity sector are separately computed and then the combined distribution is plotted. The yellow solid line corresponds to the Poisson distribution and the blue line to the Wigner-Dyson distribution. }
	\label{fig:TransStats}
\end{figure}

One can see how the distributions change  as the spectrum transitions from chaotic to integrable by considering large a sequences of values of $N$. In \figref{fig:TransStats} we plot such a sequence of distributions of spacings for $L=15$ states. For $N=15$ we find the expected Wigner-Dyson distribution while for $N=50$ and $N=100$ we find distributions between Wigner-Dyson and Poisson with $\omega\simeq 0.71$ and $\omega\simeq 0.39$ respectively while  for $N=200$  we find $\omega\simeq 0.05$ and the distribution appears to be approaching Poisson. However this is not quite the case with the value of $\omega$ further decreasing as we increase $N$ giving $\omega\simeq-0.61$ for $N=10^6$. This is due to an excess of points occurring toward $s=0$ due to the decoupling of sectors with different numbers of traces which, as explained above, should be considered separately. 

\begin{figure}   	\begin{eqnarray}
	\begin{array}{c}
	\includegraphicsbox[scale=0.3]{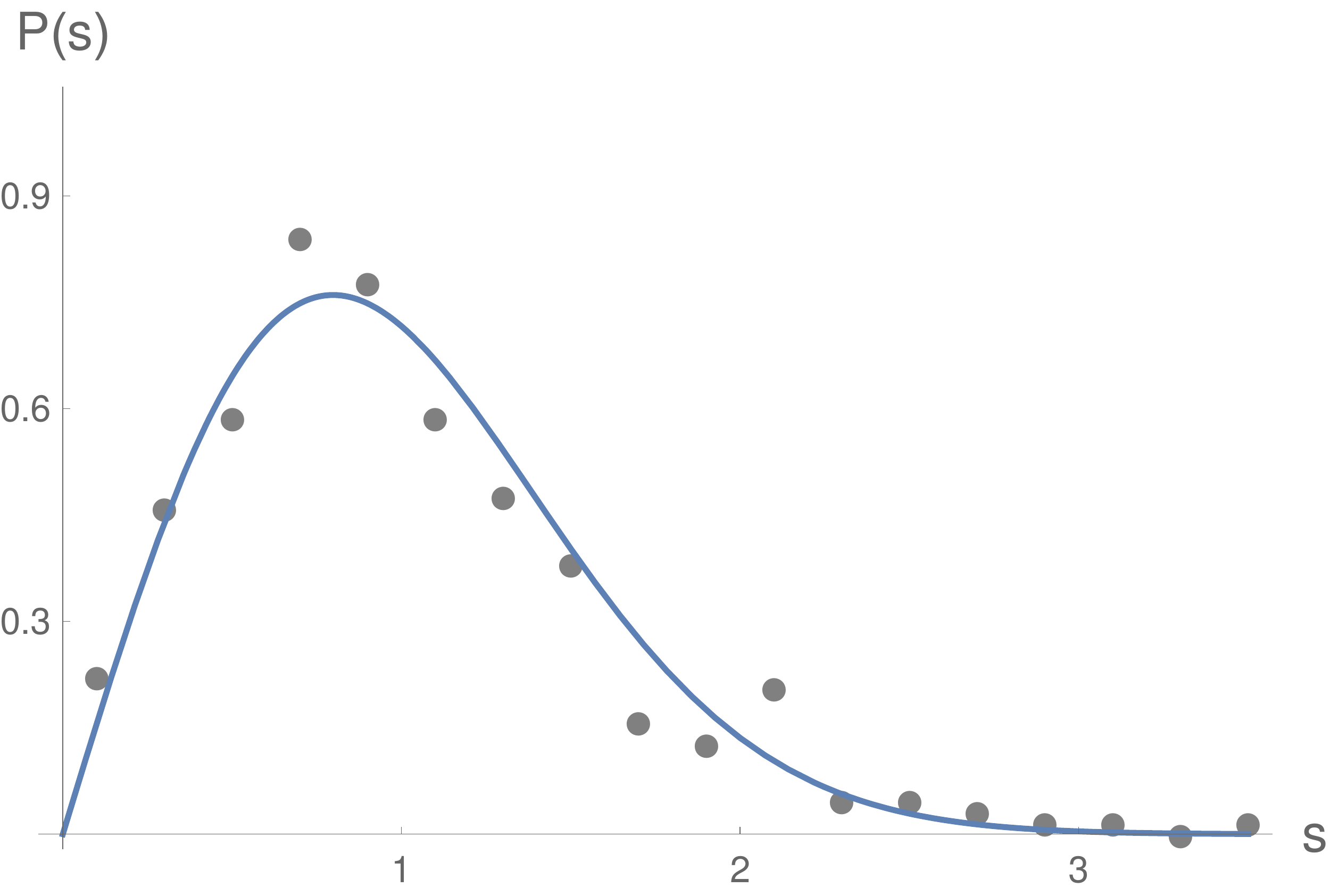}\nn
	\end{array}
	\end{eqnarray}
	\caption{The level spacing distribution for $L=16$, $M=3$ and positive parity highest weight states with $N=17$ in the undeformed theory.}
	\label{fig:Undef}
\end{figure}

We can of course ask about the statistics of the spectrum for the undeformed theory. In this case there are additional symmetries even in the non-planar theory and, in order to find the Wigner-Dyson distribution, we must carefully desymmetrize the spectrum. As mentioned previously, the parity operation, \eqref{eq:parity},
commutes with the full dilation operator in the undeformed theory and so non-planar eigenstates with different parity are uncorrelated.  Additionally the full global $\mathfrak{su}(2)$ symmetry is present in the undeformed theory and so it is necessary to work with only highest-weight states. This means that at fixed length and excitation number there are fewer available states and consequently the statistics are of poorer quality. For example, if we consider positive parity $L=16$ states with $M=3$ we find only 315 distinct states. Nonetheless, the general features of the Wigner-Dyson distribution can be seen in a plot of the level spacings, \figref{fig:Undef}, and the best fit to the Brody distribution occurs for $\omega\simeq 0.9$. This suggests that the statistics of the undeformed spectrum are similarly described by GOE random matrix theory.
%

\newpage
\section{Conclusions}

We have considered the problem of computing non-planar anomalous dimensions in $\mathcal N=4$ SYM and its deformations. The approach we have followed involves two steps: first one must obtain the mixing matrix, and then find its eigenvalues with the method of quantum-mechanical perturbation theory. In this work we have mostly focused on the first half of the problem by finding the matrix elements of the one-loop dilatation operator in terms of the Bethe rapidities. While direct application of the dilatation operator can in many cases yield the mixing matrix in a similarly efficient fashion, our formulas are given in terms of partitions of the Bethe rapidities, and therefore they are especially advantageous when the number of excitations is small. In those cases we are able to easily evaluate the overlaps, even for long operators where direct diagonalisation would be infeasible, and the bottleneck in computing anomalous dimensions is the determination of the Bethe rapidities. While there are tools for efficiently computing such rapidities, most notably the Baxter Q-function method of \cite{Marboe:2016yyn}, carrying out the sums over solutions is still non-trivial.

At a more conceptual level, we found that the matrix elements can be written in terms of Hexagon-like objects satisfying both the Watson and decoupling conditions. While our methods are not obviously related to the hexagonalization of the torus, this decomposition hints at the possibility of an approach  similar to \cite{AsymptoticFourPoint}, where four-point functions are built through the OPE, but the OPE data itself is computed within an integrable framework. Similarly, the matrix elements of the dilatation operator might have a more general description which determines their form at higher orders in the perturbative expansion. In order to study this further it would be useful to determine the overlaps at higher loops and to investigate if hexagon-like objects can be found for other sectors of the theory.

One issue with our approach to the diagonalization of the mixing matrix is that it assumes a non-degenerate spectrum of excited states. There are however many degeneracies in the planar spectrum of $\mathcal N=4$ SYM, and so we also considered the $\beta$-deformed theory where these degeneracies are lifted. A second advantage of the $\beta$-deformation is that it provides a useful regularization for the singular solutions occuring in the $\mathfrak{su}(2)$ sector of the $\mathcal N=4$ spin chain. The action of the dilatation operator in the deformed theory yields several new structures and for the purpose of evaluating $1/N^2$ corrections to the spectrum it is necessary to include an additional diagonal overlap and the contribution of the double-trace term. As an application of our method we computed the anomalous dimension of two-excitation states in the BMN limit through subleading order. We extracted the corresponding coefficients from fits to numerical data at lower lengths and found the results agreed with at least 8 digits of precision. As the problem of degeneracies occurs in other sectors of the theory additional twists will be needed. For example to study the $\mathfrak{sl}(2)$ sector it may be useful to consider the integrable dipole deformation \cite{Guica:2017mtd}.

The problem of summing over intermediate states increases with the excitation number of the operators under consideration and will rapidly become unfeasible. To compute such sums in the deformed theory it would be of great advantage to generalise the Baxter Q-function method for determining rapidities to the twisted case. It may also be possible that the sum over solutions is simpler than the individual terms and that the computational methods based on algebraic geometry discussed in \cite{Jiang:2017phk} can be fruitfully applied. It will likely be of interest to study the semi-classical limit of the non-planar corrections where both the number of excitations and the spin-chain lengths are taken to be large. For the planar theory this limit proved to be of great use in making contact with the strong-coupling classical string description. One tool to carry out the sum over intermediate states in this thermodynamic limit is the Quench Action \cite{Caux:2013ra, Caux_2016}, where the sum over Bethe solutions is replaced by a functional integral over root densities which can then be evaluated by saddle-point approximation.

There are of course alternative methods  for studying non-planar anomalous dimensions such as Hexagonalisation. There are in principle two approaches that can be taken within this formalism. On the one hand, hexagonalisation can be used to compute four-point functions on the torus \cite{Bargheer_2018} and OPE limits then used to extract anomalous dimensions. While in this method one can restrict to correlation functions of protected operators, it gives access only to sum rules of OPE data. On the other hand, it is possible to consider the two-point function on the torus \cite{Eden:2017ozn} by taking the four-point function with two identity operators, and while this approach does not solve the problem of diagonalizing the mixing matrix, pursuing it beyond tree-level might provide an alternative way of finding the matrix elements of the dilatation operator.

In addition to computing specific operator dimensions it is also of interest to understand the general properties of the spectrum. To this end we analysed the distribution of level spacings and found that at infinite $N$ the one-loop spectra of both $\mathcal{N}=4$ SYM and its deformation were well described by the Poisson distribution which is characterisic of integrable systems. This could be seen to transition at finite-$N$ to the Wigner-Dyson distribution of chaotic quantum many-body systems which suggests that the statistical properties of the finite-$N$ spectrum can be well described by a GOE random matrix model.  Quantum chaos has in recent years been studied extensively in the context of the holographic duality between the SYK-model of $N$ $(0+1)$-dimensional Majorana fermions and Jackiw-Teitelboim gravity on AdS$_2$ \cite{Sachdev:1992fk, Kitaev:2015Talk, Maldacena:2016hyu, Kitaev:2017awl}. The distribution of the level spacings for the SYK model was numerically computed in \cite{You_2017}, see also \cite{Cotler_2017, Garc_2017, Lau:2018kpa},  and it was shown that it is Wigner-Dyson with all three ensembles, GOE, GUE and GSE, occuring depending on the value of $N$. It would be naturally interesting to study this chaotic behaviour at higher loop-orders in $\mathcal{N}=4$ SYM and whether, by the holographic correspondence, we can describe the properties of interacting quantum strings on anti-de Sitter space by RMT.
\\

\noindent
\textbf{Acknowledgements}
We thank Marius De Leeuw, Sergey Frolov, and Pak Hang Chris Lau for useful conversations. We also thank Jo\~{a}o Caetano, Thiago Fleury and Pedro Vieira for comments on an earlier draft of this work.  
T.\ McL.\ would particularly like to thank Samuel Bateman, Joe Davis and Denis Murphy for their contributions to summer projects out of which this work originated. This work was supported by the Science Foundation Ireland through grant 15/CDA/3472 and has received funding from the European Union's Horizon 2020 research and innovation programme under the Marie Sk\l{}odowska-Curie grant agreement No. 764850 "SAGEX".

\appendix
\section{Overlaps from the Algebraic Bethe Ansatz}
\label{sec:ABA}
The algebraic Bethe ansatz (ABA), see \cite{Faddeev:1996iy, Levkovich-Maslyuk:2016kfv, 2018arXiv180407350S} for introductions, provides a powerful framework for studying integrable systems such as the spin chains arising in the one-loop planar dilatation operator. Of particular interest in this work are the computationally efficient formulae for scalar products of Bethe states \cite{korepin1982calculation, slavnov1989calculation}. These scalar products have previously appeared in the context of $\mathcal{N}=4$ SYM structure constants and we will mostly follow the conventions of \cite{Escobedo:2010xs}.  

Central to the ABA approach is the monodromy matrix, $\hat{T}_a(u)$, which is an operator depending on the spectral parameter, $u\in \mathbb{C}$, and acting on the tensor product of the $L$ spin-chain sites, $(\mathbb{C}^2)^{\otimes L}$, and an extra auxiliary space, $V\simeq \mathbb{C}^2$, labelled by the index $a$. Considering $\hat{T}_a(u)$ as a $2\times 2$ matrix, whose entries are operators acting on the spin chain, we can write
\<
\hat{T}_a(u)=\begin{pmatrix}
	\mathcal{A}(u) & \mathcal{B}(u) \\
	\mathcal{C}(u) & \mathcal{D}(u)
\end{pmatrix}~.
\>
The commutation relations of these entries can be found from the relations
\<
\label{eq:RTT}
R_{a_1 a_2}(u-v) \hat{T}_{a_1}(u)\hat{T}_{a_2}(v)=
\hat{T}_{a_2}(v)\hat{T}_{a_1}(u)R_{a_1 a_2}(u-v)
\>
where the R-matrix, $R_{a_1a_2}(u-v)$, is an operator acting on the two auxiliary spaces labelled by $a_1$ and $a_2$ and, for the theories we consider, depending on the difference of the spectral parameters $u$ and $v$. Viewed as a $4\times 4$ matrix, mapping $(\mathbb{C}^2)^{\otimes 2}\to (\mathbb{C}^2)^{\otimes 2}$, we can write
\<
R_{a_1 a_2}(u-v)=\begin{pmatrix}
	f(u,v) & 0 & 0 & 0 \\
	0 & 1 & g(u,v) & 0 \\
	0 & g(u,v) & 1 & 0 \\
	0 & 0 &  0 & f(u,v)
\end{pmatrix}  
\>
where we have introduced the functions
\<
f(u,v)\equiv f(u-v) =1+\frac{i}{u-v} ~~~\text{and}~~~ g(u,v)\equiv g(u-v)=\frac{i}{u-v }~.
\>
The trace of the monodromy matrix over the auxiliary space defines the transfer matrix, $\hat{T}(u)=\Tr\, \hat{T}_a(u)$, and it follows from \eqref{eq:RTT} that transfer matrices with different spectral parameters commute. The Hamiltonian of the spin chain is given by the log derivative of the transfer matrix evaluated at $u=i/2$ while the higher conserved charges can be found by further expanding the logarithm of the transfer matrix near $u=i/2$. The eigenstates of the transfer matrix thus simultaneously diagonalise the Hamiltonian and all higher charges. One can define Bethe states as 
\<
\label{eq:ABS}
\ket{\{u\}}^{\text{al}}=\prod_{i=1}^M\mathcal{B}(u_i)\ket{0}
\>
where the pseudovacuum is defined by $\mathcal{C}(u)\ket{0}=0$ and satisfies
\<
\mathcal{A}(u)\ket{0}=a(u)\ket{0}~~~\text{and}~~~ \mathcal{D}(u)\ket{0}=d(u)\ket{0}
\>
with $a(u)=(u+i/2)^L$ and $d(u)=(u-i/2)^L$. When the rapidities $\{u_i\}$ in \eqref{eq:ABS} satisfy the Bethe equations \eqref{eq:momBE}, using the parametrisations \eqref{eq:rapmom} and \eqref{eq:rapSE}, the Bethe states are eigenstates of the transfer matrix with eigenvalues
\<
\hat{T}(v)\ket{\{u\}} =T(v,\{u\})\ket{\{u\}}~~~\text{with}~~~
T(v,\{u\})=a(v)\prod_{i=1}^M f(v,u_i)+d(v)\prod_{i=1}^M f(u_i, v)~.
\>
The operators $\mathcal{B}(u_i)$ can thus be viewed as creating excited states whose relative normalisation is given by, see \cite{Escobedo:2010xs}, 
\<
\ket{\{p\}}=\frac{1}{\sqrt{ S}^{\{u\}}_<f_<^{\{u\}} d^{\{u\}} g^{\{u+i/2\} }}~\ket{\{u\}}^{\text{al}}~,
\>
where we use the product notation \eqref{eq:prodnot}. 
The dual states in the ABA are defined by 
\<
{}^{\text{al}}\bra{\{u\}}=(-1)^M \bra{0} \prod_{i=1}^M \mathcal{C}(u^\ast_i)
\>
where the dual vacuum satisfies $\bra{0}\mathcal B(u)=0$ and 
\<
\bra{0}\mathcal{A}(u)=\bra{0}a(u)~~~\text{and}~~~ \bra{0}\mathcal{D}(u)=\bra{0}d(u)~.
\>
These dual states are related to Bethe states by Hermitian conjugation using the definition
\<
\ket{0}=\bra{0}^\dagger~,~~~ \text{and}~~~\mathcal{C}(u^\ast)=-\mathcal{B}^\dagger(u)
\>
and are dual eigenstates of $\hat{T}(u)$ when the rapidities satisfy the Bethe equations.
We will be interested in the quantity $I_M(\{v\},\{u\})$ which is related to the scalar products of Bethe states by the definition
\<
I_M(\{v\},\{u\})&\equiv&\bra{0}\prod_{j=1}^M \mathcal{C}(v_j)\prod_{j=1}^M\mathcal{B}(u_j)\ket{0}\\
&=&(-1)^M~  {}^{\text{al}}\braket{ \{ v^\ast \} }{ \{u\} }^{\text{al}}~
\>
and, following \cite{Korepin:1982gg}, can be written as a sum over partitions of the excitations. The partitions are defined by splitting each set of excitations, $\{u\}$ and $\{v\}$, into subsets, 
$\alpha\cup \bar{\alpha}=\{u\}$ and $\beta\cup \bar{\beta}=\{v\}$, with the cardinality of $\alpha$ is equal to that of $\beta$. The scalar product is then given as
\begin{align}
\label{eq:OBIP}
I_M(\{v\},\{u\})
&=&{g_<}^{\{u\}}g_>^{\{v\}}
\sum_{\substack{\alpha\cup \bar{\alpha}=\{u\}\\
		\beta\cup \bar{\beta}=\{v\}}}
\text{sgn}(\alpha)\text{sgn}(\beta)d^\alpha a^{\bar{\alpha}}a^\beta d^{\bar{\beta}} k^{\alpha\beta}k^{\bar{\beta}\bar{\alpha}}k^{\alpha\bar{\alpha}}k^{\bar{\beta}\beta} \text{det}\, t^{\alpha\beta}\text{det}\, t^{\bar{\beta}\bar{\alpha}}
\end{align}
where
\<
k(u,v)=\frac{f(u,v)}{g(u,v)}=1-i(u-v)~,~~~\text{and}~~~t(u,v)=\frac{g^2(u,v)}{f(u,v)}=\frac{-1}{(u-v)(u-v+i)}
\>
and $\text{sgn}(\alpha)$ is the signature of the permutation required to put $\alpha \cup \bar{\alpha}$ into the canonical order $\{u\}$. This formula is valid for arbitrary Bethe states, even those whose rapidities do not satisfy the Bethe equations and which are thus said to be "off-shell". In the case where one set of rapidities does satisfy the Bethe equations, they are said to be "on-shell", the formula can be dramatically simplified to the calculation of a single determinant \cite{slavnov1989calculation}. There is a further simplification when both sets of rapidities are on-shell and equal. In this case, as the set of rapidities is invariant under complex conjugation, the quantity $I_M$ is related to the norm of the Bethe state and is given by Gaudin's formula:
\<
\label{eq:Gaudin}
I_M(\{u\},\{u\})&=&d^{\{u\}}a^{\{u\}}f_>^{\{u\} } f_<^{\{u\}}~\text{det}_{j,k}~\partial_{u_j}\phi_k
\>
where $\phi_k$ is defined in \eqref{eq:momBE}.

\section{Unfolding Procedure}
\label{app:Unfold}
For an ordered spectrum $E_1\leq E_2 \leq \dots \leq E_N$ we define the level density function 
\<
n(E)=\sum_{i=1}^N \delta(E-E_i)~.
\>
and the cumulative spectral function, or staircase function, 
\<
I(E)=\int_{0}^{E} n(E')dE'= \sum_{i=1}^N \Theta(E-E_i)~.
\>
We now separate these spectral functions into smooth and fluctuating parts
\<
n(E)=n_{\text av}(E)+n_{\text fl}(E)~, ~~~\text{and}~~~I(E)=I_{\text av}(E)+I_{\text fl}(E)
\>
and then define new unfolded variables 
\<
x_i=I_{\text{av}}(E_i)~, ~~~\text{for}~~~i=1, 2, \dots, N
\>
so that for small separations
\<
x_{i+1}-x_i\simeq \frac{(E_{i+1}-E_i)}{D}~,
\>
where $D=1/n_{\text{av}}(E_i)$ is the local mean spacing. These new variables thus capture the nature of the spectral fluctuations about the smoothed behaviour.

\begin{figure}
\begin{eqnarray}
	\includegraphicsbox[scale=0.3]{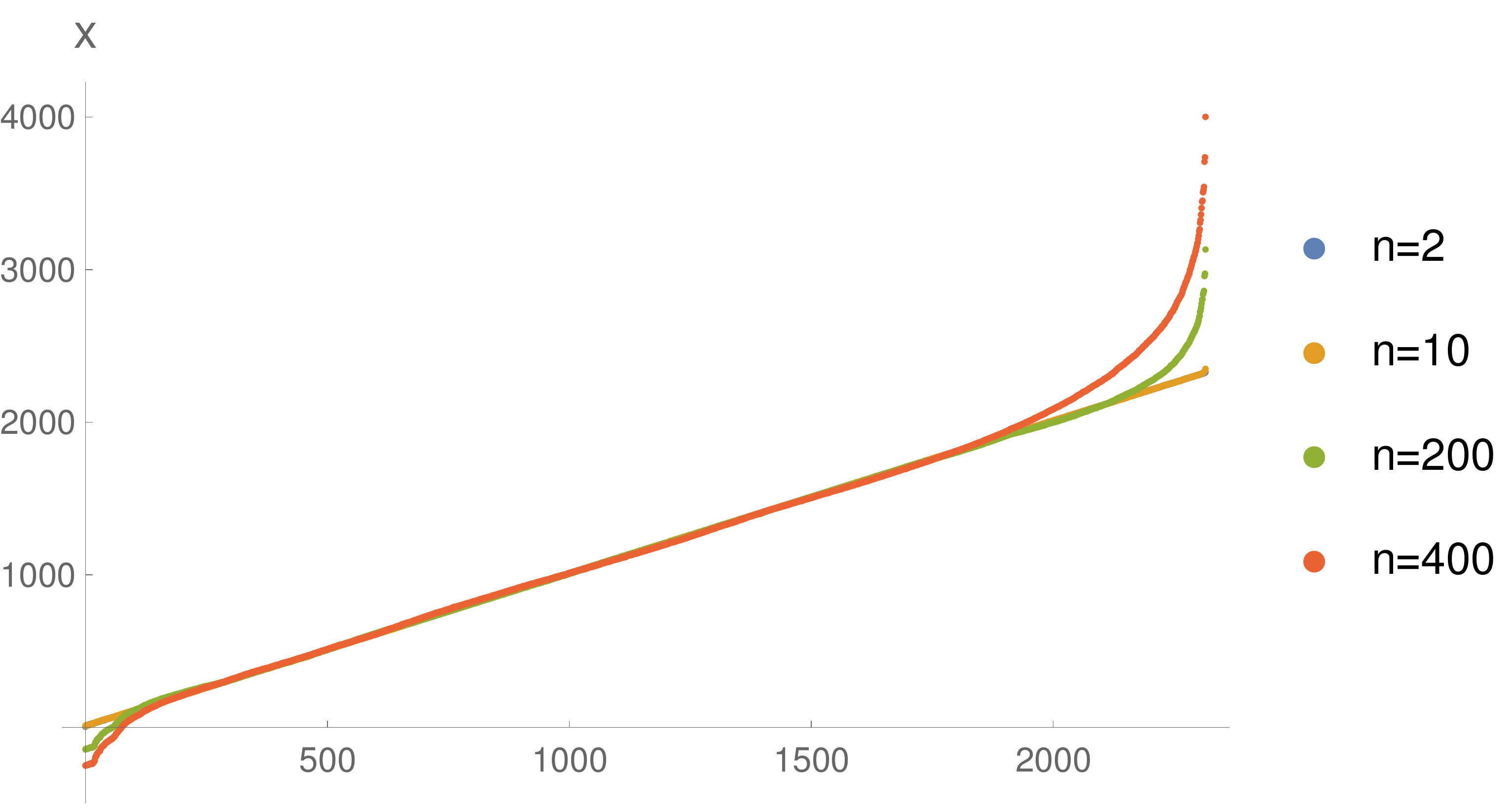}
\end{eqnarray}
	\caption{The unfolded spectrum of anomalous dimension for $L=16$, $M=4$, $N=17$, $\beta=0.9$ using linear interpolations based on choosing $n=2, 10,200, 400$. The different unfoldings are essentially identical and are all similar to the original spectrum. Here we include the $n$ extrapolated values at the each end of the spectrum where the difference in the unfolding procedure is large and which we neglect in our computations. }
	\label{fig:unf}
\end{figure}

However, without \textit{a priori} knowledge of the smooth or mean level density for a physical system we must use approximate methods to compute the unfolded spectrum. There does not appear to be an optimal procedure and so we use a relatively straightforward method. We select each $n$-th energy from the spectrum  $\{E_i\}$ and then perform a piece-wise linear interpolation  to define $I_{\text{av}}$. Fortunately the final result does not appear to be particularly sensitive to the choice of method. For example, we took $n=10$ but alternative choices such as $n= 2, 200, 400$ all give similar results, and so the values for the unfolded spectrum are likely reasonably robust, see \figref{fig:unf}. The procedure does cut-off the first and last $n$-elements and so has edge effects, however as we are interested in differences of energies the overall shift has no effect and the differences in the tails of the unfolded distribution do not modify the final results significantly.
In fact for the anomalous dimensions the unfolding process has only a very minor effect and could have been neglected. 
 
\bibliographystyle{nb}
\bibliography{Non-planar-dimensions}

\end{document}